\documentclass{article}
\pdfoutput=1

\usepackage[a4paper]{geometry}
\usepackage{graphicx}
\usepackage[textwidth=8em,textsize=small]{todonotes}
\usepackage{hyperref}
\hypersetup{colorlinks,allcolors=black}
\usepackage{amsmath}
\usepackage{natbib}
\usepackage{subfigure}
\usepackage{amsfonts}
\usepackage{amssymb}
\usepackage{mathtools}
\usepackage{multirow}
\usepackage{appendix}
\usepackage{indentfirst}
\usepackage[utf8]{inputenc}
\usepackage{float}

\usepackage{array}
\usepackage{wrapfig}
\usepackage{tabularx}
\usepackage{etoolbox}
\usepackage{bm}

\title{\bf Dynamical non-Gaussian modelling of spatial processes}

\author{Thaís C. O. Fonseca$^{1}$\footnote{{{\it Corresponding author}: Tha\'is C. O. Fonseca, Departamento de M\'etodos Estat\'{\i}sticos, Instituto de Matem\'atica, Universidade Fe\-de\-ral do Rio de Janeiro, Av. Athos da Silveira Ramos, Centro de Tecnologia, Bloco C, CEP 21941-909. \newline {\it E-mail}: {\tt thais@im.ufrj.br}. {\it Homepage}: https://sites.google.com/site/thaisf/}} , Viviana G. R. Lobo$^{1}$ and Alexandra M. Schmidt$^{2}$ \\
\textit{$^{1}$Instituto de Matem\'atica,} \\ 
\textit{Universidade Federal do Rio de Janeiro, Brazil } \\
\textit{$^{2}$Department of Epidemiology, Biostatistics and Occupational Health,} \\ \textit{McGill University, Canada }}

\date{}

\begin{document}

\maketitle

\begin{abstract}
Spatio-temporal processes in environmental applications are often assumed to follow a Gaussian model, possibly after some transformation. However, heterogeneity in space and time might have a pattern that will not be accommodated by transforming the data. In this scenario, modelling the variance laws is an appealing alternative. This work adds flexibility to the usual Multivariate Dynamic Gaussian model by defining the process as a scale mixture between a Gaussian and log-Gaussian processes. The scale is represented by a process varying smoothly over space and time which is allowed to depend on covariates.  State-space equations define the dynamics over time for both {mean} 
and variance processes resulting in feasible inference and prediction. Analysis of artificial datasets show that the parameters are identifiable and simpler models are well recovered by the general proposed model. 
The analyses of two important environmental processes,  maximum temperature and maximum ozone, 
illustrate the effectiveness of our proposal in 
 improving the uncertainty quantification in the prediction of spatio-temporal processes.\\
 
\noindent {\bf Keywords:}Forward filtering backwards sampling, Non-Gaussian models, Non-constant variance, Bayesian inference. 
 \end{abstract}


\section{Introduction}

\subsection{Variance patterns in space-time\label{mot}}


In many fields of science interest lies on extreme events such as large temperatures or ozone crossing a threshold. Often these processes are observed over space and time and common characteristics are non-normality of observations, presence of outliers or non-constant variance. These characteristics are even more noticeable if data are obtained through long temporal windows, in which case it is often unrealistic to assume that variances are constant for the whole period. In the context of environmental applications, even if seasonality is accounted for, it is rather common to observe changes in variance depending on the influence of air flows or ocean currents. This heterogeneity when not considered in the modelling might lead to poor predictions in out-of-sample locations or future time points.

\begin{figure}[H]
\centering
\begin{tabular}{cc}
\includegraphics[width=5cm]{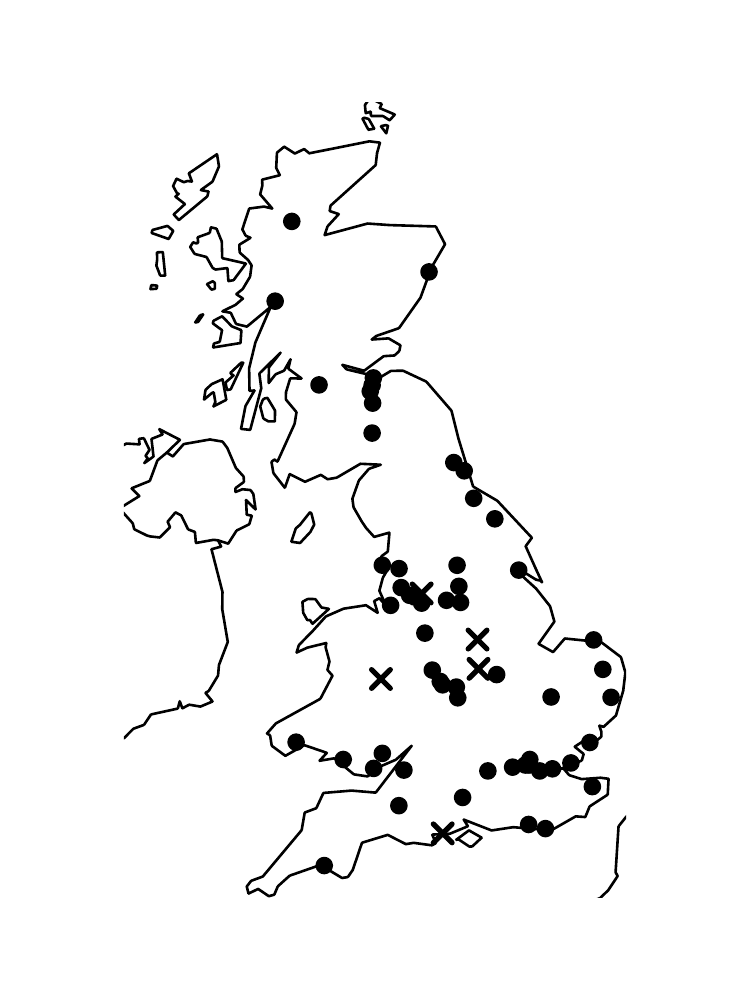} &
\includegraphics[width=5cm]{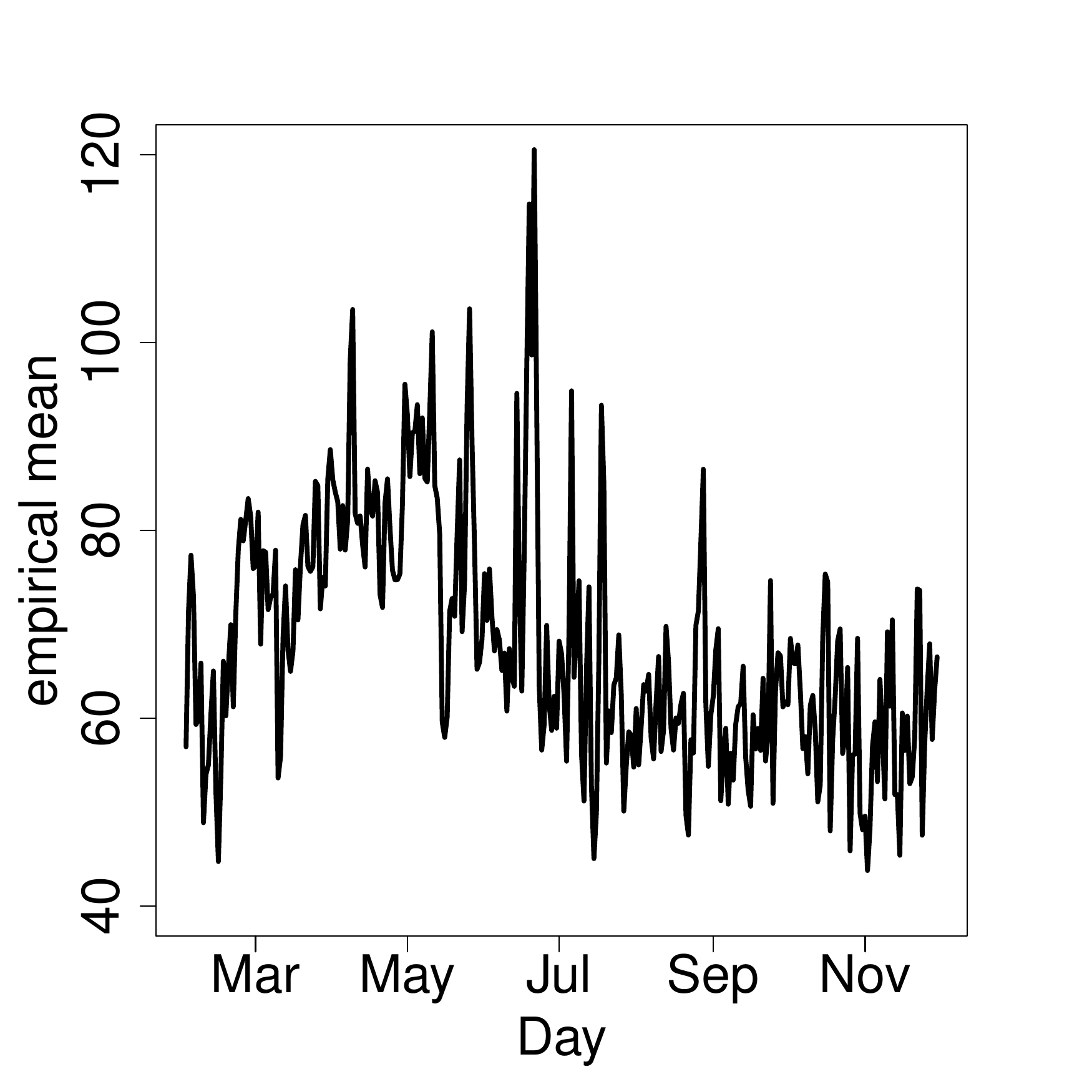} \\
(a) UK map and spatial locations. & (b) Empirical temporal mean. \\
 \includegraphics[width=5cm]{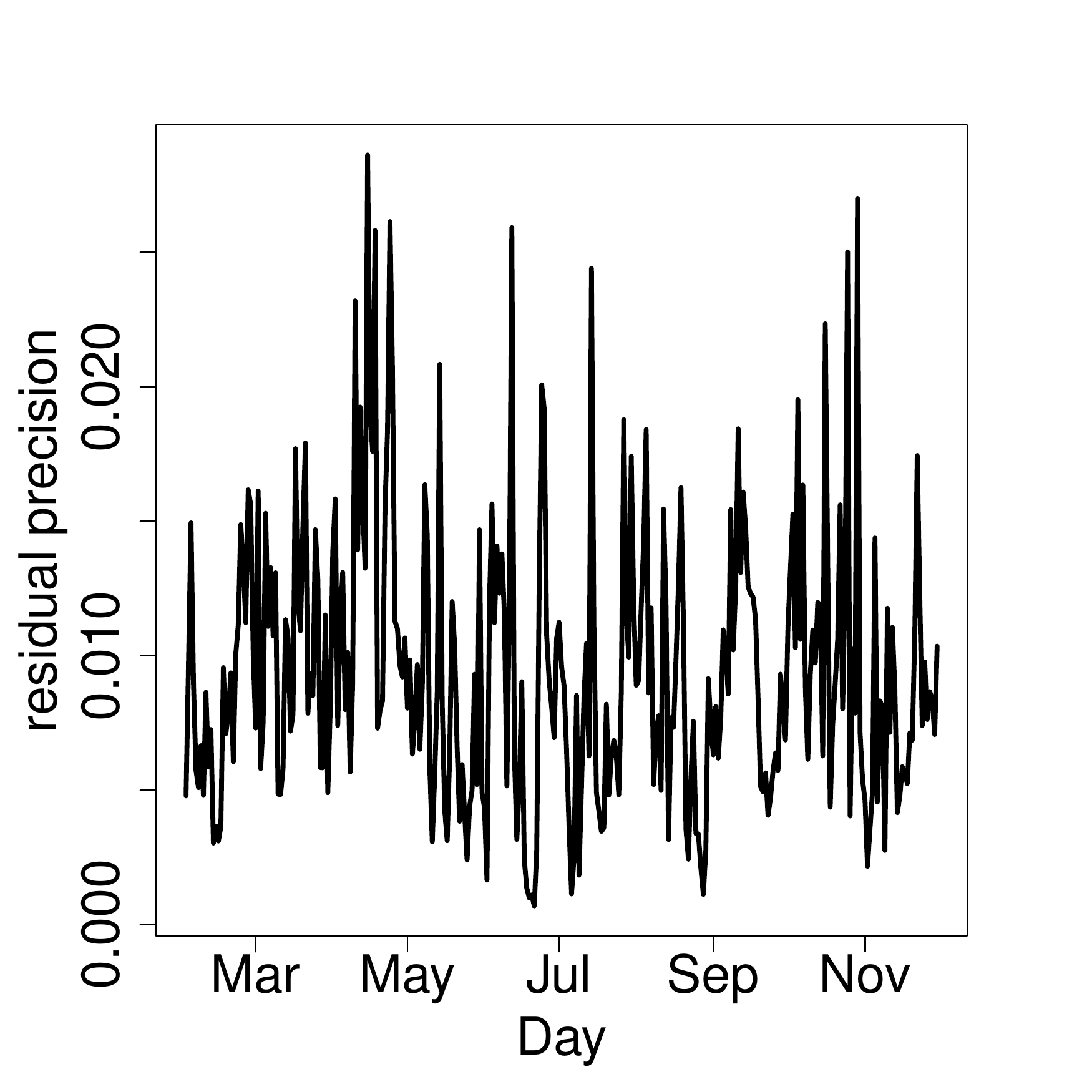}&   \includegraphics[width=5cm]{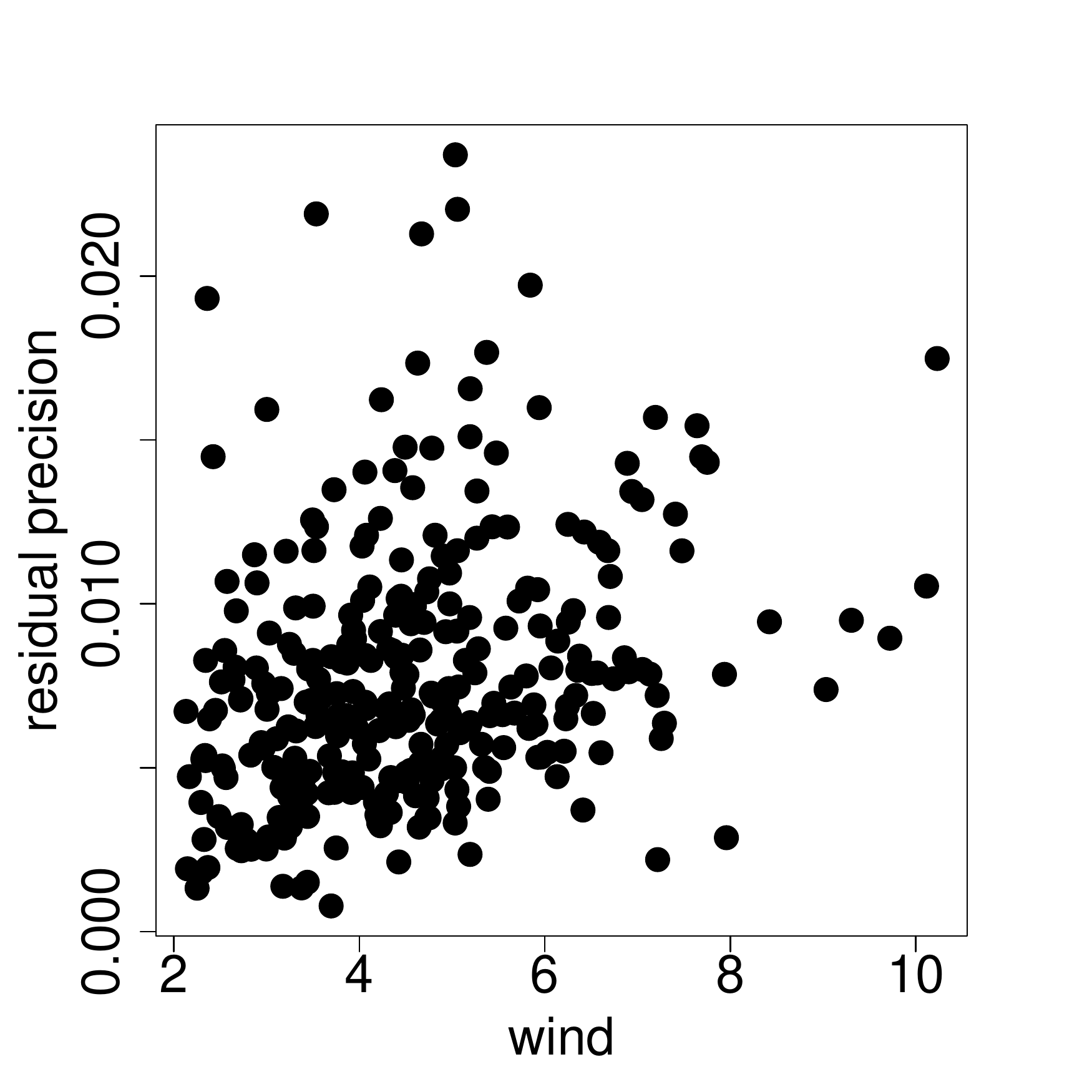}\\
\\
{(c) Residual temporal precision.} & {(d) Residual temporal precision versus wind.} \\
 \includegraphics[width=5cm]{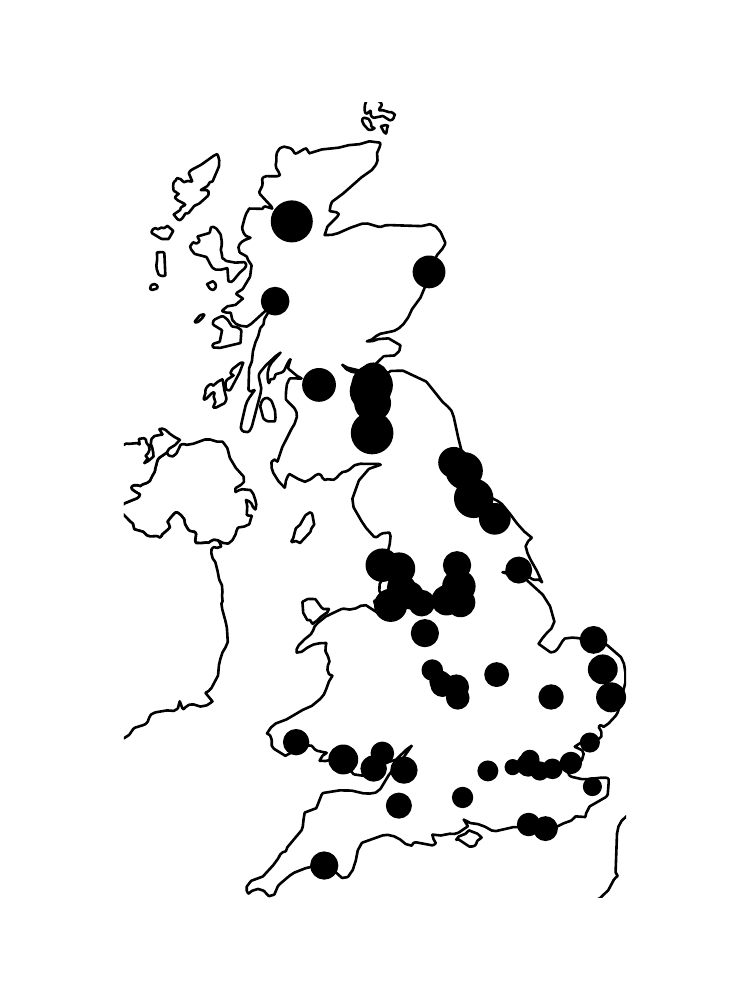}&   \includegraphics[width=5cm]{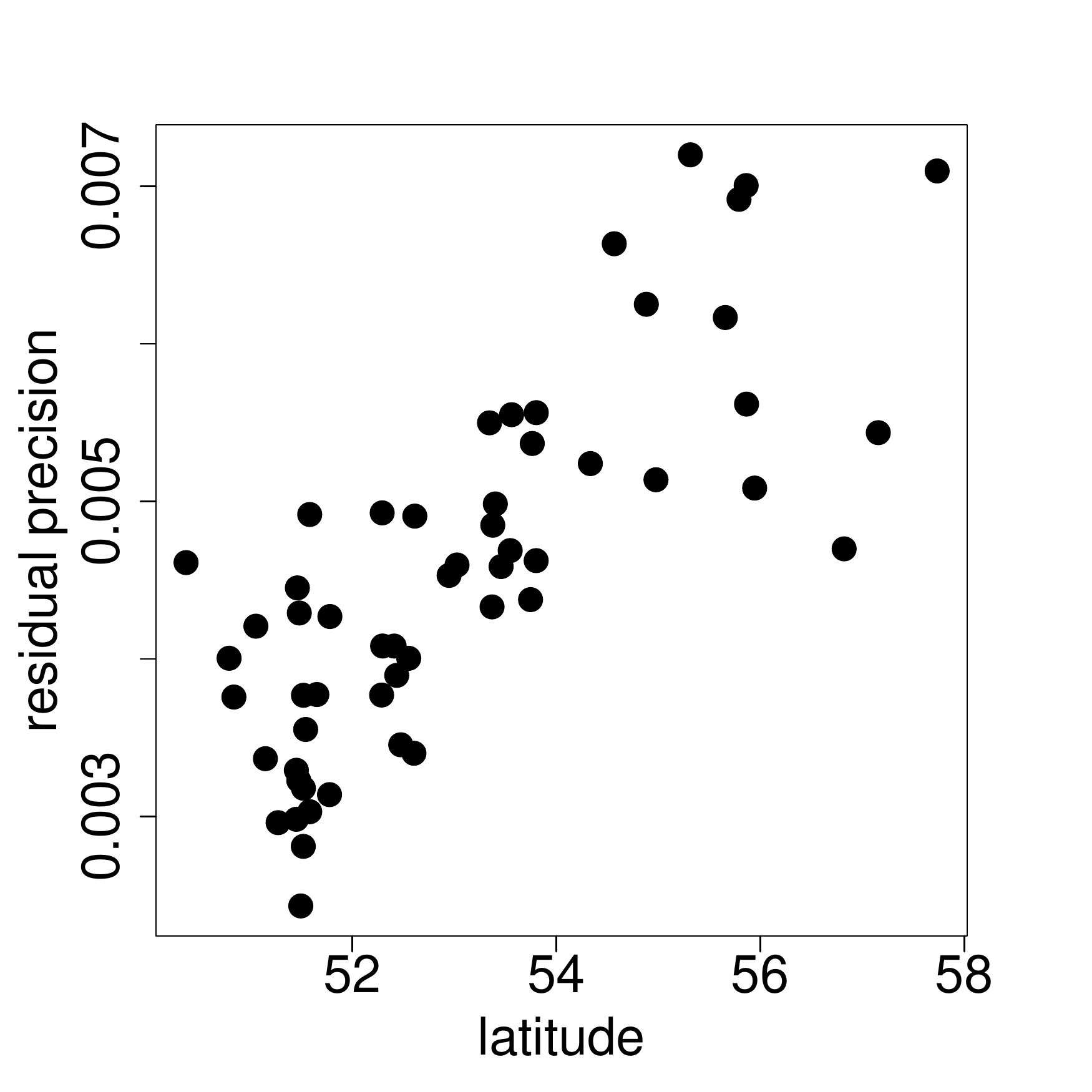}\\
\\
{(e) Residual spatial precision.} & {(f) Residual spatial precision versus latitude.} \\

\end{tabular}
\caption{Data summaries for the ozone data observed over the UK. Panel (a) displays the UK map with the training locations (solid circles) and the testing locations ($\bf{\times}$). Panel (b) presents the empirical mean over the year. Panels (c)-(f) present the precision over space and time of the residuals based on the fitting of a multivariate dynamic linear model.\label{figUK1} 
}
\end{figure}

To illustrate this characteristic of environmental processes, consider the daily maximum ozone data in the United Kingdom observed across 61 locations (Panel (a) of Figure \ref{figUK1}) from March to November of 2017. This period was chosen because it comprises  the highest levels of ozone (Panel (b) of Figure \ref{figUK1}). Ground level ozone is created by chemical reactions when pollutants emitted by cars, industry, to mention a couple of examples, react with sunlight. Moreover, high levels of ozone can also be found in rural areas due to wind transportation. It is well known that high levels of ozone can be harmful to human health and this problem has motivated several new modelling developments over the last years.
 
We  start by fitting a multivariate dynamic linear model (MDLM) to this data \citep{West97}. The mean structure of the MDLM  includes time varying effects of latitude, longitude, daily mean temperature and wind speed. In space, we assume a Cauchy correlation function \citep{Gneit00}, that is, $c(s,s')=   \left[1+ \left( ||s-s'||/\phi \right)^{\alpha} \right]^{-1}$ with 
$s,s'$ any two locations in $D$, $\phi >0 $ the spatial range parameter and $\alpha$ the shape parameter.
Panels (c) to (f) of Figure \ref{figUK1} show temporal and spatial residuals based on the MDLM fitting. Panel (c) presents the residual temporal precision, whereas panel (d) shows the scatter plot of wind speed versus the residual precision. It is clear that there is some temporal structure left in the residual of this fitted MDLM. Panels (e) and (f), on the other hand, show the spatial residual precision after fitting the MDLM. It is clear that there are smaller residual precisions in the south-eastern portion of the region, and a non-linear relationship of the spatial precision with latitude.

In this data the heterogeneity is mostly due to volatility in time with peaks of small precision (and large variance) in the months of June and July. This suggests that the proposed model should account for these patterns 
to explain the volatility of ozone observed across the different locations. 
In what follows we review some attempts to treat the volatility in spatiotemporal applications and present our proposed approach based on modelling the variance laws through a dynamic linear model.

\clearpage

\subsection{Related literature}

Several papers have investigated the presence of patterns in the variance of spatiotemporal processes and its effects on the predictive performance of the process of interest. 
\cite{Stein09} discusses the presence of peaks in the temporal variance in the modelling of atmospheric pressure even after including altitude in the mean function. In particular, the author suggests that the observed patterns is possibly due to the passage of weather fronts over the region. Often transformations such as the log or squared root are applied to the data aiming to stabilise the variance \citep{DeOliveira97,Johns03} or to account for truncated domains \citep{Allcroft03}. Recently, \cite{Gent17} proposed to add flexibility to the usual transformed Gaussian fields by considering a large family of possible transformations. However, the transformation approaches will not result in reasonable predictions if changes in variance have a pattern over time. That is, in many applications, even after fitting a Gaussian process to the data, the residuals still present varying variances which might depend on covariates which were already included in the mean function \citep[see e.g. ][]{Bueno2017}.
Moreover, 
the transformation approach may have difficult interpretations and may obscure the relationship between the response and the covariates \citep[see][for an example]{Bolin15}. In these situations, keeping the observations in their original scale and modelling the variance laws is an appealing alternative. 


\cite{Gelf2005} constructed a spatial model based on mixtures via a Dirichlet process which is non-stationary and non-Gaussian. \cite{Ge07} extend this idea to allow a random surface to be selected  in each site based on latent covariates. The approach is non parametric and replications are required for full inference, in which case dynamical models are considered to model temporal dependence. 
To account for outliers \cite{Bai15} 
propose an estimator to robustify the kriged Kalman filter, extending the spatio-temporal approach of \cite{Mardia98} which is highly affected by outlying observations. \cite{bevil20} propose a skew-t model for geostatistical data aiming to accommodate fat tails and asymmetric marginal distributions.

In the context of variance modelling, \cite{PSteel06} propose a non-Gaussian process for geostatistical data which accommodates fat tails by scale mixing a Gaussian process; the Gaussian model is a limiting case. This approach was extended by \cite{Fons11} to account for non-gaussianity in spatio-temporal processes. The proposed model for the variance is the product of two separable mixing processes, one in space by another in time and both are assumed continuous. 
\cite{Bueno2017} extended \cite{Fons11}  by allowing the use of covariate information
to explain the spatial patterns observed in the variances, and time is also assumed to vary continuously in $\mathbb{R}_+$. 
 More recently, \cite{Tadayon2018} propose a modelling approach that considers the use of covariates in the measurement error and can capture the effects of the skewness and heavy tails for datasets with non-Gaussian characteristics. \cite{Chu18} consider hierarchical modelling of Student-t processes with heterogeneous variance. The dynamic mean and variances depend on the lagged observations in time instead of past states.  

Note that, if time is assumed to be continuous in the variance model, correlation matrices will have large dimension and inference becomes too costly for reasonably long temporal windows. Thus, to allow for computational feasibility of real data applications, {different from \cite{Fons11}}, this paper considers discrete time and dynamic linear models for the spatio-temporal variance process.
This proposal modifies the well known multivariate dynamic linear model (MDLM) \citep{West97} which assumes Gaussianity to account for heterogeneity in spatio-temporal data analysis by modelling the variance laws over space and time. In the context of temporal evolution of variances, \cite{Uhlig94} extends the usual Gaussian dynamic model by including a sequential evolution for the precision matrix by assuming a Matrix-Beta evolution. An alternative specification considers the Wishart sequential filtering for the variance matrix. \cite{Liu00b} presents further discussion and model implementations for these proposals. In the context of more flexible state space models, \cite{Liu00} propose a conditional dynamical model specification which allows for non-Gaussian errors 
accounting for outliers. 
However, the model does not consider possible patterns in the variance model and the distributions are the same over time. 

\cite{West97} proposed to model variance laws  
by letting the observational variance to be a function of a known weight. In the context of spatial data, it is reasonable to assume that the weighting depends on  Euclidean distances and smoothness in space should be ensured.
In the usual multivariate dynamical modelling approach, the variance may vary stochastically as an inverse Gamma (or inverse Wishart) distribution, in which case the resulting sampling distribution for the response is Student-t. However, this extension is not flexible enough to capture spatial heterogeneity as discussed in \cite{PSteel06} and \cite{Fons11}. In our proposed solution to this issue, the variance is assumed to vary according to a log-Gaussian process \citep{PSteel06} and the mixing distribution varies in discrete-time assuming smooth transitions. Besides, the variance laws are allowed to depend on covariates. In this case, recurrence equations for filtering and smoothing are presented for the variance process allowing for feasible computations even for large temporal windows.

The remaining of the paper is organised as follows. Section \ref{sec:propmodel} describes the proposed model and its properties. In particular, Sections \ref{sec:2.2} and \ref{sec:2.3} describe the inference and prediction procedures for dynamical spatial modelling over time with stochastic variance. 
Sections \ref{sec:real} and  \ref{sec:real2} present the analysis of the maximum temperature in the Spanish Basque Country and the  maximum ozone levels in the United Kingdom, respectively. Different models are fitted and these analyses illustrate the effectiveness of our proposal in modelling varying variances over both time and space and the improvement it provides in the precision of predictions.   
Section \ref{sec:conclusion} concludes with some discussion. Some simulated examples are presented in  Appendix \ref{ApSimD} to verify that our proposed predictive comparison measures indicate the correct data generating models and do not result in overfitting.

\section{
Non-Gaussian state-space modelling }\label{sec:propmodel}

This section extends the multivariate Gaussian dynamic model by allowing for stochastic variance over space and time.

\subsection{Spatial Dynamic Linear Models with stochastic variance 
}\label{sec2.2}

Consider $\{Z_t(\mathbf{s}): \mathbf{s} \in D\subseteq \mathbb{R}^d, \, 
t\in T\subseteq \mathbb{Z}\}$ a spatio-temporal random field. We assume $Z_t(\mathbf{s})$ follows a spatial mixture model, that is,


\begin{equation}\label{model:eq2}
Z_t(\bm{s}) = \mathbf{x}_t(\mathbf{s})' \boldsymbol{\theta}_t + \sigma \frac{{\epsilon}_t(\mathbf{s})}{\sqrt{\lambda_t(\mathbf{s})}}+\tau  \rho_t(\mathbf{s}), 
\end{equation}
where $\mathbf{x}_{t}(\mathbf{s})$ is a vector of observed covariates, $\bm{\theta}_t$ is a vector of time varying regression coefficients, $\sigma$ is a scale parameter and $\tau$ is a nugget effect. The process  ${\epsilon}_t(\cdot)$ is a zero mean Gaussian process with correlation function $c(\cdot,\cdot)$, 
$\rho_t(\cdot)$ is an uncorrelated Gaussian noise with zero mean and unit variance responsible for small scale variation and $\lambda_t(\mathbf{s})$ is a mixing process. Conditionally on the mixing process $\lambda_t(\cdot)$ and on the coefficients $\boldsymbol{\theta}_t$, the process $Z_t(\cdot)$ has mean function $m_t(\mathbf{s})=\mathbf{x}_t(\mathbf{s})'\boldsymbol{\theta}_t$ and covariance function $K(\mathbf{s},\mathbf{s}')=\sigma^2c(\mathbf{s},\mathbf{s}')/\sqrt{\lambda_t(\mathbf{s})\lambda_t(\mathbf{s}')}$, $\mathbf{s},\mathbf{s}'\in D$, $t\in T$. For $\lambda_t(\mathbf{s})\not = 1$ the process $Z(\mathbf{s})$ has heterogeneous spatiotemporal variance and if $\lambda_t(s)$ is integrated out the resulting process is non-Gaussian. If $\lambda_t(\mathbf{s})=1$ and an evolution state equation is assumed for ${\boldsymbol{\theta}_t}$ then the resulting model is the usual Gaussian Dynamic Linear Model \citep{West97}. 

In the sequel, we discuss the mixing process specification which is the crucial part of this spatiotemporal mixture model. Assuming $\lambda_t(\mathbf{s}) =\lambda \sim {\rm Gamma}(v/2, v/2)$, $\thinspace \forall \thinspace \mathbf{s} \in D$ implies that the distribution of $Z_t(\cdot)$ is a Student-t process \citep{Omre06} with $v$ degrees of freedom. \cite{PSteel06} discuss the limitations of assuming a Student-t process for spatial observations. In short, the Student-t process is not able to account for spatial heterogeneity as it inflates the variance of the whole process whenever outliers or spatial heterogeneity is observed. 
On the other hand, if we assume the Gaussian-log-Gaussian (GLG) model proposed by \cite{PSteel06}, then ${\rm ln} \thinspace[ \lambda_t(\cdot)]$ is a Gaussian Process with mean $-\nu/2$ and covariance function $\nu c(\cdot,\cdot)$ such that ${\rm E}[\lambda_t(\mathbf{s}) ] = 1$, ${\rm Var}[\lambda_t(\mathbf{s})] = e^{\nu} -1$ and the kurtosis of the process $Z_t(\cdot)$ is $3e^{\nu}$ which is controlled by $\nu$. This implies that the marginal distribution of $\lambda_t(\mathbf{s})$ is concentrated around one for very small value of $\nu$ (of the order $\nu= 0.01$) and as $\nu$ increases, the distribution becomes more spread out and more right-skewed, while the mode becomes zero. 
Our proposed model extends the GLG specification by defining a model for $\lambda_t(\mathbf{s})$ through state space equations which assume that, conditionally on state parameters, the variances are independent in time, resulting in computationally efficient estimation algorithms. This approach takes advantage of the recurrence equations of DLM while accounting for more flexible variance laws for spatiotemporal data.  


Next we specify the dynamical evolution for both the mean and variance states and we discuss the connection between the usual Gaussian dynamic spatial model and the proposed non-Gaussian extension in equation (\ref{model:eq2}). Let $\mathbf{Z}_t= (Z_t(\bm{s}_1), \ldots, Z_t(\bm{s}_n))'$ be the data collected at $n$ spatial locations in $D$. Conditional on the latent variables $\Lambda_t=diag({\lambda}_t(\bm{s}_1),\ldots,{\lambda}_t(\bm{s}_n))$, the observation and system equations obtained by integrating $\epsilon_t(s)$ and $\rho_t(s)$ out are given by 
\begin{subequations}\label{eq5}
\begin{equation}\label{eq5a}
 \mathbf{Z}_t\mid \boldsymbol{\theta}_t,\Lambda_t \sim N\left ( \bm{F}_t' \boldsymbol{\theta}_t,\;\sigma^2  \Lambda_t^{-1/2}C_{{\bm \psi}}\Lambda_t^{-1/2}+\tau^2I_n\right ),
 \end{equation}
 \begin{equation}\label{eq5b}
 \boldsymbol{\theta}_t\mid \boldsymbol{\theta}_{t-1} \sim N\left ( { G}_t \boldsymbol{\theta}_{t-1} ,W_t\right ),
 \end{equation}
\end{subequations}
where $\bm{F}_t=(\mathbf{x}_t(\bm{s}_1),\ldots,\mathbf{x}_t(\bm{s}_n))$ is the $p \times n$ design matrix with observed $p$ covariates,  $\bm{\theta}_t$ is the $p-$dimensional state vector,  $C_{{\bm \psi}}$ represents the correlation matrix with elements computed by $C_{\bm {\psi},ij}=c(\bm{s}_i,\bm{s}_j)$ that depends on parameters ${\bm \psi}$ and the Euclidean distance among locations, $G_t$ represents the evolution matrix and $W_t$ is a $p$-dimensional covariance matrix of the states. 
Equation \eqref{eq5b} defines the temporal evolution of state variables in the mean function and the smoothness of this evolution is controlled by $W_t$.



We now focus on the specification of the spatio-temporal mixing process $\lambda_t(\bm{s})$, $\bm{s}\in D$, $t \in T$. To keep the model parsimonious, 
 we define $\lambda_t({\bm s}) = \lambda_1(\bm{s})\lambda_{2t}$ as a separable process. 
The mixing distributions and the evolution equation for the state space parameters in the variance model are defined as 
\begin{eqnarray}\label{eq:lambda1}
{\rm ln}(\mathbf{\boldsymbol{\lambda}}_1) & \sim &   N \left ( -\frac{\nu_1}{2}\bm{1}_n+\bm{F}_{1}'\boldsymbol{\beta},\nu_1 C_{{\bm{\xi}}}\right ), 
\end{eqnarray}
\begin{subequations}\label{eqL}
\begin{equation}\label{eq:lambda21}
{\rm ln}(\lambda_{2t}) =    \bm{F}_{2t}'\boldsymbol{\eta}_t+v_{2t},  \; v_{2t}\sim N \left ( -\frac{\nu_2}{2},\nu_2 \right ), 
\end{equation}
\begin{equation}\label{eq:lambda22}
\boldsymbol{\eta}_t  = G_{2t}\boldsymbol{\eta}_{t-1}+\mathbf{\omega}_{2t}, \; \mathbf{\omega}_{2t}\sim N \left ( 0,W_{2t} \right ),
\end{equation}
\end{subequations}
where, in equation (\ref{eq:lambda1}), ${\rm ln}(\mathbf{\boldsymbol{\lambda}}_1)= ({\rm ln}(\lambda(\bm{s}_1), \ldots, {\rm ln}(\lambda(\bm{s}_n))'$ and $C_{{\bm \xi}}$ the spatial correlation matrix that depends on parameter $\bm{\xi}$ and the Euclidean distance between locations. Note that $C_{\bm{\xi},ij}=c^*(\bm{s}_i,\bm{s}_j)$ which could differ from $c(\bm{s}_i,\bm{s}_j)$, that is, in the spatio-temporal context it is possible to estimate a different correlation structure for the process $\epsilon_t(\cdot)$ and the process ${\rm ln}[\lambda_1(\cdot)]$. In equation \eqref{eq:lambda1}, $\bm{F}_{1}=(\bm{\tilde x}(\bm{s}_1),\ldots, \bm{\tilde x}(\bm{s}_n))$  is a $p_1\times n$ design matrix that will allow for the effect of covariates in the spatial variance, and $\boldsymbol{\beta}$ is a $p_1$-dimensional vector of coefficients to be estimated. In equation \eqref{eq:lambda21}, $\bm{F}_{2t}=\bm{x}_{t}^*$ is a $p_2$-dimensional vector that will allow for the effect of covariates in the temporal variance. Equation \eqref{eq:lambda22} defines the temporal evolution of state parameters $\boldsymbol{\eta}_t$ in the variance model, with ${W}_{2t}$ controlling the temporal smoothness, and $G_{2t}$ representing the evolution matrix.  


 

The resulting covariance function of  $\left\{  Z_t(\bm{s}): \bm{s} \in D; t \in T \right\}$ 
, defined in \eqref{model:eq2}, is obtained by integrating out the mixing processes $\lambda_1(\bm{s})$ and $\lambda_{2t}$. 
If $t_1=t_2=t$ and $\bm{s}_1=\bm{s}_2=\bm{s}$ we obtain the spatio-temporal variance as \begin{equation}
    Var\left(Z_t({\bm s})\mid \boldsymbol{\eta}_{1:T},\boldsymbol{\theta}_{1:T}\right) = \sigma^2 \thinspace \exp\left\{\nu_1  +\nu_2  -\bm{F}_1'(\bm{s})\bm{\beta}-\bm{F}_{2t}'\boldsymbol{\eta}_{{t}}\right\},
\end{equation} 
with $F_1(\bm{s})=\tilde{\bm{x}}(\bm{s})$ the vector of spatial covariates at site $\bm{s}\in D$. The temporal dependence is carried out by the states $(\boldsymbol{\theta}_t,\boldsymbol{\eta}_t)$, $t=1,\ldots,T$ and the conditional spatial correlation is given by
\begin{equation}\label{eq:corr}
Corr\left[Z_{t}(\bm{s}_1), Z_{t}(\bm{s}_2)\mid \boldsymbol{\eta}_{1:T},\boldsymbol{\theta}_{1:T}\right] = C_{{\bm \psi}}(\mathbf{s}_1,\mathbf{s}_2)\exp\left\{ \frac{\nu_1}{4} \left( C_{{\bm \xi}}(\mathbf{s}_1,\mathbf{s}_2) -1 \right) \right\}. 
\end{equation}

The kurtosis in each location unconditional on $\lambda_t({\bm s})$ is given by
\begin{equation}\label{eq:kurt}
Kurt\left[Z_t({\bm s})\right] = 3 \thinspace \exp\left\{\nu_1 + \nu_2  \right\}.
\end{equation}
See Appendix \ref{ApB} for the proofs of these results. A particular case of the model proposed in equation \eqref{model:eq2} is obtained for $\lambda_t(\mathbf{s})= 1$ and, consequently, the non-Gaussian distribution converges to the Gaussian distribution for small values of $\nu_1$ and  $\nu_2$. 


\subsection{Resultant posterior distribution and inference procedure}\label{sec:2.2}

We follow the Bayesian paradigm to make inference, predictions and model comparisons that are obtained from the joint posterior distribution of the parameters. In particular, we take advantage of the hierarchical structure of our proposal in our iterative estimation algorithm to sample from the joint posterior and to make predictions. In what follows we present the prior, the joint posterior distributions and briefly describe the steps to obtain samples from the posterior distribution. 

In our motivating example, we assume a Cauchy correlation function with range parameter $\phi>0$ and shape parameter $\alpha>0$. This function is flexible allowing for long-memory dependence and also correlations at short and intermediate lags. 
We assume an exponential correlation function for the spatial mixing process $ln[\lambda_1(\bm{s})]$ given by $c^*(\bm{s},\bm{s}') = \exp\left\{ -||\bm{s}-\bm{s}'||/\gamma\right\}$, where $\gamma> 0$.  
Model specification is complete after assigning a prior distribution for the static parameters $\Phi=(\sigma^2,\tau^2,\nu_1,\nu_2, {\boldsymbol{\beta}}, {\bm \psi}=(\alpha,\phi), {\bm \xi}=(\gamma))$. We assign vague independent priors to the static parameters in $\Phi$. 
 In particular, we assume $\sigma^{-2} \sim Gamma(a_{\sigma^2},b_{\sigma^2})$ with small values for $a_{\sigma^2}$ and $b_{\sigma^2}$. 
For the range parameter $\phi$, we take into account that the prior is critically dependent on the scale of the observed distances among locations. For the Cauchy correlation function, we assign a gamma prior $\phi$, i.e. $\phi \sim Gamma\left(1, c/med(d) \right)$, with $med(d)$ representing the median of observed distances and the shape parameter follows a uniform prior, that is, $\alpha \sim U(a_{\alpha};b_{\alpha})$. For the exponential correlation function parameter, we assume $\gamma \sim Gamma(a_{\gamma}, b_{\gamma})$. For the mixing parameters $\nu_i$, $i=1,2$, we assign a $Gamma(a_\nu,b_{\nu})$ prior. Notice that very small values of $\nu_i$ (around 0.01) lead to approximate normality while large values of $\nu_i$ (of the order of say 3) suggest very thick tails. 

Following Bayes’ theorem, the posterior distribution of model parameters and latent variables given the observed data, $\mathbf{z}_t = (z_t(\bm{s}_1), \ldots, z_t(\bm{s}_n))'$, $t=1,\ldots,J$,  is proportional to 
\begin{eqnarray}\label{eq:post} \nonumber
    p\left( \boldsymbol{\theta}_{1:J}, \boldsymbol{\eta}_{1:J} \boldsymbol{\lambda}_{1},\boldsymbol{\lambda}_{2}, \Phi  \mid \mathbf{z}\right) &\propto& \prod_{t=1}^{J}  f_{N_{n}}(\mathbf{z}_t|\bm{F}'_t\boldsymbol{\theta}_t,\Sigma_t)  \nonumber \\ 
& \times &  f_{N_{n}}(\boldsymbol{\Delta}|{\bf 0},\nu_1 C_{\bm{\xi}}) \prod_{t=1}^{J} f_{N_{1}}(L_t|\bm{F}'_{2t}\boldsymbol{\eta}_t,\nu_2) \nonumber \\ 
    & \times &  f_{N_{p}}(\boldsymbol{\theta}_0\mid \bm{m}_0,C_0) \prod_{t=1}^{J} f_{N_{p}}(\boldsymbol{\theta}_t \mid \boldsymbol{\theta}_{t-1},W_t)  \\
 & \times & f_{N_{p_{2}}}(\boldsymbol{\eta}_0\mid \bm{m}_0^*,C_0^*) \prod_{t=1}^{J} f_{N_{p_{2}}}(\boldsymbol{\eta}_t \mid \boldsymbol{\eta}_{t-1},W_{2t}) \;  \pi(\Phi), \nonumber
\end{eqnarray}
\noindent where $f_{N_{p}}(\cdot\mid A,B)$ denotes the density function of a $p$-variate multivariate normal distribution with mean A and covariance matrix B, $\boldsymbol{\Delta}= ln (\boldsymbol{\lambda}_1) +  \nu_1/2 \,  {\bf 1}_n  {- \bm{F}_1'\boldsymbol{\beta}}$, ${L}_t= ln \lambda_{2t} + \nu_2/2$, $\Sigma_t = \sigma^2 \Lambda_t^{-1/2} C_{\bm{\psi}} \Lambda_t^{-1/2} + \tau^2 I_n$, $\Lambda_t = diag(\lambda_1(\bm{s}_1), \ldots, \lambda_J(\bm{s}_n))$ and  $\pi(\cdot)$ the prior distribution of static parameters. Finally, $f_{N_{p}}(\boldsymbol{\theta}_0\mid \bm{m}_0,C_0)$ and $f_{N_{p_{2}}}(\boldsymbol{\eta}_0\mid \bm{m}_0^*,C_0^*)$ are the densities for the initial prior information at time $t=0$ for $\boldsymbol{\theta}_0$ and $\boldsymbol{\eta}_0$, respectively.

The resultant posterior distribution does not have closed form and we resort to Markov chain Monte Carlo methods \citep{gamerman} to obtain samples from the posterior. In particular, posterior samples are obtained through a Gibbs sampler algorithm with  steps of the Metropolis-Hastings algorithm for $\phi$, $\alpha$, $\gamma$ and $\nu_i$, $i=1,2$  which are based on random walk proposals.



\paragraph{Brief description of the MCMC algorithm} Conditional on the latent variables $\boldsymbol{\lambda}_1$ and $\boldsymbol{\lambda}_2$, Gaussianity is preserved and samples from the posterior full conditional distributions for the state vectors $\boldsymbol{\theta}_t$ in the mean are obtained through the usual forward filtering and backward smoothing recursions (FFBS) proposed by \cite{Sylvia1994} and \cite{Carter1994}.
Analogously, conditionally on $\boldsymbol{\lambda}_2$, the posterior distribution of states $\boldsymbol{\eta}_t$ are also obtained through the FFBS algorithm. 
Appendices \ref{ApA1} and \ref{ApA2} provide the equations to run the FFBS for  $\boldsymbol{\theta}_t$ and $\boldsymbol{\eta}_t$, respectively. Note that, 
different from \cite{Bueno2017} and \cite{Fons11}, as we assume time to be discrete, we do not need to rely on computing the inverse of high dimensional  covariance matrices at each iteration of the MCMC. 
Conditional on the regression coefficients $\bm{\beta}$, the spatial latent mixing variable  $\boldsymbol{\lambda}_{1}=(\lambda_{1}(\bm{s}_1),\cdots,\lambda_{1}(\bm{s}_n))'$ is sampled as part of a Gibbs algorithm using blocks of random walks or the independent sampler proposed in \cite{PSteel06}. To sample $\bm{\beta}$ the Gibbs step is given by
\begin{equation}
    \bm{\beta}\mid \bm{\lambda}_1,\nu_1,\bm{\xi}\sim N_{p_1}((\bm{F}_1'C_{\bm{\xi}}^{-1}\bm{F}_1)^{-1}\bm{F}_1'C_{\bm{\xi}}^{-1}(ln \bm{\lambda}_1+\nu_1/2\; \bm{1}_n),\nu_1(\bm{F}_1'C_{\bm{\xi}}^{-1}\bm{F}_1)^{-1}).
\end{equation}
A summary of our proposed sampling algorithm is described in Appendix \ref{ApPostComp}. The algorithm was coded in {\tt R} using RStudio 
Version 1.1.442 \citep{RCoreTeam} and the source code can be obtained at: https://github.com/thaiscofonseca/DynGLG.

\subsection{Predictions in space-time}\label{sec:2.3}


For spatial interpolation {for given observed times}, consider the vector $(\mathbf{Z}_t^{obs}, \mathbf{Z}_t^{pred})$, with $\mathbf{Z}_t^{obs}$ and $\mathbf{Z}_t^{pred}$ representing, respectively, observed and out-of-sample values of $Z_t(\bm{s})$, at each time $t= 1, \ldots, J$. Let $\Phi= ( \sigma^2,\tau^2, \nu_1,\nu_2, \bm{\beta}, \bm{\psi},\bm{\xi})$ be the static parameters in the proposed model in equation \eqref{eq5}. 
In order to obtain samples from the posterior predictive distribution $p(\mathbf{Z}_t^{pred} \mid \mathbf{Z}_t^{obs})$ we resort to composition sampling;  assume that $\Phi$, 
$\bm{\lambda}_{1}^{obs}=(\lambda_{1}(\bm{s_1}),\ldots,\lambda_{1}(\bm{s}_n))'$, $\bm{\lambda}_{2}^{obs}=(\lambda_{2,1},\ldots,\lambda_{2J})'$, $\boldsymbol{\theta}^{obs}=(\bm{\theta}_{1},\ldots,\bm{\theta}_J)'$, $\boldsymbol{\eta}^{obs}=(\bm{\eta}_{1},\ldots,\bm{\eta}_J)'$ were sampled from the joint posterior distribution $p(\Phi, \bm{\lambda}_{1}^{obs},\bm{\lambda}_{2}^{obs},\boldsymbol{\theta}^{obs},\boldsymbol{\eta}^{obs} \mid \mathbf{Z}_t^{obs})$. 
Thus, samples from $p(\mathbf{Z}_t^{pred} \mid \mathbf{Z}_t^{obs})$ may be obtained by sampling
\begin{enumerate}
    \item[(i)] ${\rm ln}(\bm{\lambda}_{1}^{pred})\mid \bm{\lambda}_{1}^{obs},\nu_1, \bm{\xi} $ and \item[(ii)] $\bm{Z}_t^{pred}\mid \bm{Z}_t^{obs}, \bm{\lambda}_{1}^{pred},\bm{\lambda}_2^{obs},\bm{\theta}^{obs},\Phi$. 
\end{enumerate}
Both distributions in (i) and (ii) are Gaussian, the second is the observational model and the first is given by
\begin{equation}\label{eqpredL1}
{\rm ln}(\bm{\lambda}_1^{pred}) \mid \bm{\lambda}_1^{obs} , \nu_1, \bm{\xi}  \sim  N_n \left[ -\nu_1/2 \; {\bf 1}_n+{C}_{o,p} {C}_{o,o}^{-1} \mathbf{a} ;   \nu_1 \left({C}_{p,p} - {C}_{p,o}{C}_{o,o}^{-1}{C}_{o,p} \right)\right] 
\end{equation}
\noindent with $\mathbf{a} = \left(ln(\bm{\lambda}_1^{obs}) +\nu_1/2 \; {\bf 1}_n - {\bm{F}'_1 \boldsymbol{\beta}}\right)$ and
${C}_{\bm{\xi}} = \begin{pmatrix} 
C_{p,p} & C_{p,o} \\
C_{o,p} & C_{o,o} 
\end{pmatrix}$. {This result follows from the properties of the partition of the multivariate normal distribution.}


Suppose now that interest lies in forecasting future observations at a set of locations given historical data $\bm{Z}^{obs}=(Z_{1}^{obs},\ldots,Z_J^{obs})'$. Consider that at time $J$ we want to predict $h$ instants ahead and $h>0$. We define  $\bm{\lambda}_{2}^{pred}=(\lambda_{2,J+1},\ldots,\lambda_{2,J+h})'$. Thus, samples of $\bm{Z}_t^{pred}$ may be obtained by sampling from
\begin{enumerate}
    \item[(i)] $\bm{\eta}^{pred},\mid \bm{\eta}^{obs},\bm{\lambda}_{2}^{obs} $ and $\bm{\theta}^{pred}\mid \bm{\theta}^{obs},\bm{Z}^{obs} $,
    \item[(ii)] $ln(\bm{\lambda}_{2}^{pred})\mid \bm{\eta}^{pred},\nu_2 $ and 
    \item[(iii)] $\bm{Z}_t^{pred}\mid \bm{Z}_t^{obs}, \bm{\lambda}_{1}^{obs},\bm{\lambda}_2^{pred},\bm{\theta}^{pred},\Phi$. 
\end{enumerate}
If we are predicting in the future for ungauged locations we replace $\bm{\lambda}_1^{obs}$  with $\bm{\lambda}_1^{pred}$ obtained using equation (\ref{eqpredL1}). Steps (ii) and (iii) are performed simulating from the variance and observational models which are all conditionally Gaussian distributions. Step (i) depends on the usual forecast distributions available for the Gaussian Multivariate Dynamical Model \citep{West97}. 


\paragraph{Model Comparison}
To check the predictive accuracy of competing models, measures based on scoring rules are considered. Scoring rules provide summaries for the evaluation of probabilistic forecasts by comparing the predictive distribution with the actual value which is observed for the process \citep{GneitRaf07}. In particular, we consider the Interval Score, the Logarithmic Predictive Score and the Variogram Score. Note that the Logarithmic Predictive Score and the Variogram Score are multivariate measures for a $d$-dimensional vector. { We briefly describe how to compute each of these criteria. }\\

\noindent \textit{Interval Score:}  Interval forecast is a crucial special case of quantile prediction \citep{GneitRaf07}. It compares the predictive credibility interval with the true observed value (validation observation), and it considers the uncertainty in the predictions such that the model is penalised if an interval is too narrow and misses the true value. The Interval Score is given by 
\begin{equation}\label{eq:IS}
IS(u,l;z) = (u-l) + \frac{2}{\gamma}(l - z)I_{\left[z< l\right]}  + \frac{2}{\gamma}( z - u)I_{\left[z> u\right]},
\end{equation}
\noindent where $l$ and $u$ represent for the forecaster quoted $\frac{\gamma}{2}$ and $1-\frac{\gamma}{2}$ quantiles based on the predictive distribution and $z$ is the validation observation. If $\gamma=0.05$ the resulting interval has 95\% credibility.\\

\noindent \textit{Log Predictive Score:} The log predictive score evaluates the predictive density at the observed validation value $\bm{z}$. It is given by  
\begin{equation}\label{eq:LPS}
    LPS(\bm{z})= -log\left\{p(\bm{z}\mid \bm{z}^{obs})\right\}.
\end{equation}
The smaller the log predictive score, the better the model does at forecasting $\bm{z}^{obs}$.\\

\noindent \textit{Variogram Score:}
The variogram score of order $p$ \citep{Sche15} was proposed to evaluate forecasts of multivariate quantities. It depends on a matrix $w$ of non-negative weights specified subjectively that allow to emphasize or downweight pairs of observations, for instance, based on Euclidean distances. It is defined as 
\begin{equation}
   \mbox{VS-p}(\bm{z},\bm{z}^{obs}) = \sum_{i,j=1}^{d} w_{ij}\left (|\bm{z}^{obs}_{i}-\bm{z}^{obs}_j|^p-\frac{1}{m}\sum_{k=1}^M|\bm{z}_i^{(k)}-\bm{z}^{(k)}_j|^p\right )^2,
\end{equation}
where $\{\bm{z}^{(k)};\, k=1,\ldots,M\}$ are simulated values from the predictive distribution. The smaller the variogram score, the better the model does at forecasting $\bm{z}^{obs}$. Empirical studies presented in \cite{Sche15} suggest that $p=0.5$ leads to good model discrimination, however, if the predictive distribution is skewed, then values of $p<0.5$ may lead to better results.\\


\section{Data analysis}\label{sec4}

This section presents two data analyses relevant in the discussion about extremes in environmental applications: the first application considers the maximum temperature data in the Spanish Basque Country. These data have been previously analysed by \cite{PSteel06}, \cite{Fons11} and \cite{Bueno2017}. As our proposal is able to account for longer temporal windows than \cite{Fons11}, the analysis shown in Section \ref{sec:real} considers one year of daily observations instead of one month as in \cite{Fons11} and \cite{Bueno2017}. The second application focuses on the maximum ozone data described in Section \ref{mot}, which illustrates the use of spatial and temporal covariates in the variance model.
We define $\lambda_t(\bm{s})=\lambda_1(\bm{s})\lambda_{2t}$ and based on equations (\ref{eq5}) and (\ref{eqL}) we fit the models described in Table \ref{tabmodels} which are particular cases of the general model proposed in the previous section.
    

\begin{table}
  \caption{Competing models fitted to data applications: Gaussian (G), Student-t (ST), Spatial GLG (GLG), Dynamical (Dyn), Dynamical with covariates (CovDyn), Dynamical GLG (DynGLG), Dynamical GLG with covariates (CovDynGLG) and the complete model (Full).  \label{tabmodels}}
\centering
  \fbox{  \begin{tabular}{lll}
    \hline
  Model  & $\lambda_1(\bm{s})$ & $\lambda_{2t}$\\
    \hline
G & $1$ & $1$ \\
ST & $\lambda \sim Gamma(\nu_1/2,\nu_1/2)$ & $1$\\
GLG & 
    $ln (\bm{\lambda}_{1})\sim N\left (-\frac{\nu_1}{2} \; \mathbf{1}_n , \nu_1 C_{\bm{\xi}}\right )$ & $1$\\ 
Dyn & $1$ & $ln(\lambda_{2t})\sim N\left (-\frac{\nu_2}{2} +\eta_{0t}, \nu_2\right )$\\
CovDyn & $1$ &  $ln(\lambda_{2t}) \sim N\left (-\frac{\nu_2}{2}+\bm{F}_{2t}'\bm{\eta}_t, \nu_2\right )$\\
DynGLG &  
     $ln (\bm{\lambda}_{1})\sim N\left (-\frac{\nu_1}{2} \; \mathbf{1}_n, \nu_1 C_{\bm{\xi}}\right )$ & $ln(\lambda_{2t})\sim N\left (-\frac{\nu_2}{2}+\eta_{0t}, \nu_2\right )$\\
CovDynGLG & $ln (\bm{\lambda}_{1})\sim N\left (-\frac{\nu_1}{2} \; \mathbf{1}_n, \nu_1 C_{\bm{\xi}}\right )$ & $ln(\lambda_{2t}) \sim N\left (-\frac{\nu_2}{2}+\bm{F}_{2t}'\bm{\eta}_t, \nu_2\right )$\\
 Full & $ln (\bm{\lambda}_{1})\sim N\left (-\frac{\nu_1}{2}\; \mathbf{1}_n+\bm{F}_1'\bm{\beta} , \nu_1 C_{\bm{\xi}}\right )$ & $ln(\lambda_{2t}) \sim N\left (-\frac{\nu_2}{2}+\bm{F}_{2t}'\bm{\eta}_t, \nu_2\right )$\\
\hline
    \end{tabular}}
\end{table}

\subsection{Application to temperature data in the Spanish Basque Country}\label{sec:real}

\begin{figure}
\centering
\begin{tabular}{cc}
\includegraphics[width=5.5cm]{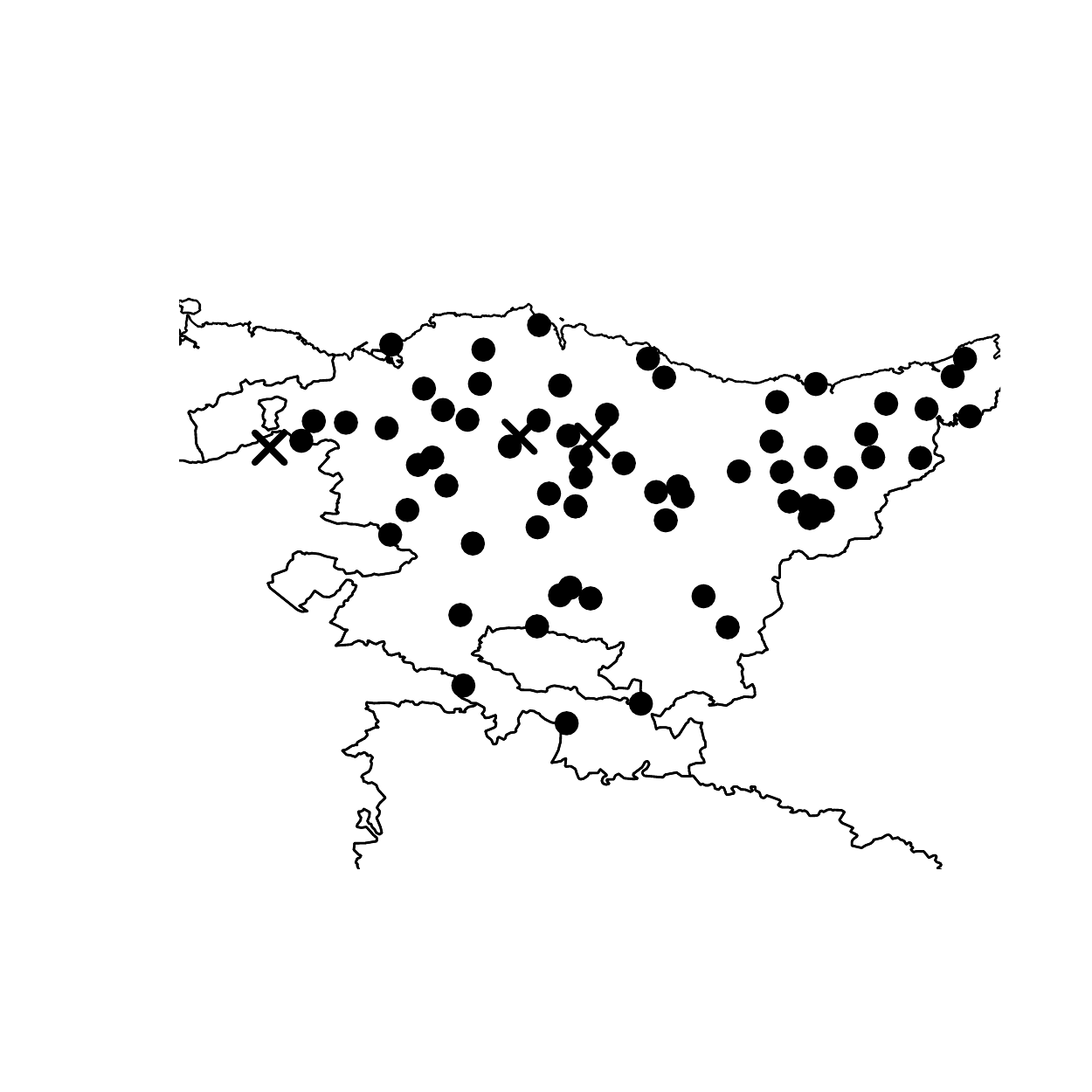} 
&
\includegraphics[width=4.6cm]{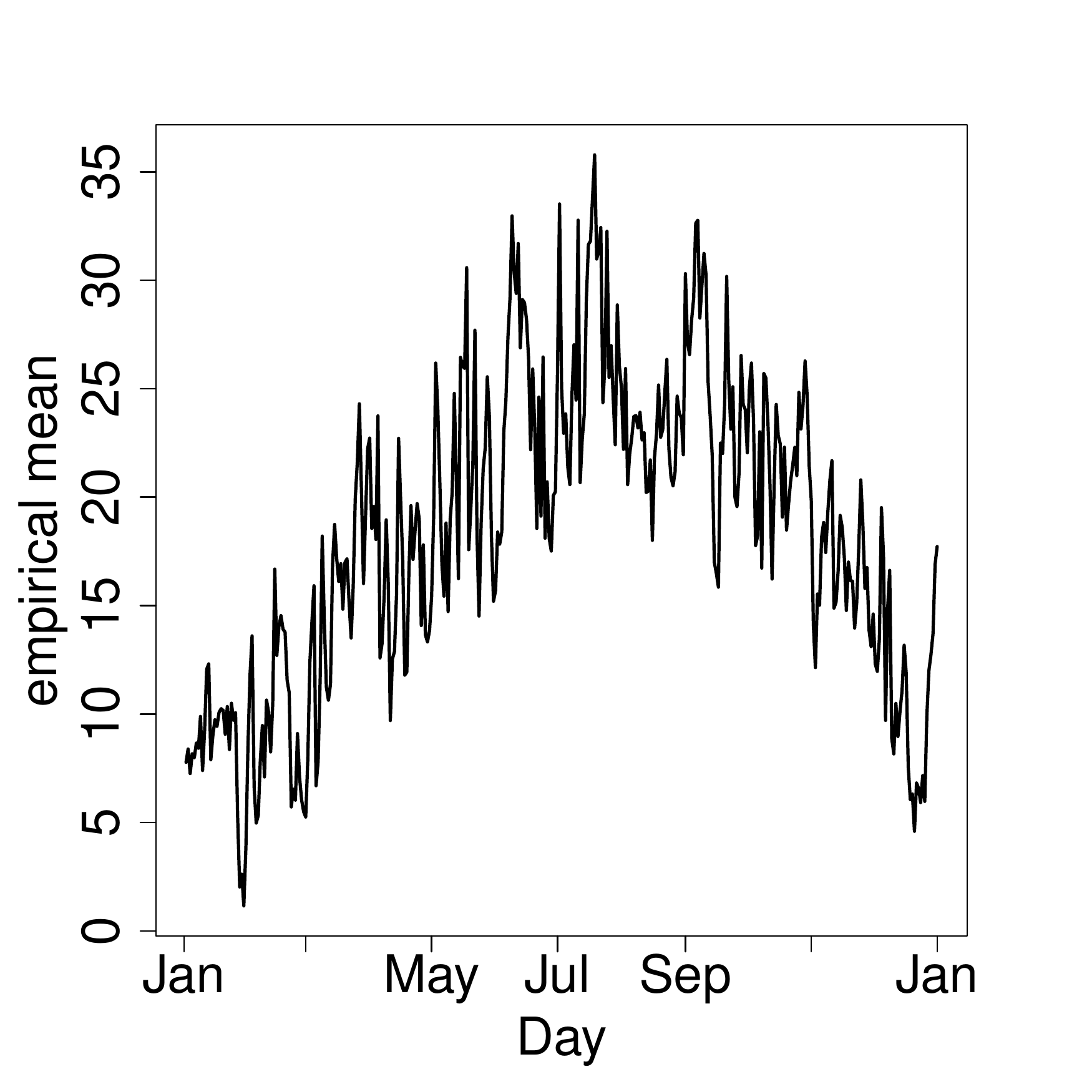} \\
(a) Spain Map and spatial locations & (b) Empirical temporal mean  \\ 

  \includegraphics[width=4.6cm]{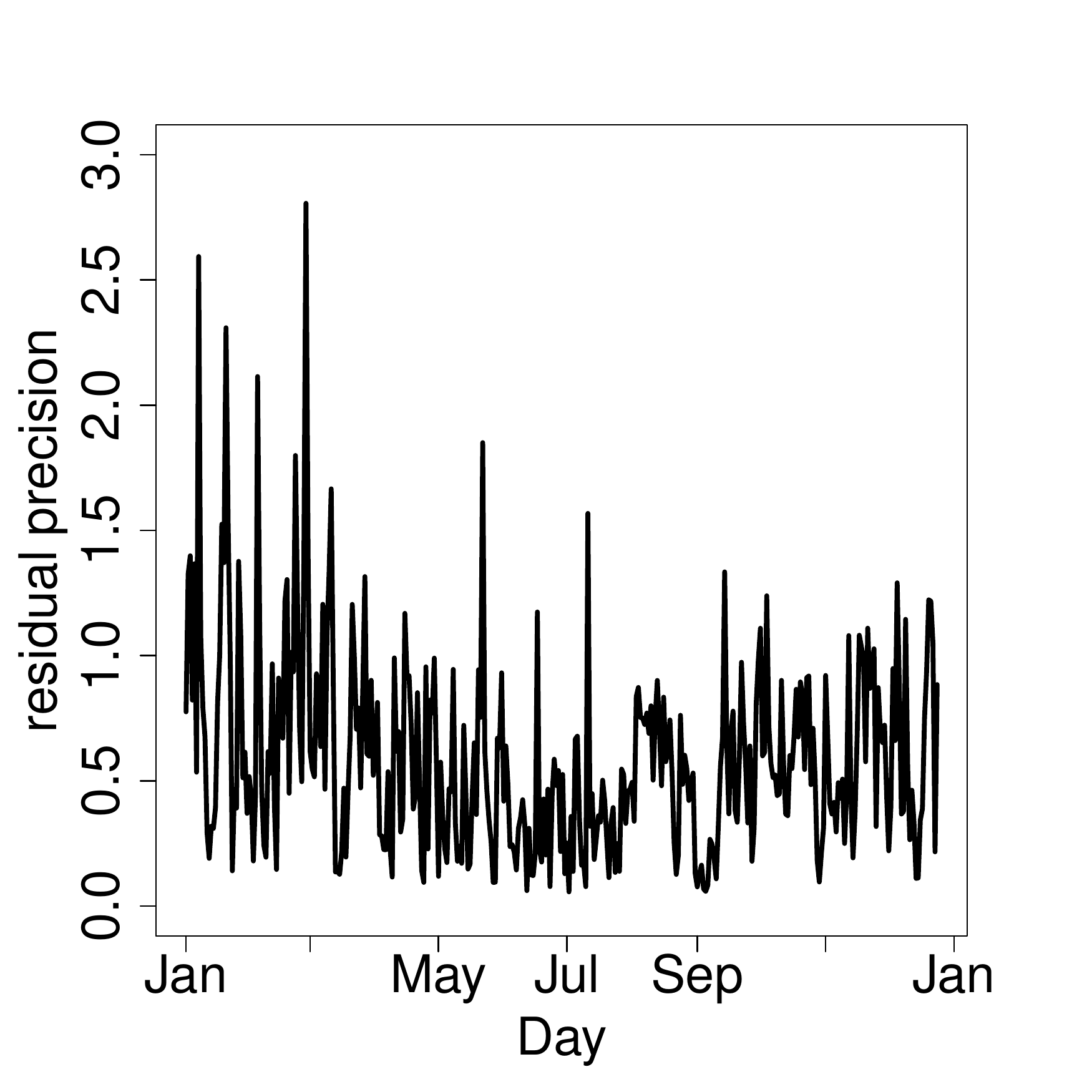} &\includegraphics[width=4.6cm]{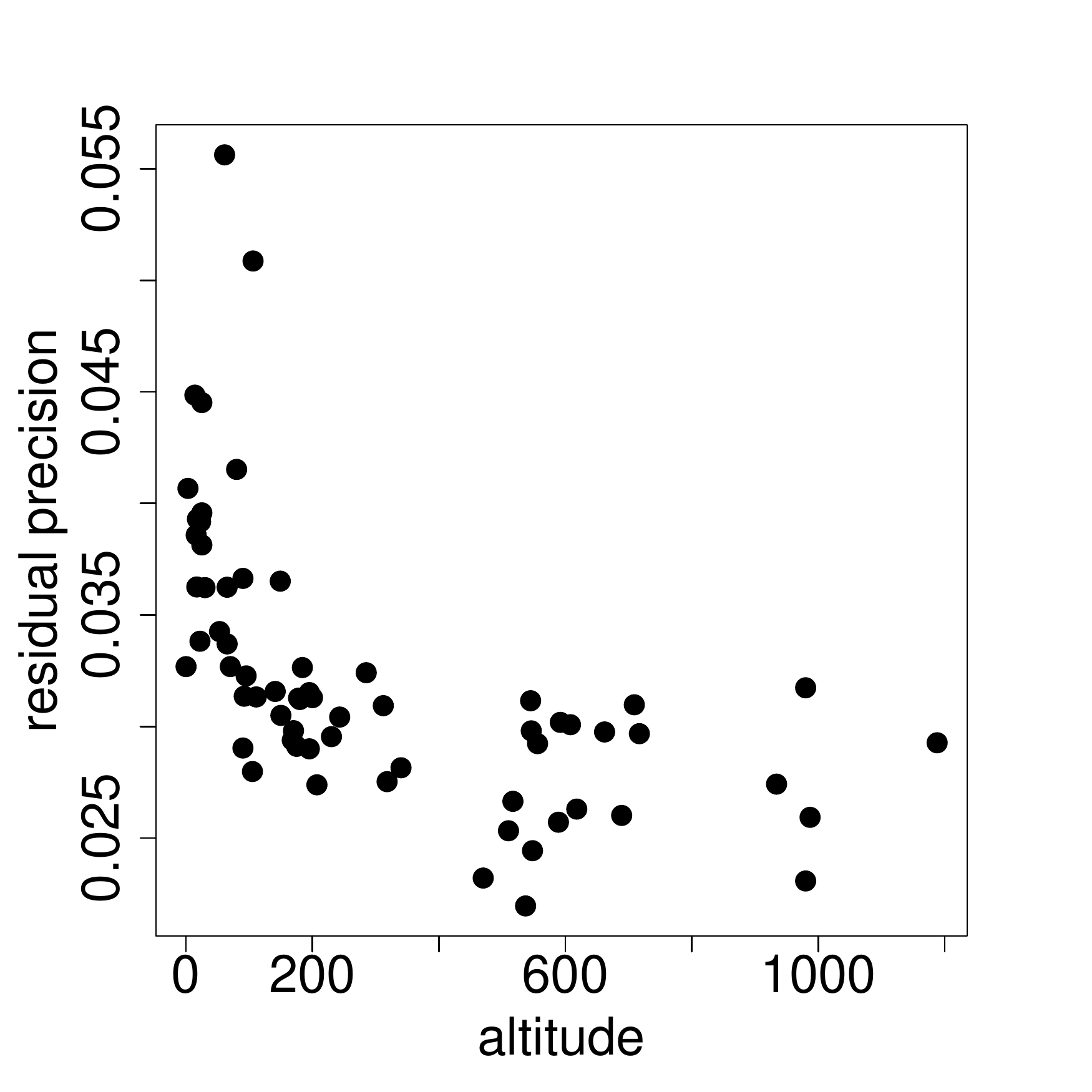} 
    \\
{(c) Residual temporal precision }& (d) Residual spatial precision versus altitude \\
    \includegraphics[width=5.5cm]{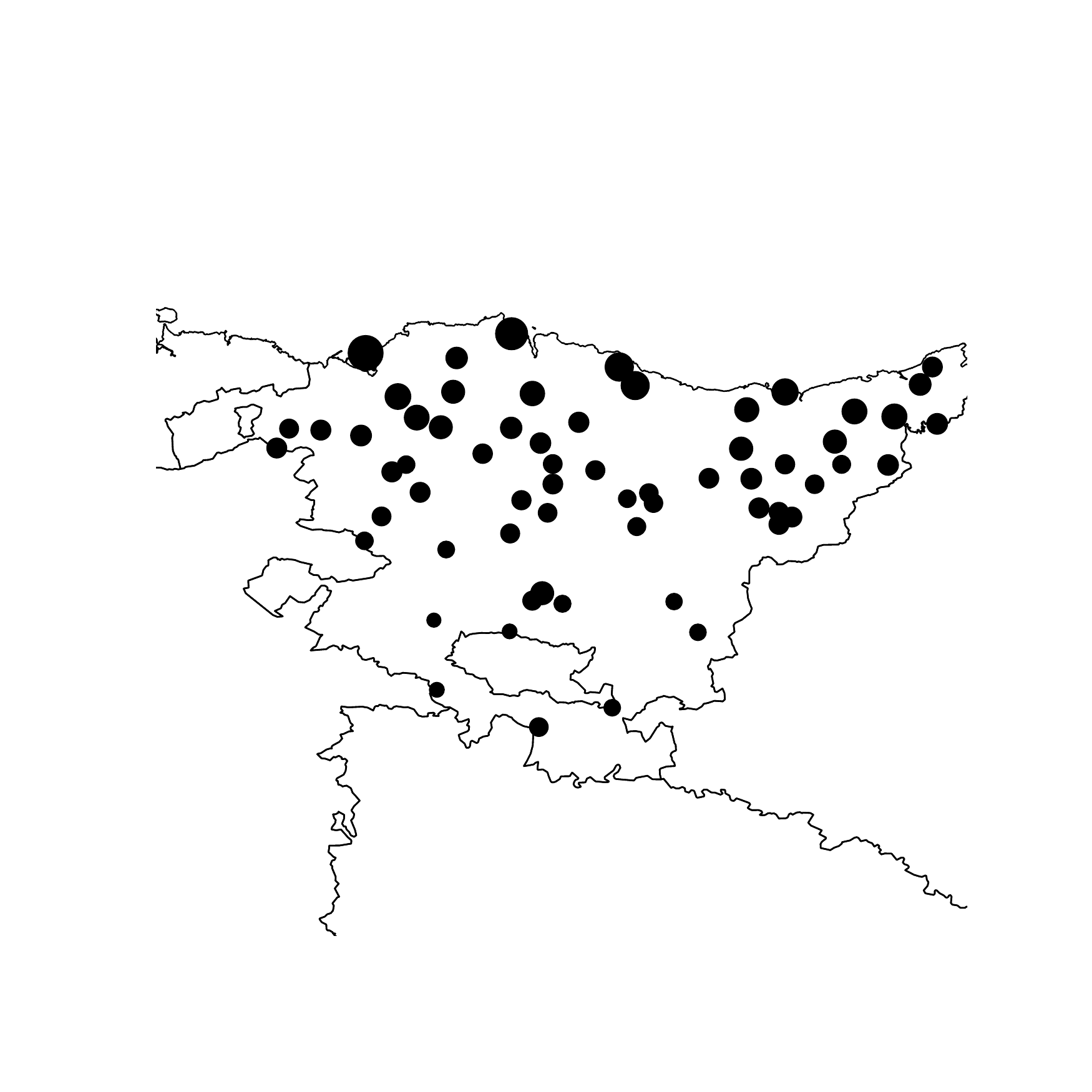} &
\includegraphics[width=4.6cm]{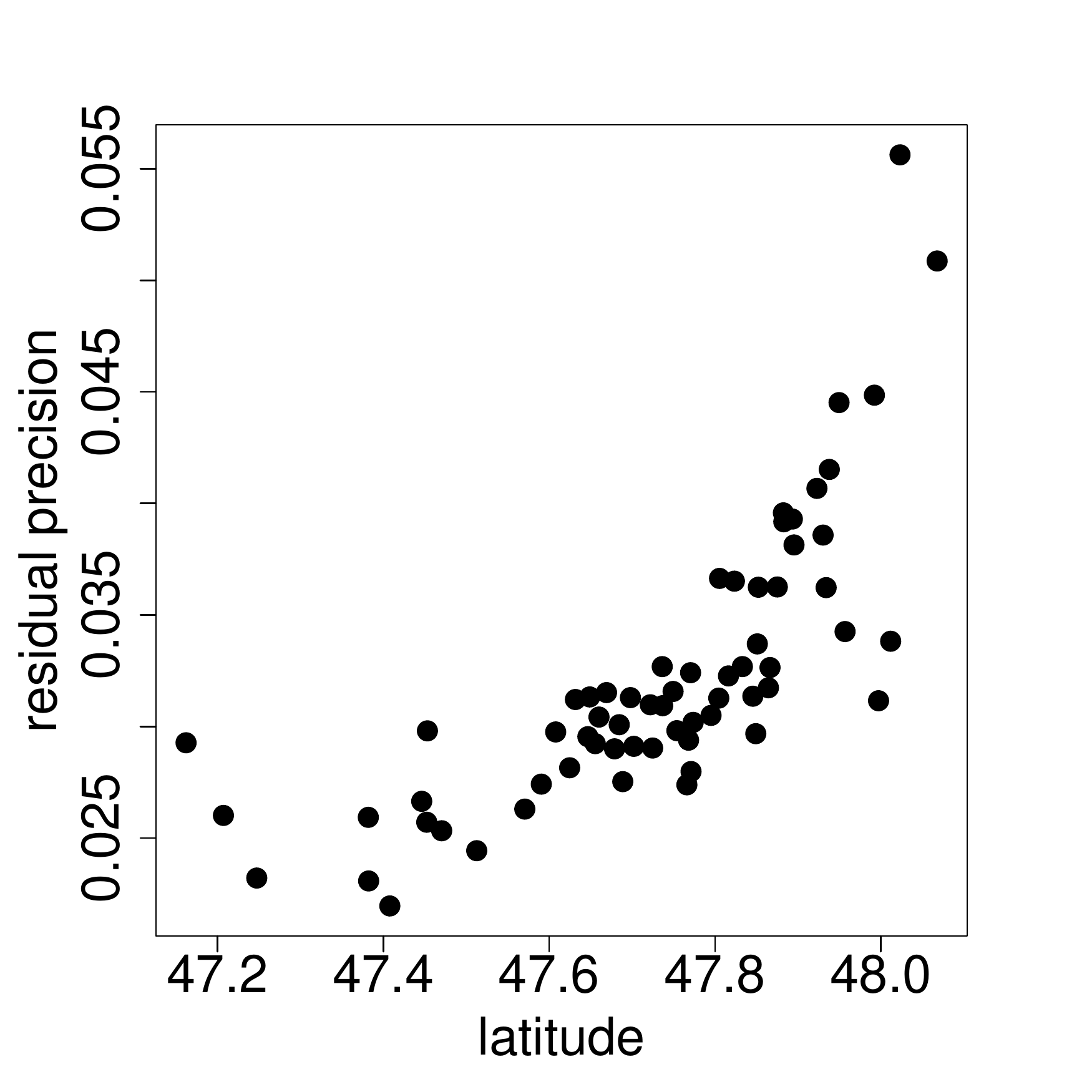} \\
(e) Residual spatial precision  & (f) Residual spatial precision versus latitude \\
\end{tabular}
\caption{Data summaries { for the maximum temperature data observed in the Spanish Basque Country}. Panel (a) displays the map with spatial locations (solid circles) and the crosses are the ones left out from the inference procedure to check the predictive ability of the different models. Panel (b) presents the empirical mean over the year. Panels (c)-(f) represent the empirical precision of the maximum temperature observed data for the residuals of a Gaussian (G) model.\label{sec6fig1}} 
\end{figure}

This dataset refers to the maximum temperature recorded in 2006 in the Spanish Basque Country (Figure \ref{sec6fig1}(a)). 
{Part of these data was analysed by} \cite{PSteel06}, \cite{Fons11} and \cite{Bueno2017} where they only used the maximum temperature recorded in July 2006 at 70 locations. 
\cite{PSteel06} considered only spatial data while \cite{Fons11} and \cite{Bueno2017} considered spatio-temporal data. As this region is quite mountainous, with altitude of the monitoring stations lying between 0 and 1188 meters, altitude is included as an explanatory variable in the dynamic mean of the process that also depends on the spatial coordinates, that is, $m_{t}(\mathbf{s}) = \theta_{0t} + \theta_{1t}\thinspace lat(\mathbf{s}) + \theta_{2t}\thinspace long (\mathbf{s}) + \theta_{3t} \thinspace alt(\mathbf{s})$, $\forall \thinspace t=1, \ldots J$.
For the stations with missing observations, data were inputted using a random forest algorithm  \citep{Stek12}. We considered stations with no more than 5\% missing data resulting in 68 locations. 

Panels of Figure \ref{sec6fig1}(c)--(e) show that the empirical mean, empirical precision over time and space for the residuals of a Gaussian dynamical model is far from constant, suggesting that a spatial model with constant variance might be unsuitable. Panel {(e)} of Figure \ref{sec6fig1} shows the behaviour of the variability across the region. The diameter of the solid circles is proportional to the value of the residual precision at the respective location.  The map suggests that there is a spatial trend left in the variance of the residuals.

Given the results from Figure \ref{sec6fig1}, we move forward and fit models: G, ST, GLG, Dyn, DynGLG and Full as described in Table \ref{tabmodels} with covariates in the spatial variance. We leave out three locations for predictive comparison (represented by `$\times$' in Panel (a) of Figure \ref{sec6fig1}). The parameters to be estimated are the dynamic coefficients ($\theta_{0t}, \theta_{1t}, \theta_{2t},\theta_{3t}$), $\boldsymbol{\eta}_t$, the covariance parameters ($\sigma^2, \tau^2, \phi, \alpha, \gamma$), the mixing parameters ($\nu_1, \nu_2$),  the latent mixing variables ($\lambda_{1}(\bm{s})$, $\lambda_{2t}$) and the variance regression coefficients $\bm{\beta}$. As already mentioned in section \ref{sec2.2} the variances $W_t$ and $W_{2t}$ are estimated through discount factors. We must specify two discount factors referring to the structure of the mean process $\mathbf{Z}_t$ and the mean of the variance process ${\lambda}_{2t}$, respectively. In a general context, the value of the discount factor is usually fixed between 0.90 and 0.99, or it is chosen by model selection diagnostics, e.g, looking at the predictive performance of the model for different values of $\delta = (\delta_1, \delta_2)$ using some comparison criteria \citep{petris2009dynamic}. To illustrate the performance of the competing models, we fixed $\delta_{1}=0.99$ (for all competing models) and $\delta_{2}=0.99$ (for the Dyn, the DynGLG and the Full models) for evaluating the behaviour and goodness of fit. 

 { Following the values of the different model comparison criteria shown in Table \ref{sec4tab1}, the G model is the one with the worst predictive performance.} As mentioned previously, the G model is not able to accommodate the uncertainty for some observations which presented larger maximum temperature values. { Under LPS the Full and the DynGLG models provide quite similar values, whereas DynGLG results in the smallest value of VS-0.25. Note that the LPS under DynGLG is similar to the one under the Full model. Therefore, in what follows we discuss the main results obtained for the DynGLG model where we do not consider spatial covariates.}

\begin{table}
   \caption{Model comparison based on the Interval Score (IS), the Log Predictive Score (LPS) and the Variogram Score of order 0.25 (VS-0.25) criteria for the predicted observations at the out-of-sample locations under all fitted models for the maximum temperature dataset. The smallest values are highlighted in boldface.  
\label{sec4tab1}}
     \centering
  \fbox{  \begin{tabular}{lcccccc}
    \hline
    & G  & ST & GLG & Dyn & DynGLG &{Full} \\
    \hline
IS &6.85 & 4.62  & 4.54 &  {\bf 4.21}& 4.34 & { 4.25}\\
LPS & 1565 & 1355  & 1206 & 1286& { 1097}& {\bf 1095}\\
VS-0.25 & 1658 & 1304 & 1222 & 1283 & {\bf 1194} & 1240\\
\hline
    \end{tabular}}
\end{table}

\begin{figure}
\centering
\begin{tabular}{cc}
\includegraphics[width=5cm]{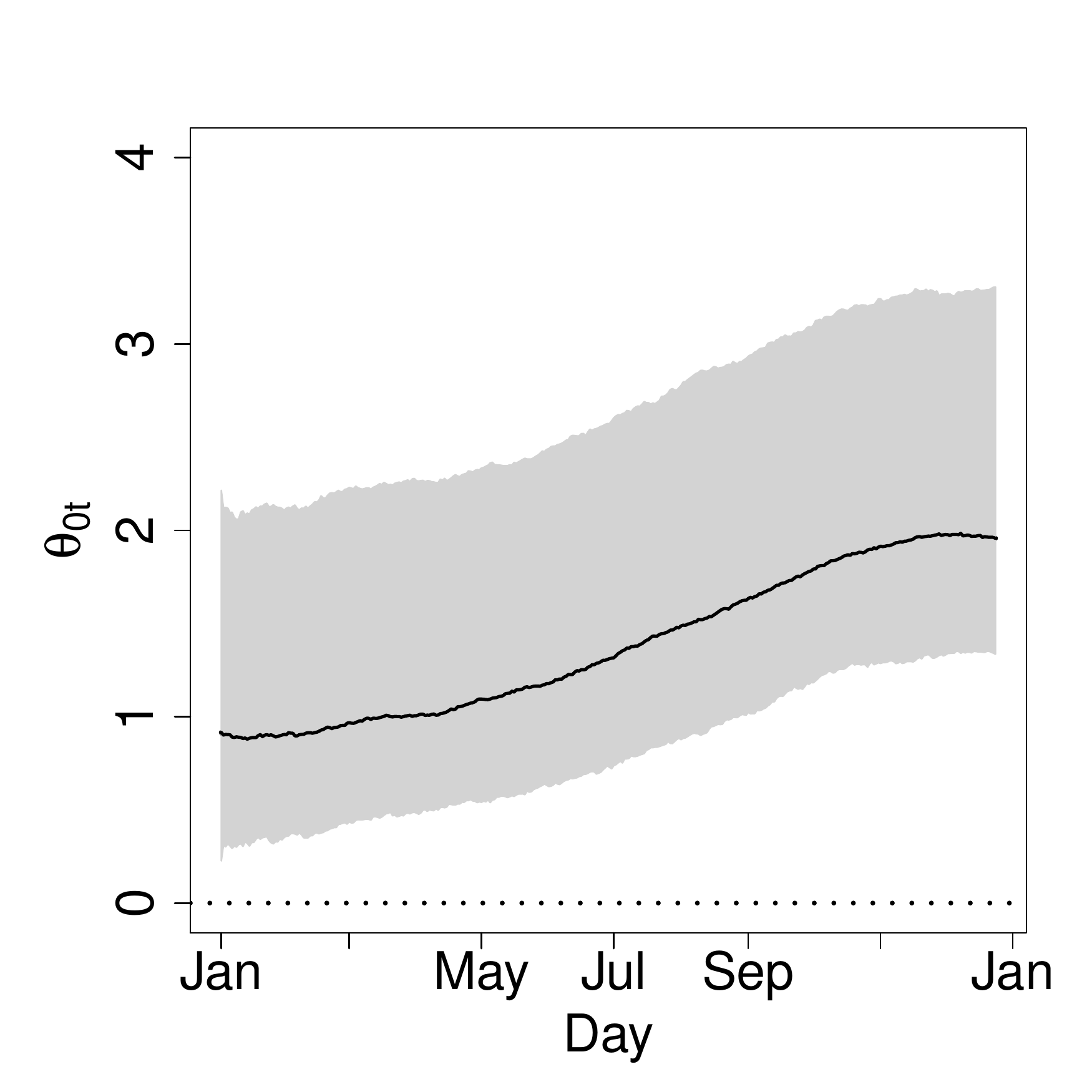}
&
 \includegraphics[width=5cm]{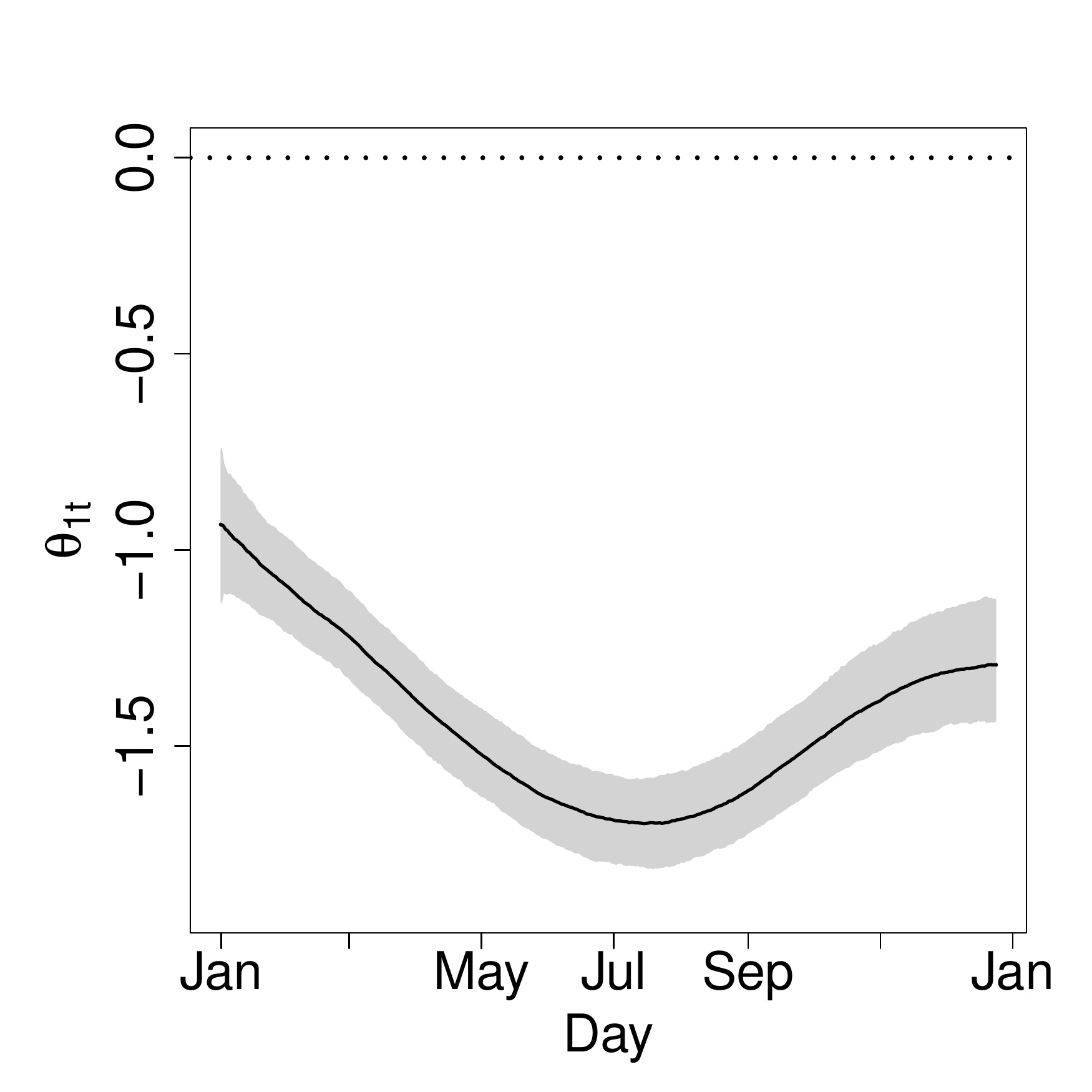} \\ {(a) Intercept }  &{(b) Latitude effect} \\ 
 \includegraphics[width=5cm]{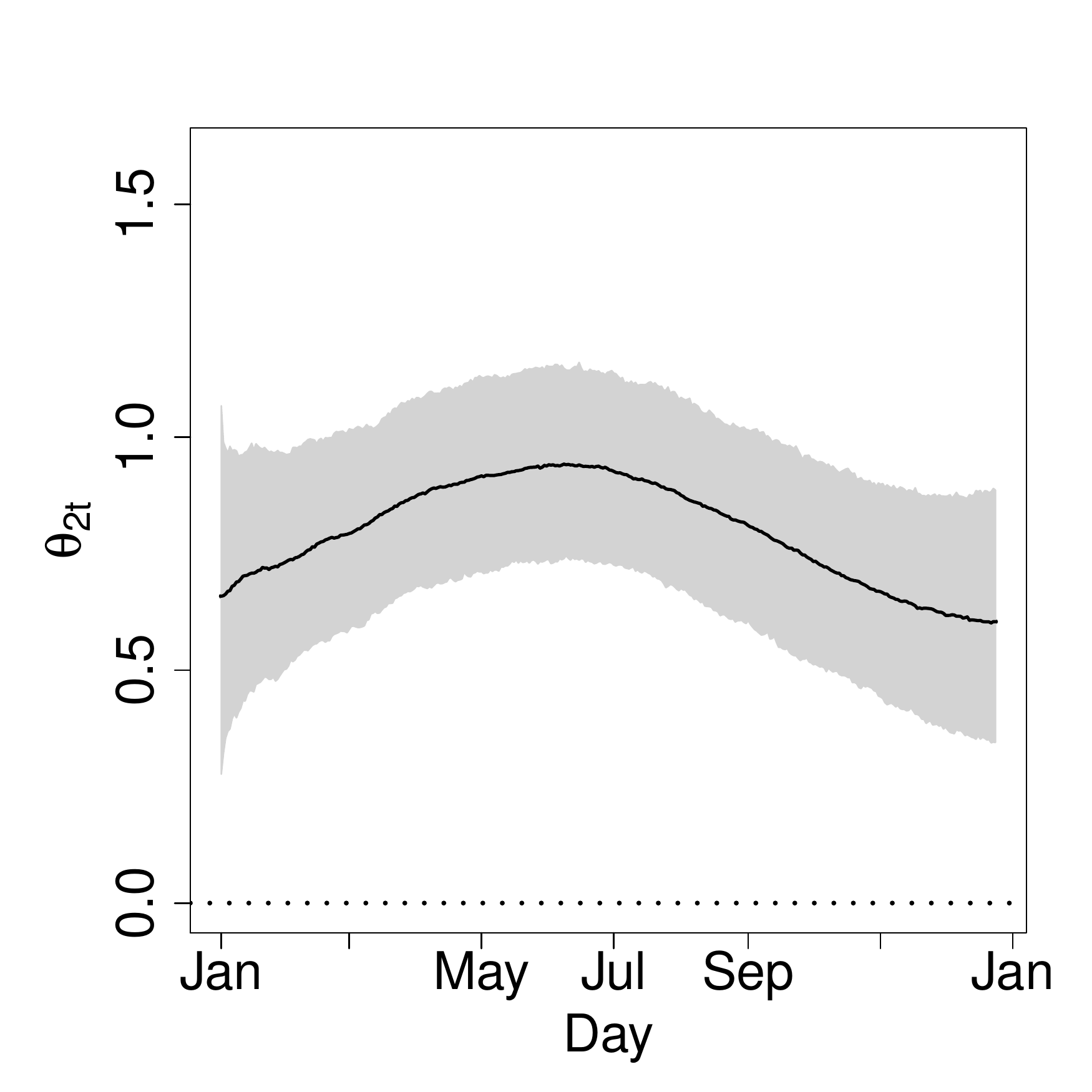} &
 \includegraphics[width=5cm]{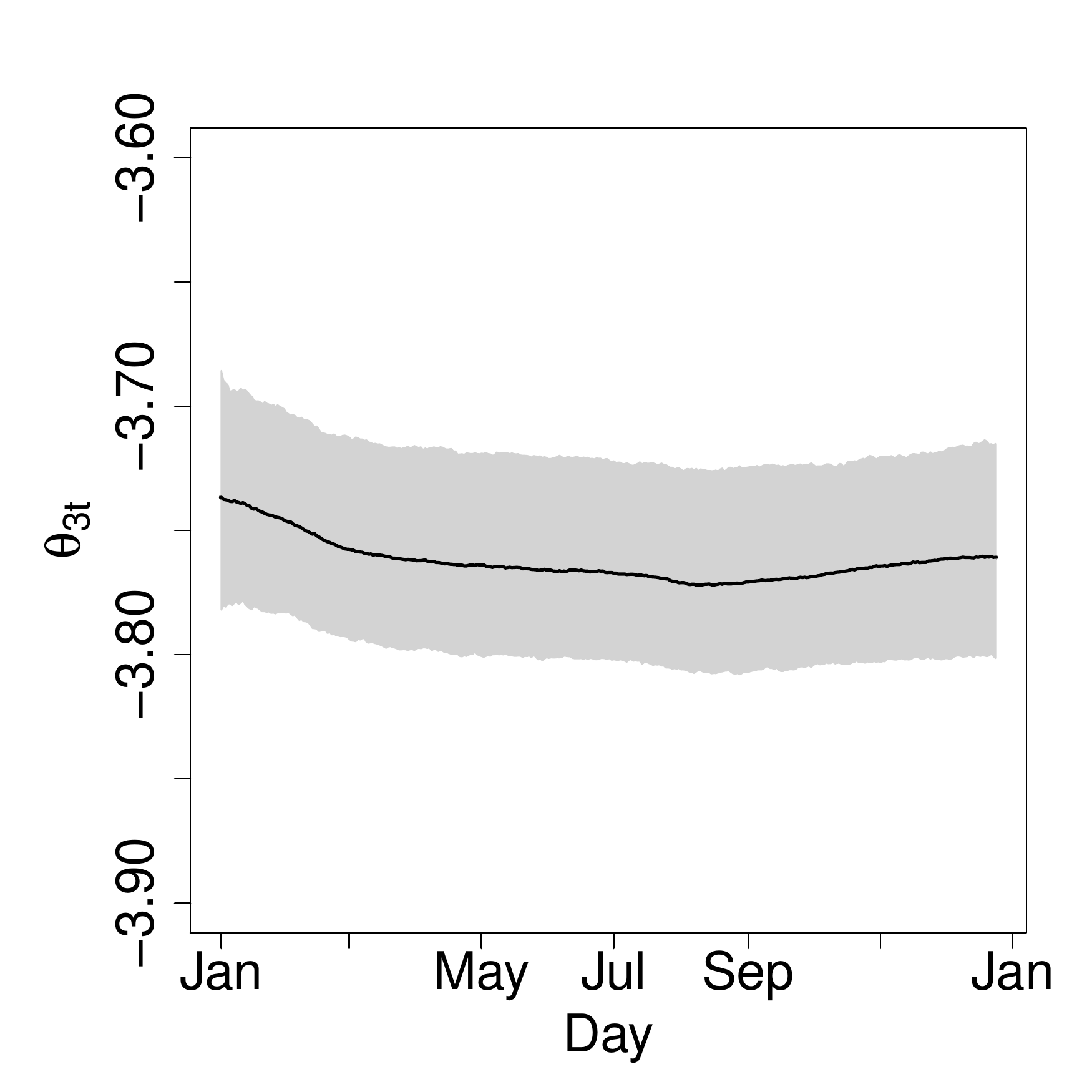}\\
{(c) Longitude effect} & {(d) Altitude effect} \\
\end{tabular}
\caption{Temperature data: posterior summaries for the dynamic mean effects $\theta_t$ under the DynGLG model.\label{sec6fig2a}}
\end{figure}

Posterior summaries (limits of the 95\% posterior credible intervals) of the time varying coefficients (see Figure \ref{sec6fig2a}) in the mean of the process do not include zero suggesting that latitude, longitude and altitude are important covariates to explain levels of temperature. In particular, and as expected, the coefficient associated with altitude (see panel (d) of Figure \ref{sec6fig2a}) is negative over time, suggesting that the maximum temperature decreases as the altitude increases.

\begin{figure}
\begin{center}
\begin{tabular}{ccc}
\includegraphics[width=4.6cm]{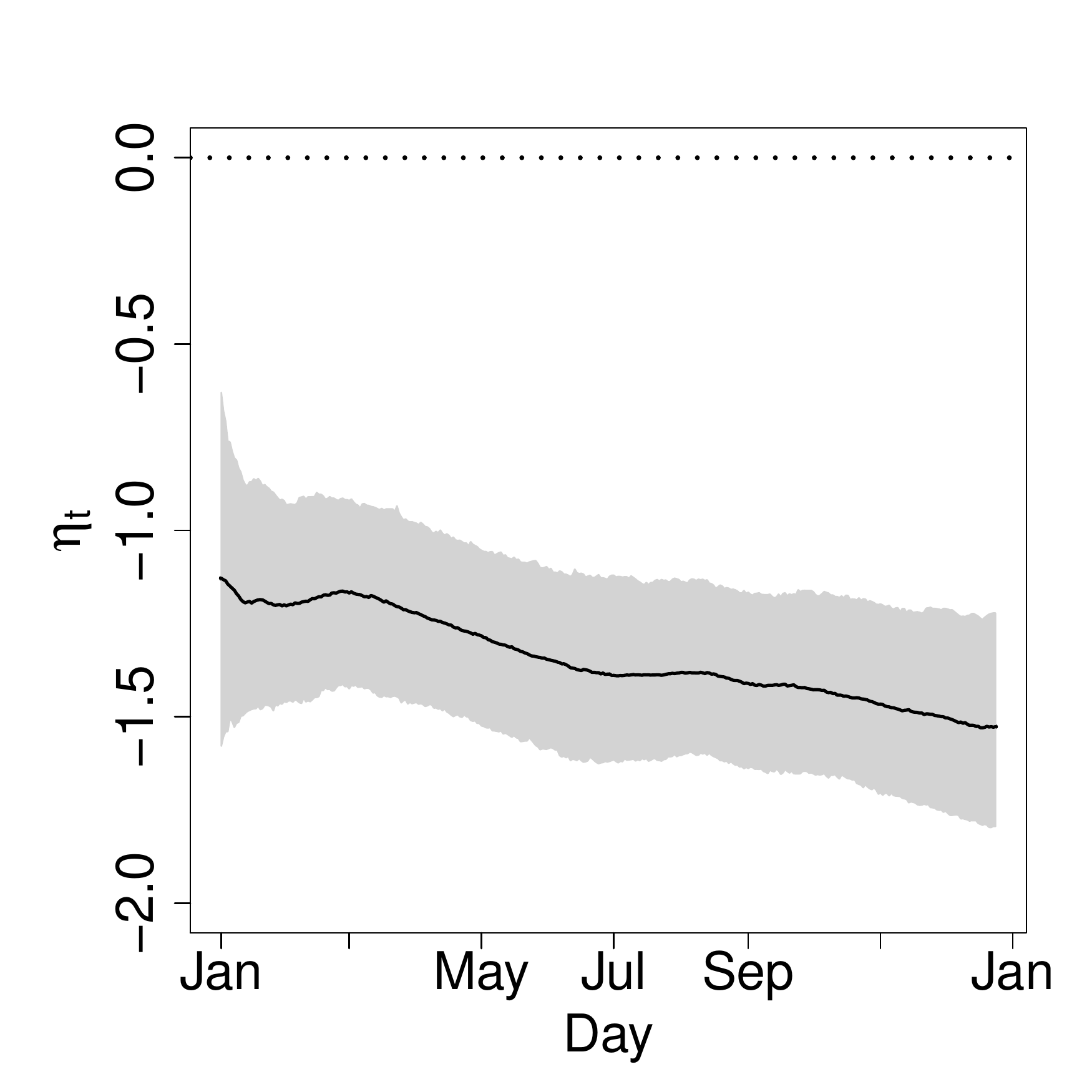} &
 \includegraphics[width=4.6cm]{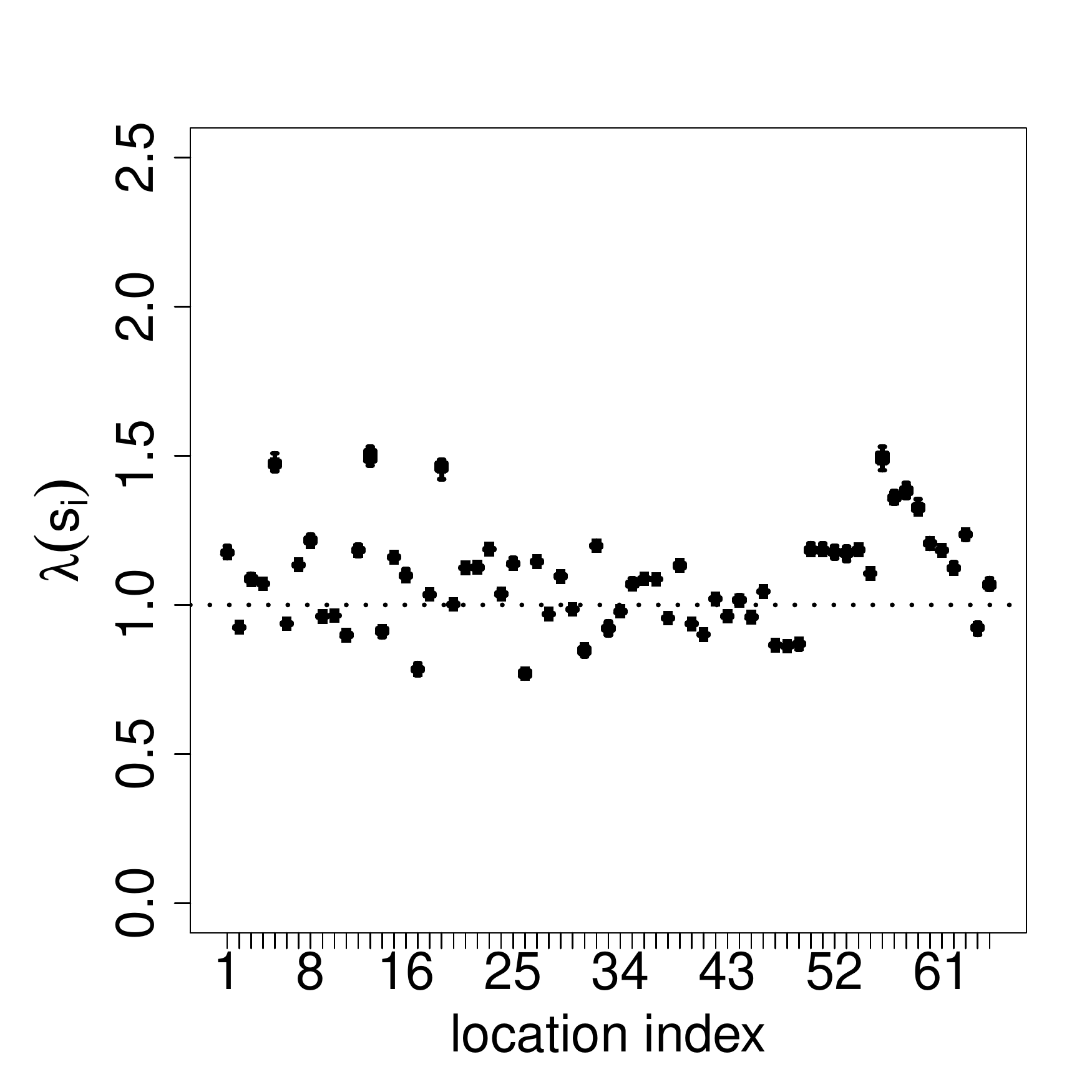} &
 \includegraphics[width=4.6cm]{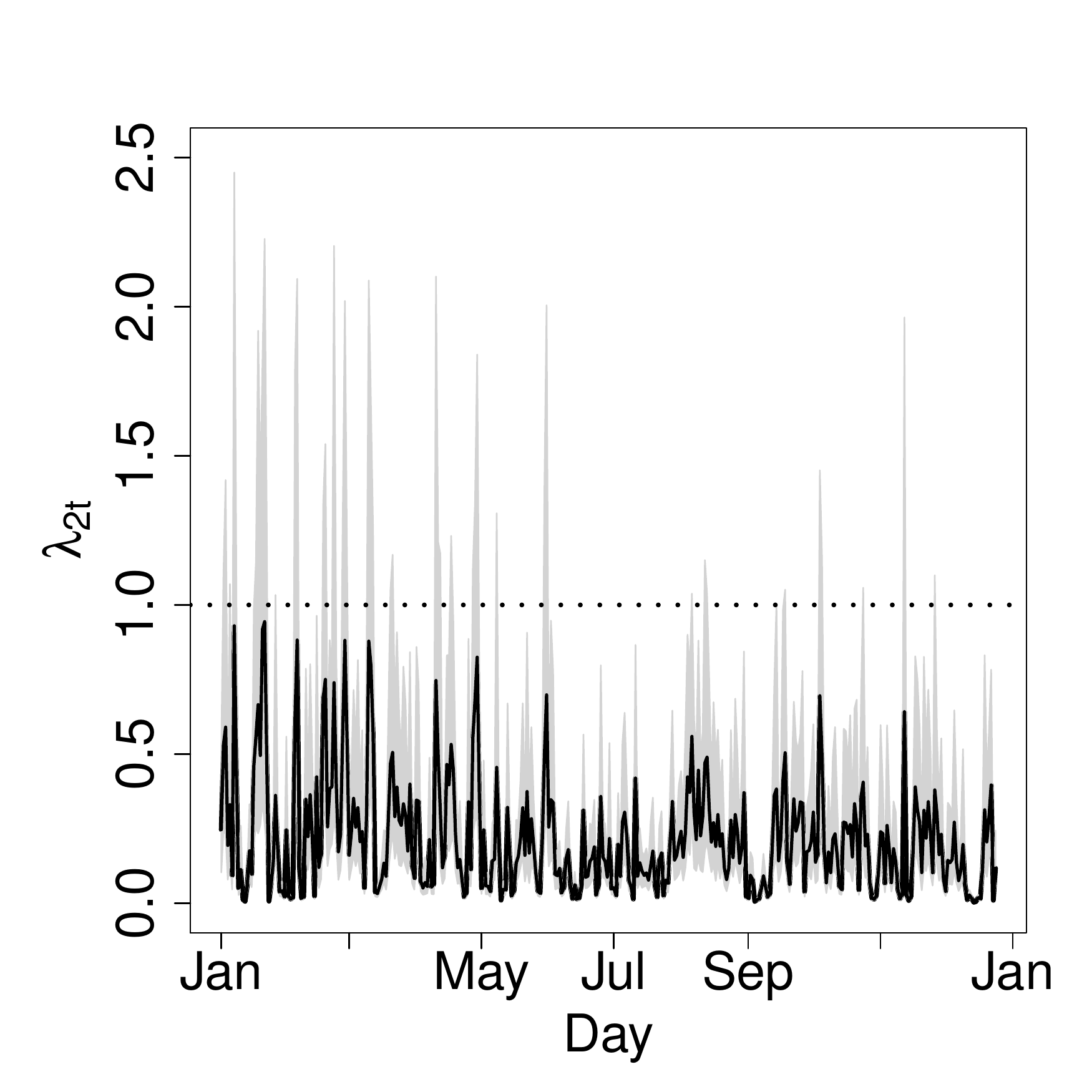}\\
 (a) $\eta_{0t}$    &{(b) $\lambda_{1}(\bm{s})$}  & (c) $\lambda_{2t}$  \\
\end{tabular}
\caption{Temperature data: posterior summaries for the DynGLG model: (a) dynamic variance mean (solid line), (b) mixing space and (c) mixing temporal.}
\label{sec6fig3}
\end{center}
\end{figure}

Panels (a)-(c) of Figure \ref{sec6fig3} present the dynamic mixing effect indicating that the model captures variability over time and it is able to identify some stations that are potential outliers over space and time. Clearly, the variance is not constant over space-time as previously suggested by Figure \ref{sec6fig1}.

\begin{figure}[!ht]
\begin{center}
\begin{tabular}{c}
 \includegraphics[width=11cm]{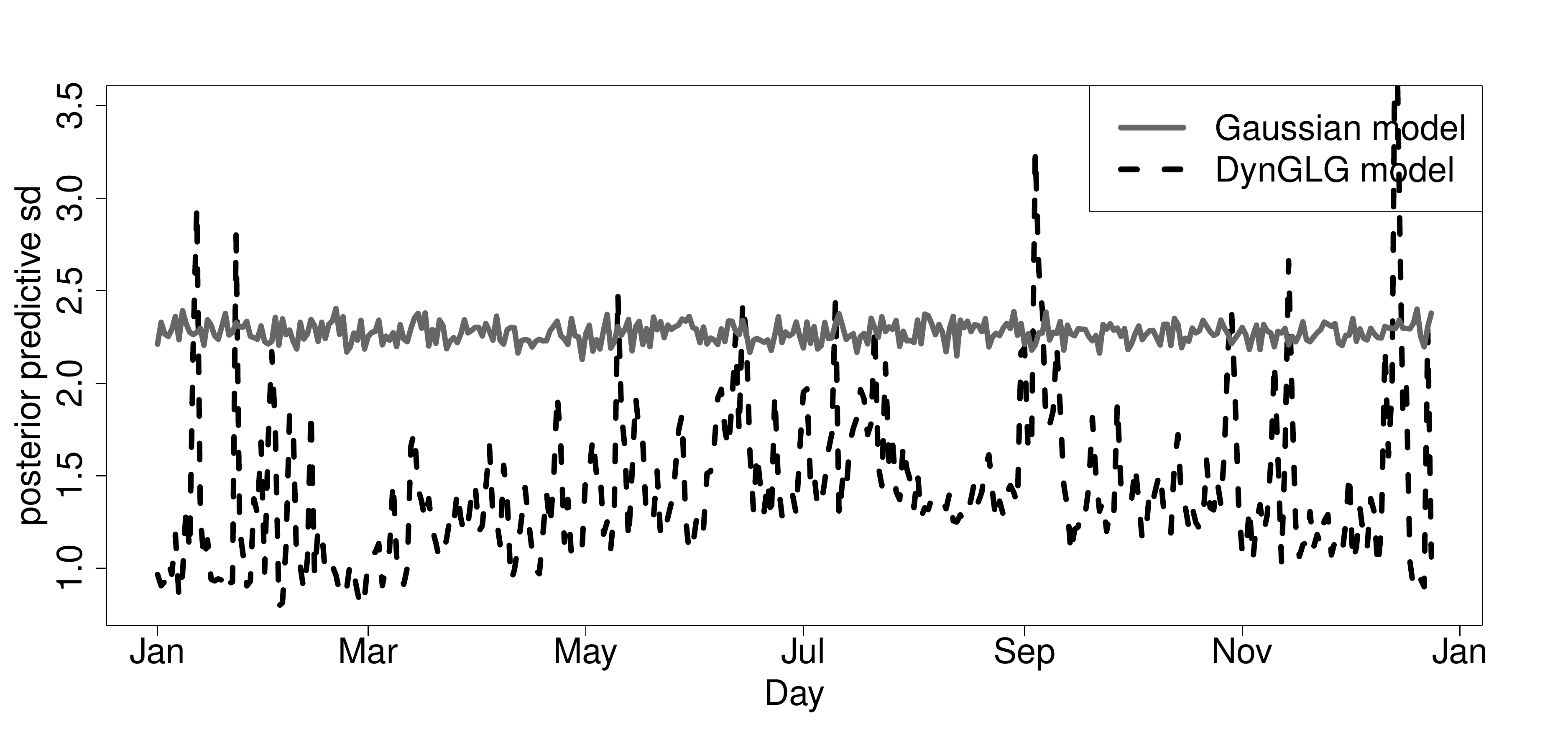} \\
 \end{tabular}
\caption{Temperature data: Approximated posterior predictive standard deviation over time for the DynGLG model and the Gaussian model for ${\bf s}=(43.16, -3.28)$. }
\label{figTempvar1}
\end{center}
\end{figure}

\begin{figure}[!ht]
\begin{center}
\begin{tabular}{c}

\includegraphics[width=12cm]{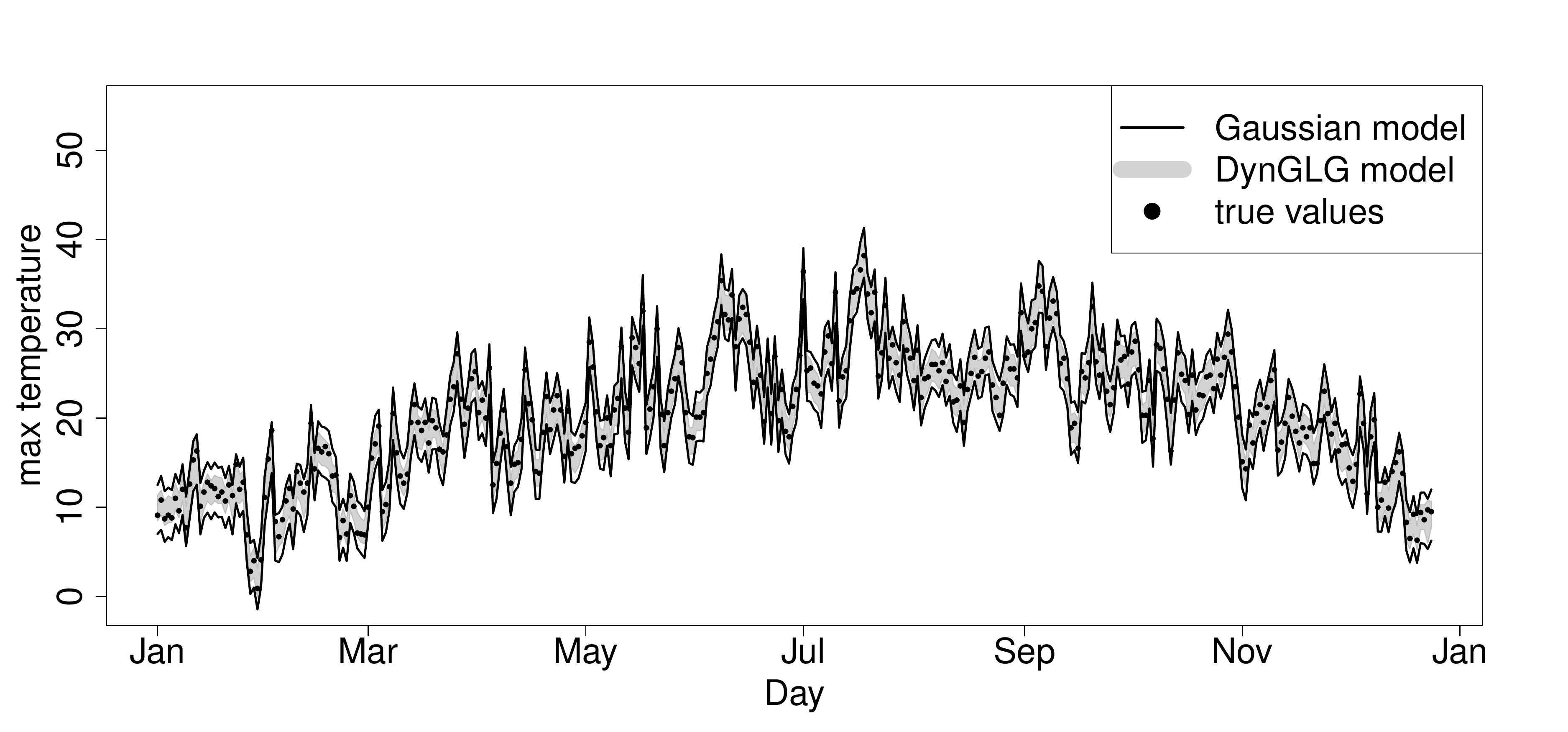} \\
 \\ 
 (a) ${\bf s}= (43.18, -2.77)$\\
 \includegraphics[width=12cm]{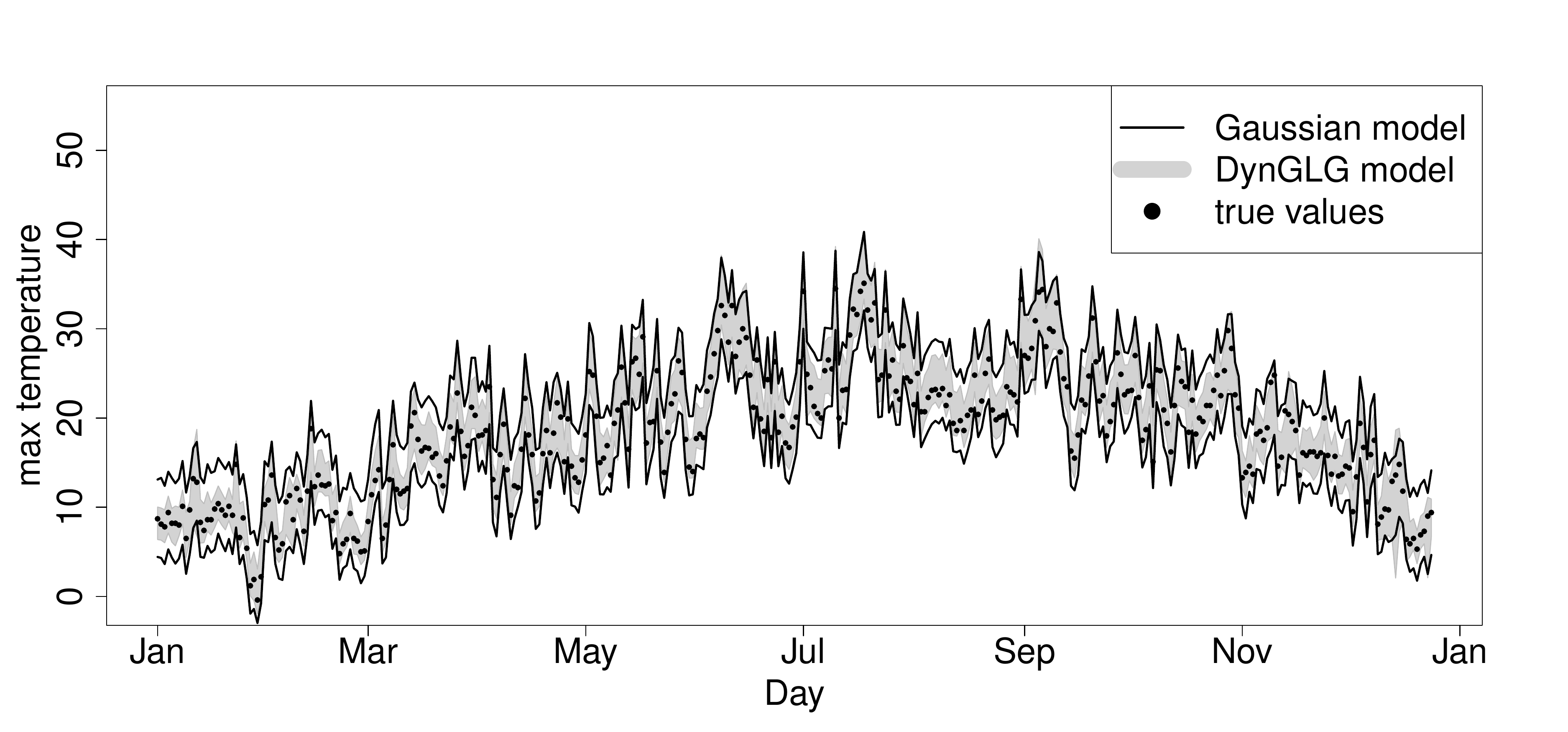} \\
(b) ${\bf s}=(43.16, -3.28)$\\
\end{tabular}
\caption{Temperature data: Predictive posterior distribution ($95\%$ interval) over time for the DynGLG model and the Gaussian model for {${\bf s}= (43.18,-2.77 )$ and ${\bf s}=(43.16, -3.28)$.}} 
\label{figTempvar2}
\end{center}
\end{figure}

 Figure \ref{figTempvar1} presents the posterior summaries for the predictive standard deviation of $z_t(s)$, ${ s}=(43.16, -3.28)$ conditional on the latent mixing variables for the DynGLG model compared to the Gaussian model. The posterior predictive standard deviation is obtained numerically by composition sampling that simulates replicated observations from the observational model and computes the empirical standard deviation of these artificial data. Note that the variance is non-constant with some peaks over time. This behaviour cannot be captured by the G model which estimates the standard deviations as almost constant over time. The advantage of our proposed model is clear from  panels (a)-(b) of Figure \ref{figTempvar2}. For this application, the DynGLG model tends to have shorter ranges of the 95\% posterior predictive intervals whereas uncertainty seems small and it presents larger intervals in periods of more volatile behaviours. 

As the Full model provided a similar value of LPS and a smaller value of IS in comparison to the DynGLG we briefly discuss the posterior summaries of the parameters in the model for $\lambda_1(\bm{s})$.
The Full model includes covariates in spatial variance $\lambda_1(\bm{s})$ and the regression coefficients indicate that latitude and longitude do not impact on the variability over space with the 95\% posterior credible interval for $\beta_1$ being $(-0.0093 , 0.0132 )$ and, for $\beta_2$ $(-0.1156,  0.0834 )$, respectively. On the other hand, the effects of altitude $IC(95\%, \beta_3)= (-0.0001, 0.0000)$ show that it influences spatial heterogeneity not only in the dynamical mean but also in the variability of the process. Note that the range of $\beta_3$ is very small and it does not improve the predictive performance  substantially. 


\clearpage


\subsection{Application to ozone data in the UK}\label{sec:real2}

This section analyses the ozone data presented in Section \ref{mot}. The proposed mean function is $m_t(\mathbf{s}) = \theta_{0t} + \theta_{1t}\thinspace lat(\mathbf{s}) + \theta_{2t}\thinspace long (\mathbf{s}) + \theta_{3t} \thinspace temp_t(\mathbf{s})+ \theta_{4t} \thinspace wind_t(\mathbf{s})$, $\forall \thinspace t=1, \ldots J$. For the stations with missing observations, data were inputted using a random forest algorithm \citep{Stek12}. We considered stations with less than $5\%$ of missing data in all variables resulting in 61 stations, with 56 stations used for model fitting and 5 stations used for prediction comparison. 
The parameters to be estimated for the complete model are the dynamic coefficients $\bm{\theta}_{t}$ and $\bm{\eta}_t$, the covariance parameters ($\sigma^2, \tau^2, \phi, \alpha, \gamma$), the mixing parameters $\nu_1$, $\nu_2$, the latent mixing processes  $\lambda_{1}(\bm{s})$ and $\lambda_{2t}$, and the variance regression coefficients $\bm{\beta}$. Analogous to the temperature application, smooth evolutions are assumed for the temporal evolution of trend and variance coefficients with discount factors $\delta_1=0.99$ and $\delta_2=0.99$, respectively.  
In what follows we discuss the main results obtained for the best model (CovDynGLG) according to our predictive comparison measures (see Table \ref{tabozone1}). The different criteria indicate that the Gaussianity is unlikely to hold for this dataset. The most complete models with dynamical effects in the variance have superior predictive performances under all criteria.

\begin{table}
     \caption{Model comparison based on the Interval Score (IS), the Log Predictive Score (LPS) and the Variogram Score of order 0.25 (VS-0.25) criteria for the predicted observations at the out-of-sample locations under all fitted models for the maximum ozone dataset.\label{tabozone1}}
     \centering
 \fbox{   \begin{tabular}{lccccccc}
    \hline
&  G  & ST & GLG & CovDyn & DynGLG & CovDynGLG & Full\\
    \hline
 IS &  78 & 75 & 76 &  77 & 70 & {\bf 68} & 71\\
 LPS & 5960  & 5883 & 5696 & 5940  & 5631 & 5563 & {\bf 5343}\\
VS-25 & 10116  & 9760 & 10081 & 9842 &  9698 & {\bf 9518} & 9535\\
         \hline
    \end{tabular}}
\end{table}

\begin{figure}
\begin{center}
\begin{tabular}{ccc}
\includegraphics[width=4.6cm]{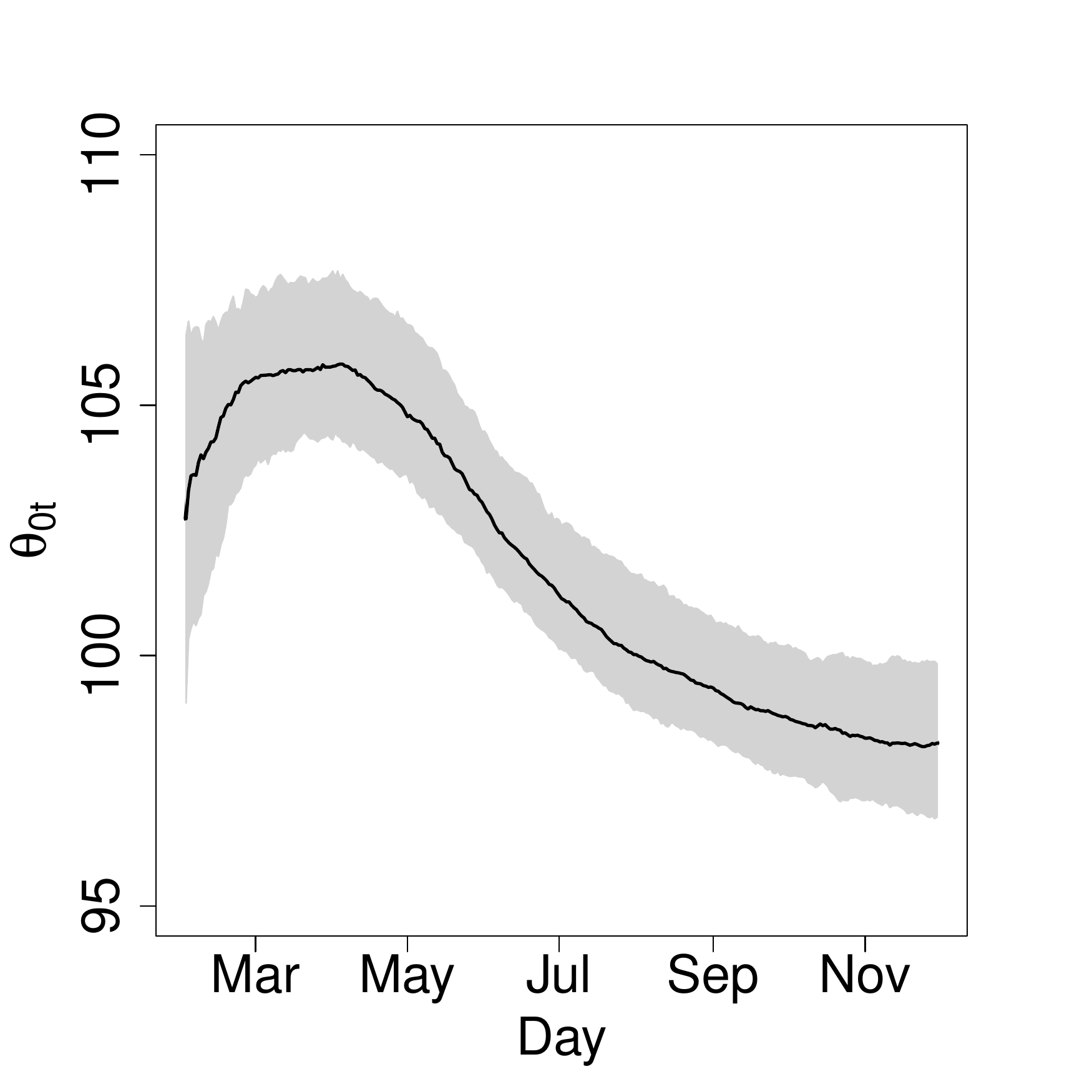}
&
 \includegraphics[width=4.6cm]{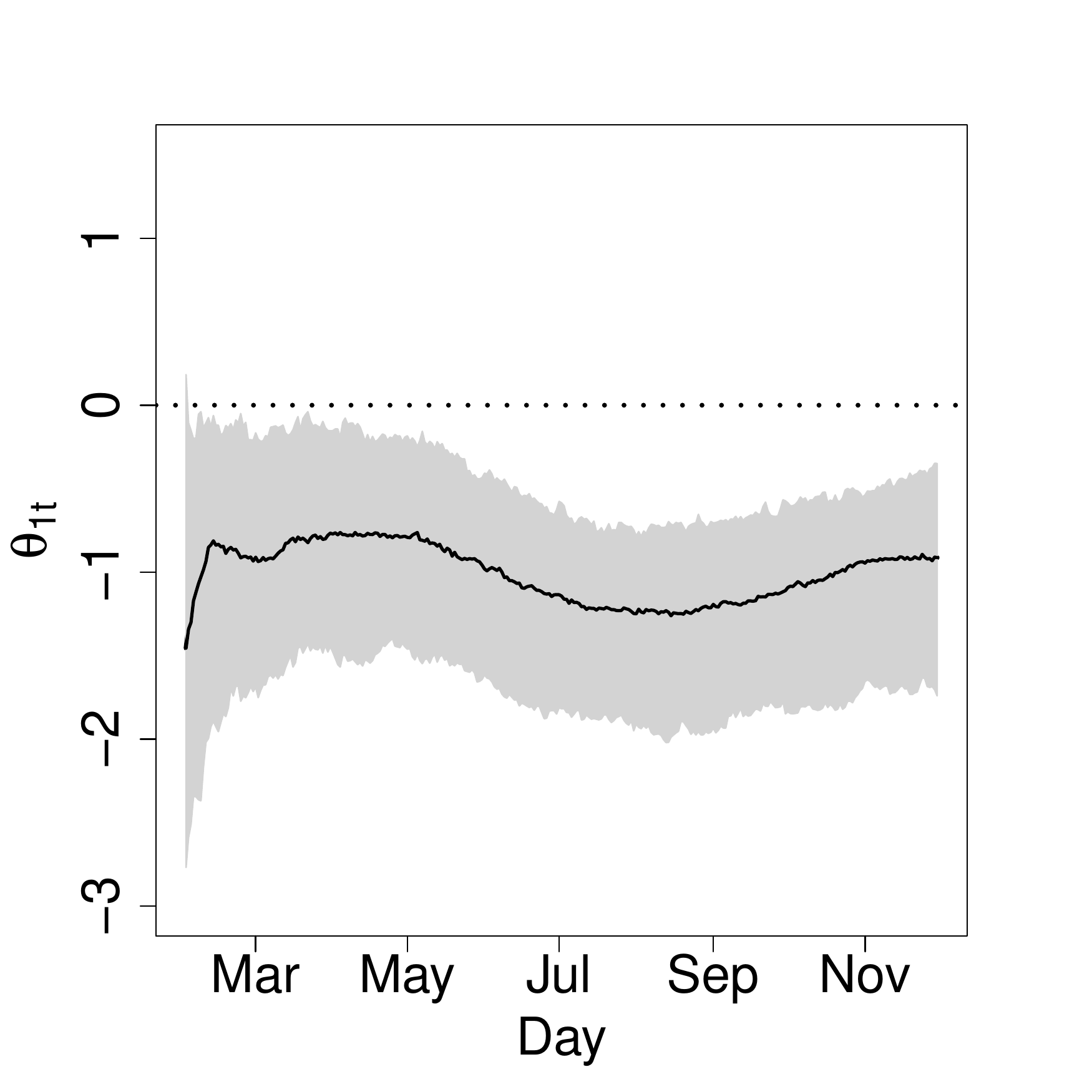} &
 \includegraphics[width=4.6cm]{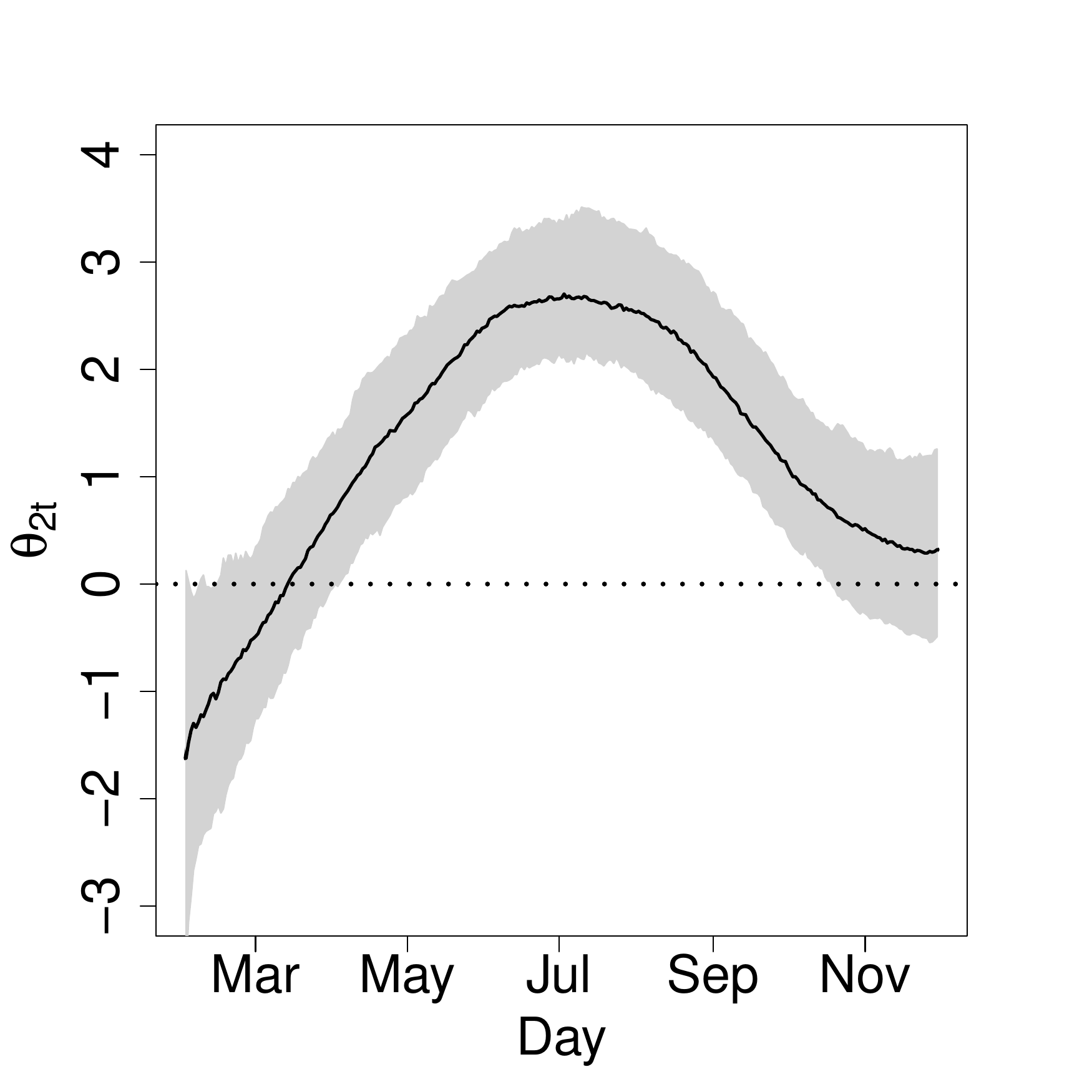}\\ {(a) Intercept. } &{(b) Latitude effect} & {(c) Longitude effect.} \\  
 \includegraphics[width=4.6cm]{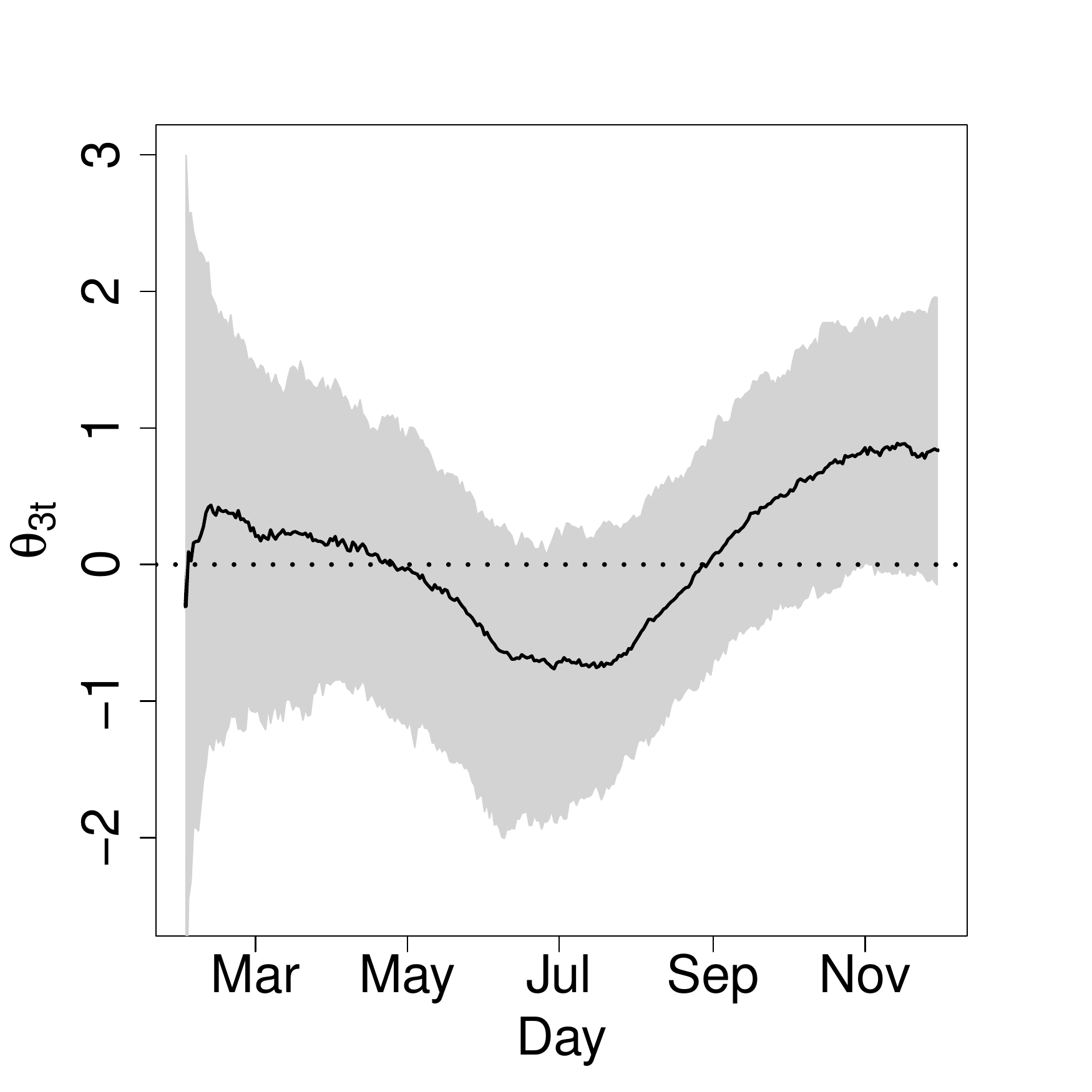} &
 \includegraphics[width=4.6cm]{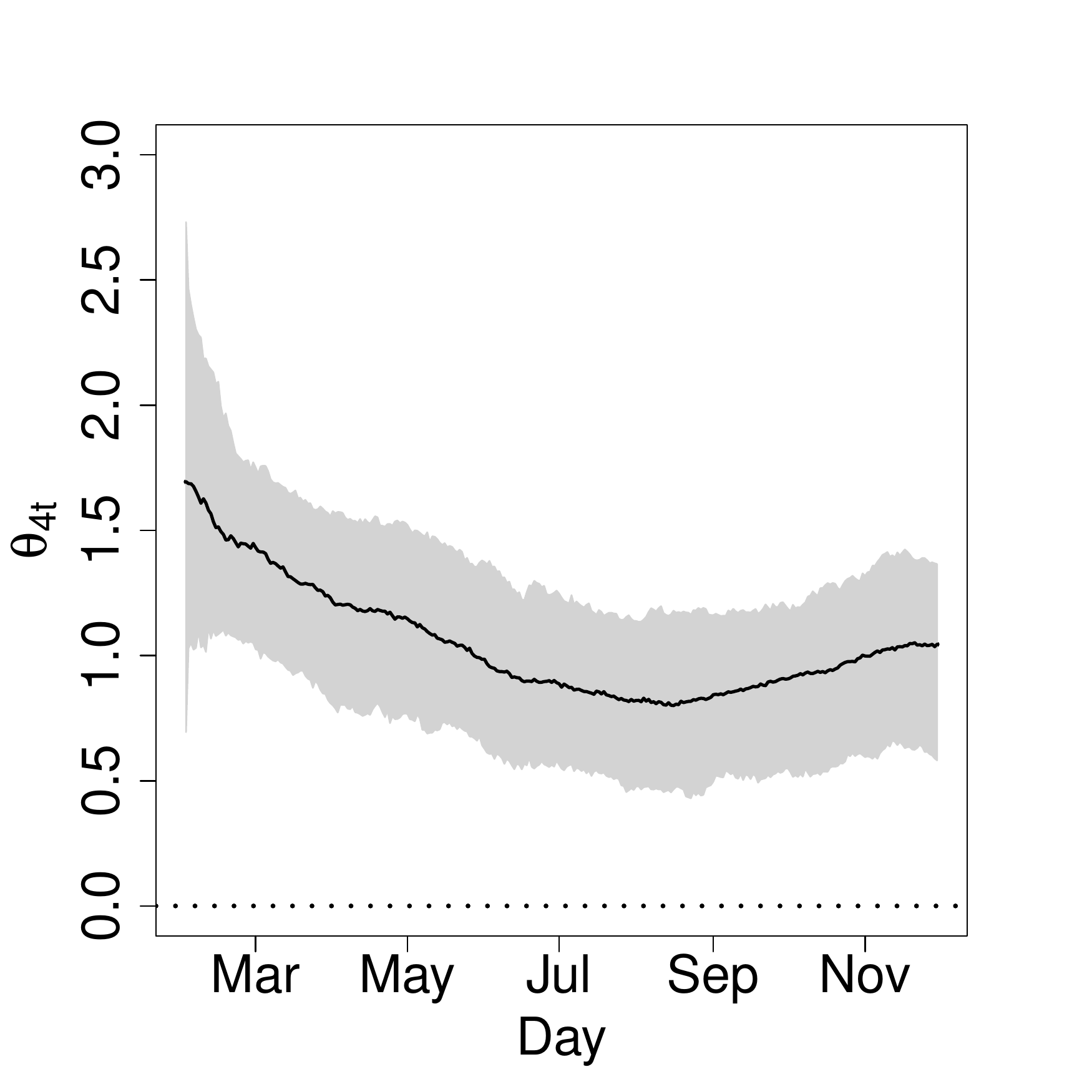} & \\
{(d) Temperature effect.} & {(e) Wind effect.} & \\
\end{tabular}
\caption{Ozone data: Posterior summaries for the dynamic mean effects, $\mathbf{\theta}_t$ in equation (\ref{eq5b}), under the CovDynGLG model.}
\label{figUK2}
\end{center}
\end{figure}

Panels of Figure \ref{figUK2} present the posterior summaries of the time varying coefficients in the mean 
for the CovDynGLG model indicating that the maximum ozone mean changes substantially from March to November (Panel (a)). Latitude, longitude and wind are associated with ozone levels resulting in a non-constant behaviour across time, while temperature is mostly not associated with ozone levels as $0$ is within the limits of the 95\% posterior credible interval.

\begin{figure}
\begin{center}
\begin{tabular}{cc}
\includegraphics[width=5cm]{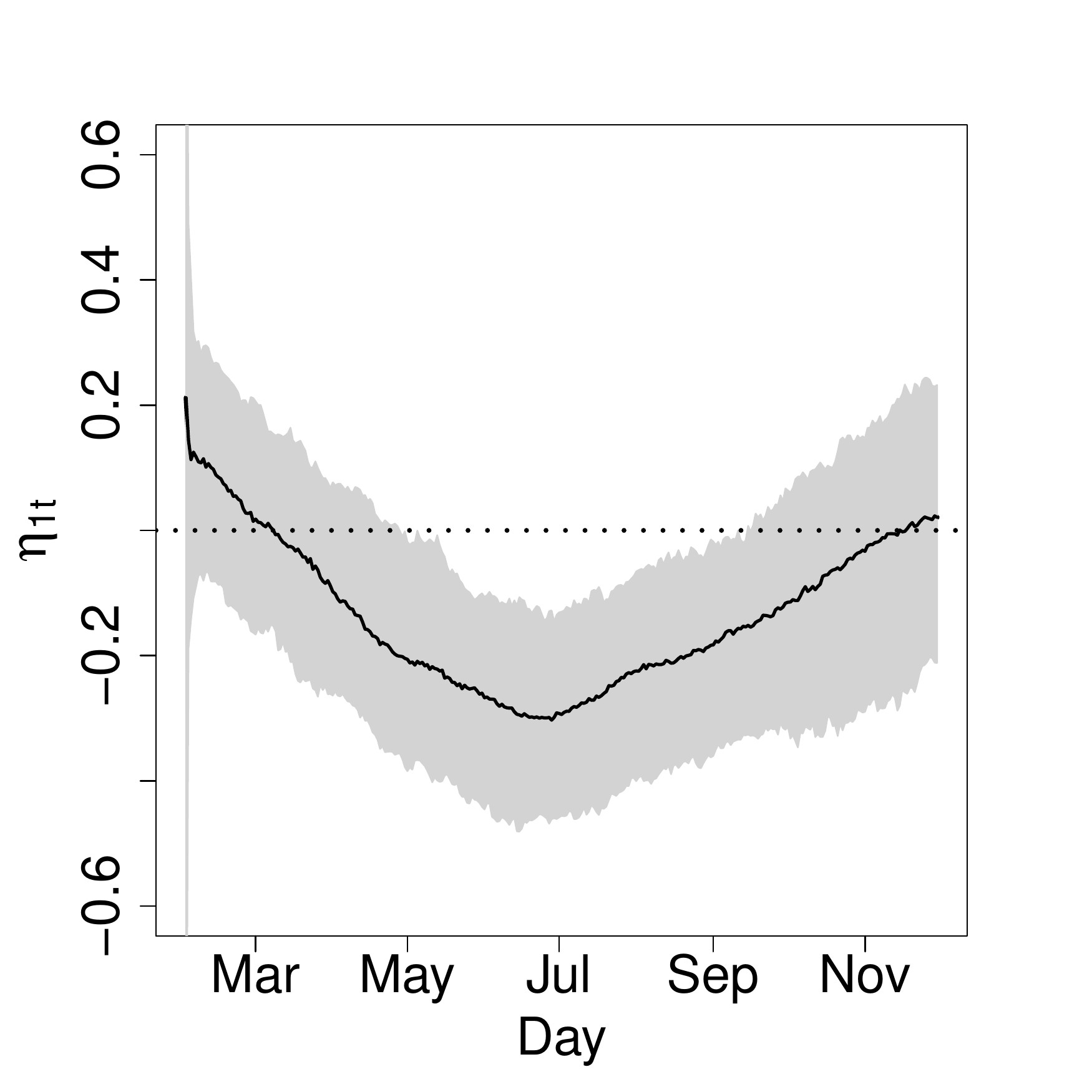}
&
 \includegraphics[width=5cm]{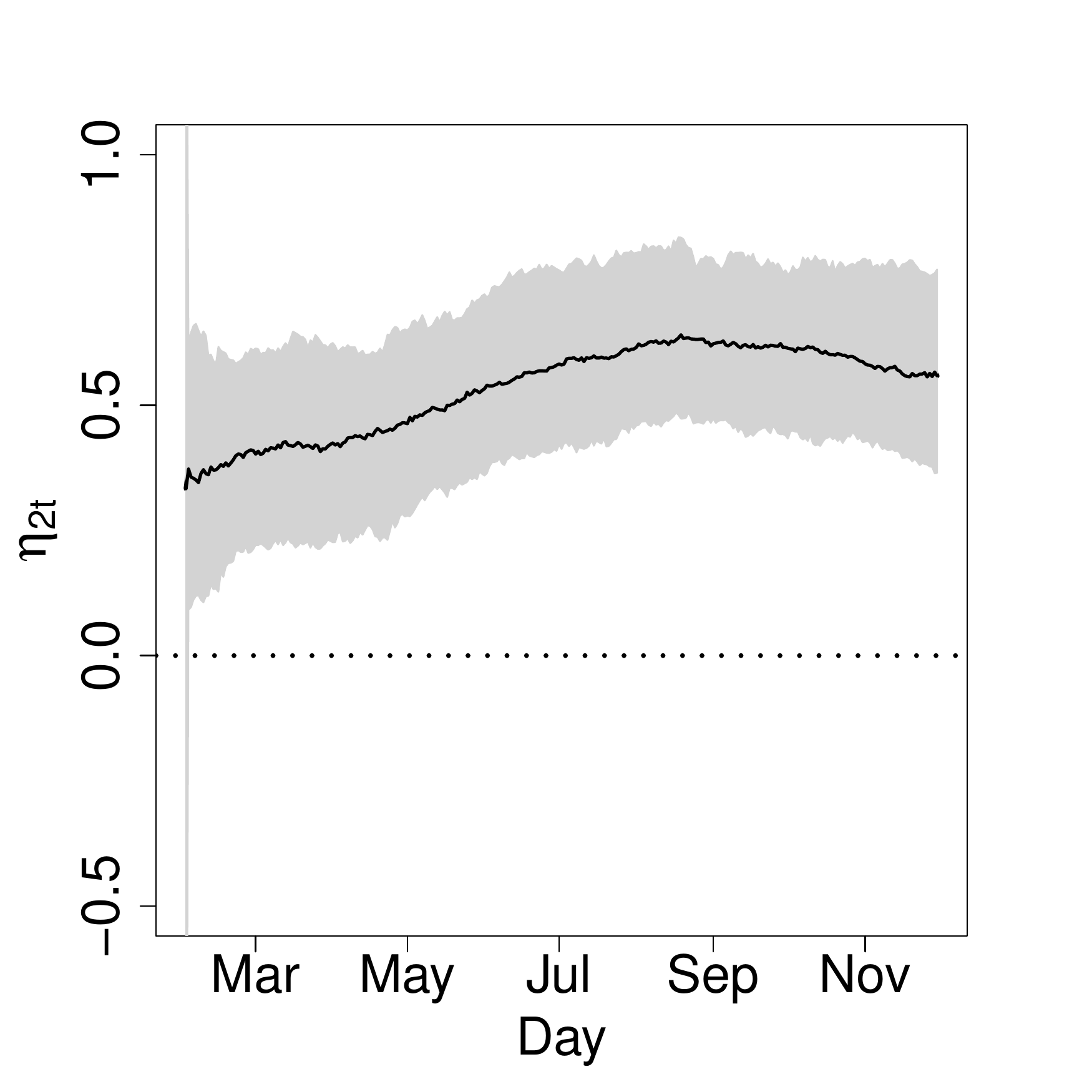} \\ 
 {(a) Temperature effect. } &{(b) Wind effect}  \\  
\end{tabular}
\caption{Ozone data: Posterior summaries of the coefficients included the equation for the time-varying variance (CovDynGLG model).}
\label{figUK3}
\end{center}
\end{figure}

 Panels of Figure \ref{figUK3} present the time varying coefficients for temperature and wind in the precision model.
Note that the coefficient for temperature is mostly negative in the precision model while in the mean model the 95\% posterior credible interval contains zero for most of the instants in time. 
Specifically, the temperature effect in the precision is negative in July, indicating smaller precision in the exponential scale, when indeed we observe the largest empirical temporal volatility of maximum ozone.  For the time-varying coefficients of wind we observe a positive association both in the mean and variance models, however, in the mean model the coefficient has a decreasing pattern (Figure 7 (e)), whereas in the variance model it has an increasing pattern with time (Figure 8 (b)).

\begin{figure}[!ht]
\begin{center}
\begin{tabular}{c}
 \includegraphics[width=12cm]{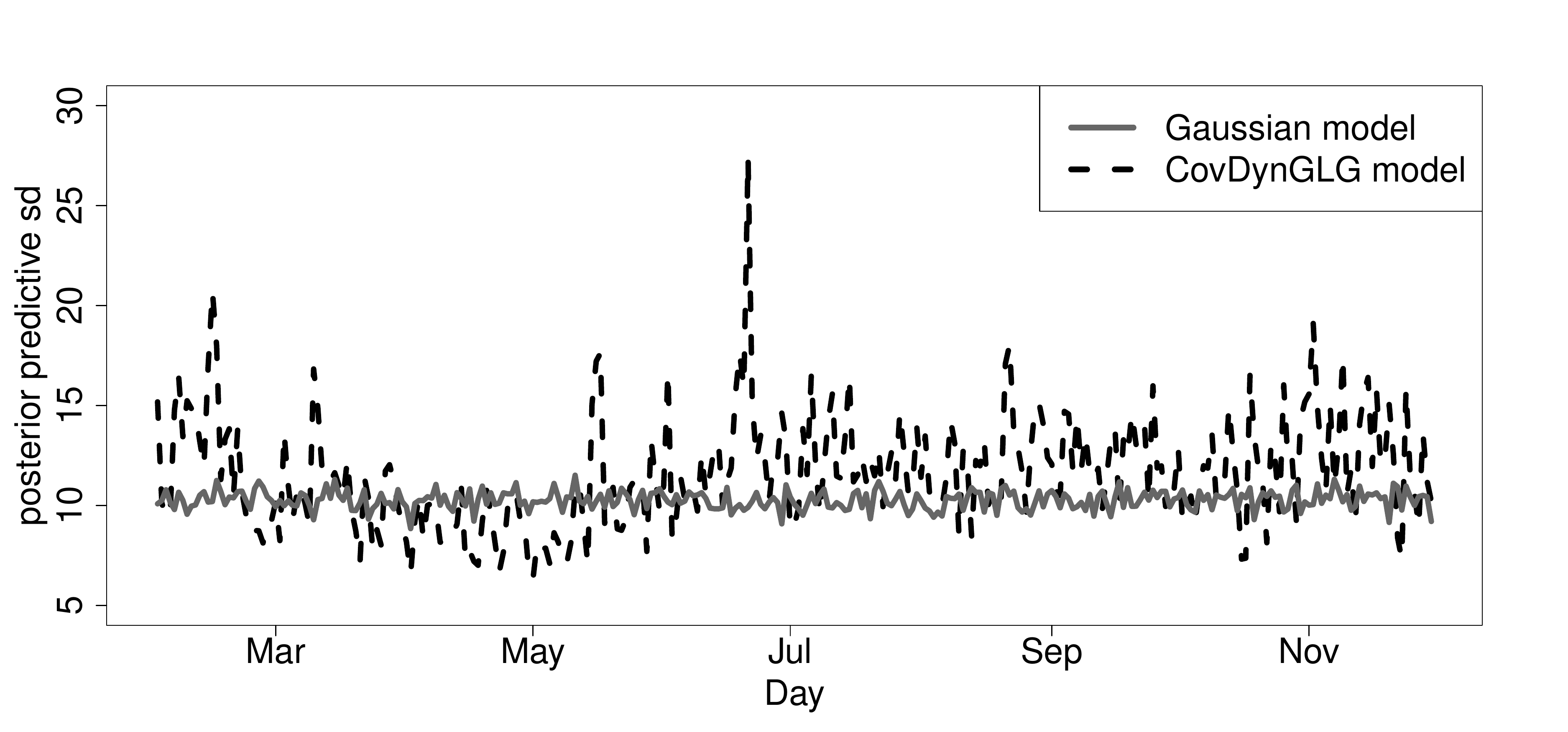} \\
 \end{tabular}
\caption{Ozone data: Approximated predictive standard deviation over time for the CovDynGLG model and the Gaussian model for ${\bf s}=(50.74,-1.83)$.}
\label{figUKvar1}
\end{center}
\end{figure}

\begin{figure}[!ht]
\begin{center}
\begin{tabular}{c}
\includegraphics[width=12cm]{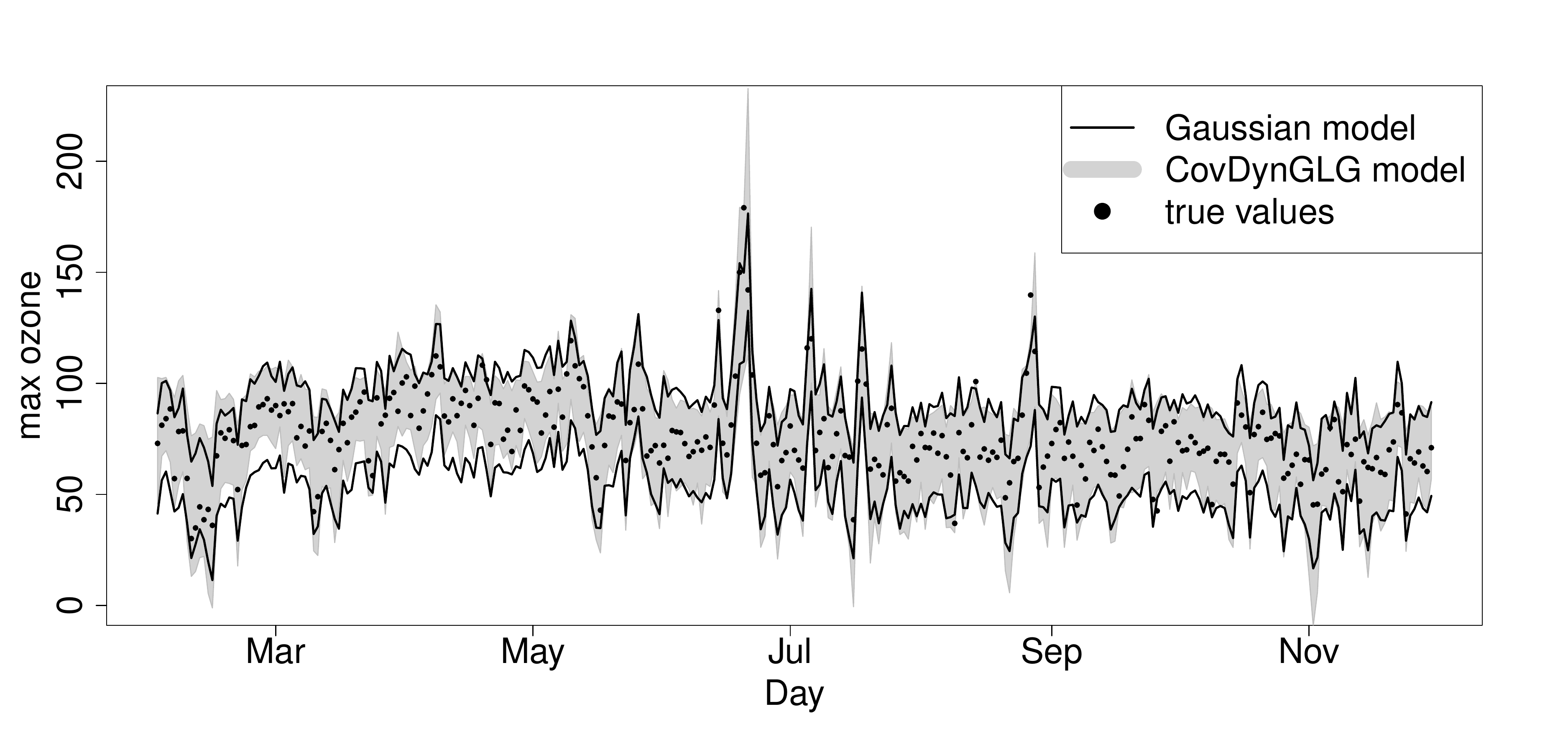} \\
(a) ${\bf s}=(50.74,-1.83)$\\
\includegraphics[width=12cm]{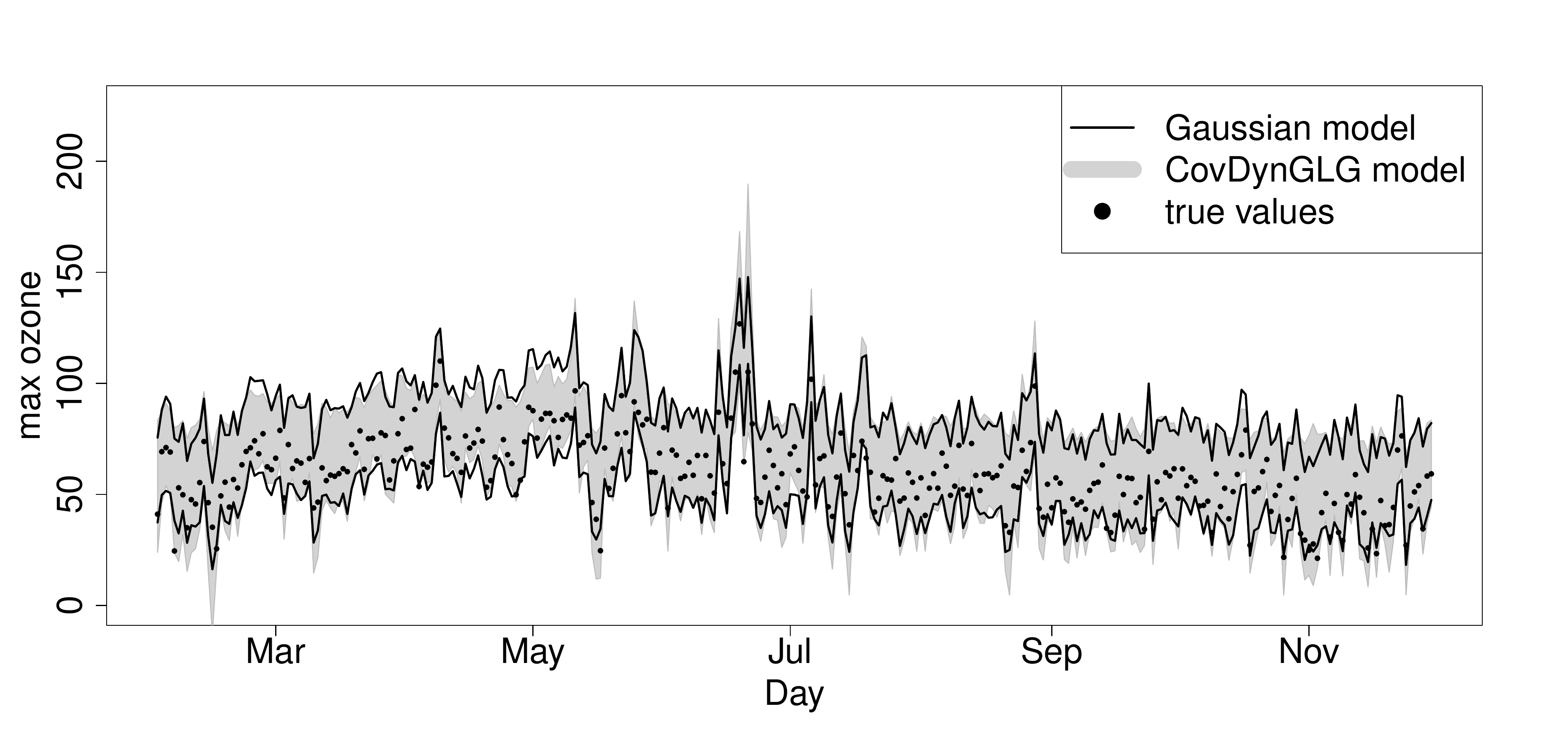} \\
(b) ${\bf s}=(52.95,-1.15)$\\
\end{tabular}
\caption{Ozone data: Predictive distribution ($95\%$ interval) over time for the CovDynGLG model and the Gaussian model for ${\bf s}=(50.74,-1.83)$ and ${\bf s}=(52.95,-1.15)$. }
\label{figUKvar2}
\end{center}
\end{figure}

Figure \ref{figUKvar1} presents the posterior summaries for the standard deviation of $z_t(\mathbf{s})$, $\mathbf{s}=(50.74,-1.83)$ for the CovDynGLG model compared to the Gaussian model. Note that the variance is non-constant with large peaks in June and July. Differently, the Gaussian dynamic model suggests a nearly constant standard deviation across time.
This pattern has a direct effect on the predictive uncertainty of the Gaussian Model which does not capture many extreme observations and tend to have greater variability across the observed period, which is clear from panels of Figure \ref{figUKvar2}.  Note that model CovDynGLG captures the periods of extreme values of ozone while it has shorter ranges of the 95\% credible intervals for those periods that observations do not change much across time. Regarding the complete model that includes the regression components in the equation for $\lambda_1(\bm{s})$, the 95\% posterior credible interval for latitude is $(-0.0107,-0.0043)$ and for longitude is $(-0.1778,-0.0442)$, suggesting that both variables have a negative association with the precision over space. This results in smaller predictive precision for south-eastern locations.



\newpage
\clearpage

\section{Conclusions}\label{sec:conclusion}
We have proposed a flexible dynamical non-Gaussian spatio-temporal model that extends the well known multivariate dynamic linear model and accommodates both outliers and regions in space or time with larger observational variance. The dynamic evolution in the variance model proposed in equation (\ref{eq:lambda21}) is able to account for different regimes of variability over time, which is a desirable feature when modelling environmental data in large temporal windows. For instance, the most complete models with covariates aiding in the representation of uncertainty over space and time presented the best performances in predicting the maximum ozone in the UK. This result indicates that patterns in periods of large variability could be explained by changes in wind and temperature that not only influence the mean but also have an impact on the description of the variance of the process. This results in a better description of the uncertainty associated with temporal predictions and spatial interpolations of interest. As inference is performed under the Bayesian paradigm using MCMC methods, we proposed an efficient sampling algorithm for inference and prediction. It takes advantage of the conditionally Gaussian distributions obtained when we condition the distribution of $Z_t({\bf s})$ on the mixing latent variables. 

As shown in Section \ref{sec2.2}, the proposed model allows the resultant variance structure to change across space and time depending on the effect of covariates. Moreover, the kurtosis depends on the mixing scales $\nu_1$ and $\nu_2$, which reflect the inflation in the tails when necessary. The correlation structure, on the other hand will not change with the covariates but will have the effect of the correlation structure assumed for the variance model. The model generalizes the well known Gaussian model for spatio-temporal data and adds flexibility to the alternative Student-t model. Although the Student-t model allows for variance inflation, it increases the variance of the process in every location and does not allow for local changes in variability as our proposal does.


We performed extensive simulation studies to investigate the ability of the proposed model to capture different structures of the spatio-temporal process of interest. Our simulated examples in Section D of the Supplementary Material indicate that the correct model is selected with the complete model having worse performance when the data does not have the effect of covariates in the spatial mixing process. The non-Gaussian proposals have equivalent performance when fitted to the Gaussian simulated data. Thus, complexity is not always preferred suggesting that our model does not lead to overfitting. Moreover it seems  that the predictive scoring rules used to compare the models are adequate measures of good predictive performance. 

We conclude that allowing for a flexible model for the variance of the process provides coherent posterior predictive credible intervals that accommodates well the structure of the spatio-temporal process under study. A possible drawback of the proposed approach is that prediction of the process to future instants in time depend on covariates that are themselves spatio-temporal processes that need to be predicted. One possible solution is to consider a multivariate spatio-temporal process which is subject for future research.



\section*{Acknowledgments}
Schmidt is grateful for ﬁnancial support from the Natural Sciences and Engineering Research Council (NSERC) of Canada (Discovery Grant RGPIN-2017-04999).


\newpage

\appendix

\section{Proofs of properties of the proposed model }\label{ApB}
In this appendix, we prove the results shown in Section \ref{sec2.2}. Consider the spatio-temporal model in \eqref{model:eq2} 
 with nugget effect $\tau=0$. The covariance function of the spatio-temporal process, conditional on the state parameters $\boldsymbol{\eta}_{1:T}$, and $\boldsymbol{\theta}_{1:T}$, is given by $C(\bm{s}_1,\bm{s}_2,t_1,t_2)=Cov\left[Z_{t_1}(\bm{s}_1), Z_{t_2}(\bm{s}_2) \mid  \boldsymbol{\eta}_{1:T}, \boldsymbol{\theta}_{1:T}\right]$ with

\begin{eqnarray}\nonumber
 C(\bm{s}_1,\bm{s}_2,t_1,t_2)&=& 
\sigma^2 Cov \left[ \ \frac{\epsilon_{t_{1}}(\bm{s}_1)}{\sqrt{\lambda_{t_1}(\bm{s}_1)}},  \frac{\epsilon_{t_{2}}(\bm{s}_2)}{\sqrt{\lambda_{t_2}(\bm{s}_2)}}  \right] \\ \nonumber
 &=& \sigma^2 E \left[ \ \frac{\epsilon_{t_{1}}(\bm{s}_1)}{\sqrt{\lambda_{t_1}(\bm{s}_1)}} \times  \frac{\epsilon_{t_{2}}(\bm{s}_2)}{\sqrt{\lambda_{t_2}(\bm{s}_2)}}  \right]\\ \nonumber  
 &=& \sigma^2 E\left[\epsilon_{t_{1}}(\bm{s}_1)\epsilon_{t_{2}}(\bm{s}_2)\right]E\left[\lambda^{-1/2}_{t_1}(\bm{s}_1) \lambda^{-1/2} _{t_2}(\bm{s}_2)\right]\\ \nonumber
 &=& \sigma^2 C_{\bm{\psi}}(\mathbf{s}_1, \mathbf{s}_2) E\left[\lambda^{-1/2} _{t_1}(\bm{s}_1)\lambda^{-1/2}_{t_2}(\bm{s}_2) \right] \\ \nonumber 
 &=& \sigma^2 C_{\bm{\psi}}(\mathbf{s}_1, \mathbf{s}_2) E\left[ \exp{\left\{ -\frac{1}{2} ln \thinspace \lambda_{t_1}(\bm{s}_1) -\frac{1}{2} ln \thinspace \lambda_{t_2}(\bm{s}_2)\right\}}\right]\\ \nonumber
 &=& \sigma^2 C_{\bm{\psi}}(\mathbf{s}_1, \mathbf{s}_2) \exp\left\{ \frac{\nu_1}{4} \left(C_{{\bm{\xi}}}(\mathbf{s}_1, \mathbf{s}_2) +3\right) \right. + \\ \nonumber
 & &\left. \frac{3}{4}\nu_2 -\bm{F}_1'\bm{\beta} -\frac{1}{2}\left[ {\bm F}
 '_{2t_1}{\bm \eta}_{t_1} + {\bm F}'_{2t_2}{\bm \eta}_{t_2}  \right] +\frac{1}{4} Cov\left( ln \lambda_{2t_1} , ln \lambda_{2t_2}\right)\right\}. \nonumber
\end{eqnarray}
Let $U = -\frac{1}{2} \thinspace ln \thinspace \lambda_{t_1}(\bm{s}_1) - \frac{1}{2} \thinspace ln \thinspace \lambda_{t_2}(\bm{s}_2)$. Then, $U$ follows a Gaussian distribution with
$$ E (U) =\frac{\nu_1}{2} +\frac{\nu_2}{2} -\bm{F}_1'\bm{\beta} - \frac{1}{2}\left({\bm F}'_{2{t_1}}{\bm \eta}_{{t_1}} +{\bm F}'_{2{t_2}}{\bm \eta}_{{t_2}}\right)$$ and $$ Var(U)=   \frac{\nu_1}{2}\left( C_{\bm{\xi}}(\mathbf{s}_1,\mathbf{s}_2) +1 \right) +\frac{\nu_2}{2} +\frac{1}{2}Cov\left( ln \lambda_{2t_1} , ln \lambda_{2t_2}\right),$$
Note that, $\lambda_{t_1}(\bm{s}_1) = \lambda_1(\bm{s}_1)\lambda_{2t_{1}}$ and $\lambda_{t_2}(\bm{s}_2) = \lambda_1(\bm{s}_2)\lambda_{2t_{2}}$. In this case, $ln \lambda_{t_1}(\bm{s}_1) = ln \lambda_1(\bm{s}_1) + ln \lambda_{2t_{1}}$ and  $ln \lambda_{t_2}(\bm{s}_2) = ln \lambda_1(s_2)+ ln \lambda_{2t_{2}}$, respectively. Notice that, if $t_1 \neq t_2$ and conditional on the state variables $\boldsymbol{\eta}_{1:T}$, and $\boldsymbol{\theta}_{1:T}$, we have $Cov\left( ln \lambda_{2t_1} , ln \lambda_{2t_2}\right)=0$. As we assume a DLM for $ln \lambda_{2t_1}$ so if $t_1 = t_2=t$ then $Cov\left( ln \lambda_{2t_1} , ln \lambda_{2t_2}\right)=Var( ln \lambda_{2t})$. \\

Conditionally on $\boldsymbol{\eta}_{1:T}, \boldsymbol{\theta}_{1:T}$, the spatio-temporal variance is
\begin{eqnarray}\nonumber
Var\left[Z_t(\mathbf{s})\mid \boldsymbol{\eta}_{1:T}, \boldsymbol{\theta}_{1:T}\right] &=& Cov\left[Z_t(\mathbf{s}), Z_t(\mathbf{s})\mid  \boldsymbol{\eta}_{1:T}, \boldsymbol{\theta}_{1:T} \right] =\sigma^2 \thinspace exp\left\{\nu_1 -\bm{F}_1'\bm{\beta} +\nu_2  -\bm{F}_{2t}'\bm{\eta}_{{t}}\right\} ,\\ \nonumber
\end{eqnarray}
and the conditional correlation function is given by  $\rho(\bm{s}_1,\bm{s}_2,t_1,t_2)=Corr\left[Z_{t_1}(\bm{s}_1), Z_{t_2}(\bm{s}_2) \mid  \boldsymbol{\eta}_{1:T}, \boldsymbol{\theta}_{1:T}\right]$ with
\begin{small}
\begin{eqnarray} \nonumber
\rho(\bm{s}_1,\bm{s}_2,t_1,t_2) &=& \frac{Cov\left[Z_{t_1}(\bm{s}_1), Z_{t_2}(\bm{s}_2)\right]}{\sqrt{Var\left[Z_{t_1}(\bm{s}_1)\right]} \sqrt{Var\left[Z_{t_2}(\bm{s}_2)\right]}}  \\ \nonumber
&=& \frac{\sigma^2 C_{\bm{\psi}}(\mathbf{s}_1, \mathbf{s}_2) exp\left\{ \frac{\nu_1}{4}\left(C_{\bm{\xi}}(\mathbf{s}_1, \mathbf{s}_2) +3\right) +\frac{3}{4}\nu_2 -\bm{F}_1'\bm{\beta} -\frac{1}{2}\left[ \bm{F}'_{2t_1}\bm{\eta}_{t_1} + \bm{F}'_{2t_2}\bm{\eta}_{t_2}  \right] \right\}}{\sigma^2 exp\left\{\nu_1 + \nu_2  -\bm{F}_1'\bm{\beta}  -\frac{1}{2}\left[ \bm{F}'_{2t_1}\bm{\eta}_{t_1} + \bm{F}'_{2t_2}\bm{\eta}_{t_2}  \right]\right\} } \\ \nonumber
&=& C_{\bm{\psi}}(\mathbf{s}_1, \mathbf{s}_2) exp\left\{ \frac{\nu_1}{4}\left(C_{\bm{\xi}}(\mathbf{s}_1,\mathbf{s}_2) -1\right)  -\frac{1}{4}\nu_2\right\}. \nonumber
\end{eqnarray}
\end{small}
The expression of the  kurtosis unconditional on the mixing process is given by
\begin{eqnarray} \nonumber
Kurt\left[Z_t(\mathbf{s})\right]& =& \frac{E\left[\left(Z_t(\bm{s}) - E(Z_t(\bm{s}))\right)^4\right]}{[Var(Z_t(\bm{s}))]^2}= \frac{E\left[\left(\mathbf{x}_t(\bm{s})'\boldsymbol{\theta}_t + \sigma \frac{\epsilon_t(\bm{s})}{\sqrt{\lambda_t(\bm{s})}}-  \mathbf{x}_t(\bm{s})'\boldsymbol{\theta}_t\right)^4 \right]}{[Var(Z_t(\bm{s}))]^2} \\ \nonumber 
& =& \frac{E\left[\left(\sigma \frac{\epsilon_t(\bm{s})}{\sqrt{\lambda_t(\bm{s})}}\right)^4 \right]}{[Var(Z_t(\bm{s}))]^2} =  \frac{\sigma^4 E\left[\epsilon_t^4(\bm{s}) \right] E\left[ \lambda_t^{-2}(\bm{s})\right]}{[Var(Z_t(\bm{s}))]^2} = 3 \thinspace exp\left\{\nu_1 +\nu_2 \right\}.  \\ \nonumber 
& =& 3 \thinspace exp\left\{\nu_1 +\nu_2 \right\}.  \nonumber 
\end{eqnarray}
Note that  $-2 \thinspace ln \thinspace \lambda_t(\bm{s}) \sim N\left(\nu_1 +\nu_2 -2\bm{F}_1'\bm{\beta}-2 \bm{F}'_{2t} \bm{\eta}_{t}; 4(\nu_1 + \nu_2)\right)$, so it follows that
$$ E\left[ \lambda^{-2}_t(\bm{s})\right] =  E\left[  exp \left\{ {-2} \thinspace ln \thinspace \lambda_t(\bm{s}) \right\}\right] = exp \left\{ 3(\nu_1 + \nu_2) - 2\bm{F}_1'\bm{\beta} -2 \bm{F}'_{2t} \bm{\eta}_{t}\right\}.$$ 

In a model without nugget effect, the kurtosis of a Gaussian distribution is equal to 3. If $\nu_1$ and $\nu_2$ take small values in our proposed model this implies that $Kurt[Z_t(\mathbf{s})] = 3$ and therefore we have a Gaussian process. In a Student-t process with $v$ degrees of freedom, the expression of the kurtosis is given by
\begin{eqnarray} \nonumber
Kurt\left[Z_t(\mathbf{s})\right]& =& \frac{E\left[\left(Z_t(\bm{s}) - E(Z_t(\bm{s}))\right)^4\right]}{\left[Var(Z_t(\bm{s}))\right]^2}  = \frac{\frac{3v^2\sigma^4}{(v-2)(v-4)}}{\left[Var(Z_t(\bm{s}))\right]^2} \\ \nonumber 
& =& \frac{\frac{3v^2\sigma^4}{(v-2)(v-4)}}{\frac{v^2 \sigma^4 }{(v-2)^2}} = 3 \; \frac{(v-2)}{(v-4)}, \hspace{0.3cm} v>4. \nonumber 
\end{eqnarray}
For $v \rightarrow \infty$ we have kurtosis equals 3, that is, the convergence to a Gaussian process and it will do not depend over time.

\section{Description of the MCMC algorithm}\label{ApPostComp}

Below we describe the general steps to sample from the posterior full conditional distributions of the model proposed in Section 2 of the manuscript.

\begin{itemize}
    \item [step 1.] Sample $\left(\boldsymbol{\theta}_{0:J}^{(k)} \mid \sigma^{2{(k-1)}},\tau^{2{(k-1)}}, \bm{\psi}^{(k-1)}, \boldsymbol{\lambda}^{(k-1)}_{1,1:n}, \boldsymbol{\lambda}^{(k-1)}_{2,0:J}, D_t \right)$ using Forward Filtering Backward Sampling (FFBS);
    \item [step 2.] Sample $\left(\boldsymbol{\eta}_{0:J}^{(k)} \mid \boldsymbol{\theta}_{0:J}^{(k)}, \boldsymbol{\lambda}^{(k-1)}_{2,0:J}, \nu_2^{(k-1)} , D^{*}_t\right)$ using Forward Filtering Backward Sampling (FFBS);
    \item [step 3.] Sample $\left(\boldsymbol{\lambda}_{1,1:n}^{(k)} \mid \boldsymbol{\theta}^{(k)}_{0:J}, \sigma^{2{(k-1)}},\tau^{2{(k-1)}}, \bm{\xi}^{(k-1)}, \nu_1^{(k-1)}, D_t\right)$ from their posterior full conditional distributions, defined from equations (\ref{eq5a}) and (\ref{eq:lambda1}), using Metropolis-Hastings steps with random walk proposals.
    \item [step 4.] Sample $\left(\boldsymbol{\lambda}_{2, {0:J}}^{(k)} \mid \boldsymbol{\theta}^{(k)}_{0:J},\boldsymbol{\eta}_{0:J}^{(k)}, \nu_2^{(k-1)}, D_t \right)$ from their posterior full conditional distributions, defined from equations (\ref{eq5a}) and (\ref{eq:lambda21}), using Metropolis-Hastings steps with random walk proposals.
\item [step 5.] Sample the static parameters ($\Phi$) from their posterior full conditional distributions using Metropolis-Hastings steps with random walk proposals. 
\end{itemize}

The inference of the state parameter vectors present in dynamic linear models follows the usual steps in Bayesian inference. Two main operations are considered: the evolution to build up the prior and the updated to incorporate the new observation arrived at time $t$. The following subsections describe in detail the FFBS algorithm for sampling the state parameter vectors $\bm{\theta}_t$ and $\bm{\eta}_t$ present, respectively, in the equations of the mean function of the outcome and of the variance component of the model.

\subsection{FFBS for the state parameter vector in the mean function of the outcome}\label{ApA1}
\textit{Forward filtering equations:}
\begin{itemize}
\item[--] Posterior distribution at time $t-1$: $
\boldsymbol{\theta}_{t-1}|\textbf{D}_{t-1} \sim N(\textbf{m}_{t-1},C_{t-1}); 
$

\item[--] Prior distribution at time $t$: $
\boldsymbol{\theta}_{t}|\textbf{D}_{t-1} \sim N(\textbf{a}_{t},R_{t}),$
\noindent
with $\textbf{a}_{t} = G_{t}\textbf{m}_{t-1}$ and $R_{t}=G_{t}C_{t-1}G'_{t}+W_{t} ;$

\item[--] One step ahead prediction:
$ \textbf{Z}_{t}|\textbf{D}_{t-1} \sim N(\textbf{f}_{t},Q_{t}), $
\noindent
with $\textbf{f}_{t} = \textbf{F}'_{t}\textbf{a}_{t}$ and $Q_{t}=\textbf{F}'_{t}R_{t}\textbf{F}_{t}+V_t^{(1)}$, $V_t^{(1)}=\sigma^2\Lambda_t^{1/2}V\Lambda_t^{1/2}$;

\item[--] Posterior distribution at time $t$: $\boldsymbol{\theta}_{t}|\textbf{D}_{t} \sim N(\textbf{m}_{t},C_{t}),$
\noindent
with $\textbf{m}_{t} = \textbf{a}_{t}+\textbf{A}_{t}\textbf{e}_{t}$ and $C_{t}=R_{t}-A_{t}Q_{t}A'_{t}$ and $A_{t}=R_{t}\textbf{F}_{t}Q_{t}^{-1}$, $\textbf{e}_{t}=\textbf{z}_{t}-\textbf{f}_{t}$.
\end{itemize}

\textit{Backward Sampling equations:}\\

This step is computed retrospectively, using the following decomposition:
$$p(\boldsymbol{\theta}_0,...,\boldsymbol{\theta}_T|\textbf{D}_T)=p(\boldsymbol{\theta}_T|\textbf{D}_T)\prod_{t=0}^{T}p(\boldsymbol{\theta}_t|\boldsymbol{\theta}_{t+1},\textbf{D}_t).$$

From Bayes Theorem, for $t=T-1,...,0$:
$$ p(\boldsymbol{\theta}_t|\boldsymbol{\theta}_{t+1},\textbf{D}_T) \propto p(\boldsymbol{\theta}_{t+1}|\boldsymbol{\theta}_{t},\textbf{D}_T)p(\boldsymbol{\theta}_t|\textbf{D}_T),  $$
with $\boldsymbol{\theta}_t|\boldsymbol{\theta}_{t+1},\textbf{D}_T \sim N(\textbf{h}_t,H_t)$, with 
$$\textbf{h}_t = \textbf{m}_t+C_t G'_{t+1}R_{t+1}^{-1}(\boldsymbol{\theta}_{t+1}-\textbf{a}_{t+1}),$$ 
$$H_t = C_t - C_tG'_{t+1}R_{t+1}^{-1}G_{t+1}C_t,$$ and $\textbf{h}_T = \textbf{m}_T$ e $H_T = C_T$, the initial values. \\

For $W_t$ it is possible to use a discount factor $\delta_1 \in (0,1)$ subjectively evaluated, controlling the loss of information. In this case $R_t$ is recalculated according to a discount factor $\delta_1$ such as $W_t = \frac{1- \delta_1}{\delta_1} G'_t C_{t-1} G_t$. Notice that $R_t$ can be rewritten as $R_t=  G'_t C_{t-1}G_t/ \delta_1$. For more details, see \cite{West97}.

\subsection{FFBS for the state parameter vector in the variance model\label{ApA2}}

\textit{Forward filtering equations:}
\begin{itemize}
\item[--] Posterior distribution at time $t-1$: $
\boldsymbol{\eta}_{t-1}|\mathbf{D}_{t-1}^{*} \sim N(\bm{m}_{t-1}^{(\eta)},C_{t-1}^{(\eta)}); 
$

\item[--] Prior distribution at time $t$: $
\boldsymbol{\eta}_{t}|\mathbf{D}_{t-1}^{*} \sim N(\bm{a}_{t}^{(\eta)},R_{t}^{(\eta)}),$
\noindent
with $\bm{a}_{t}^{(\eta)} = \bm{m}_{t-1}^{(\eta)}$ and $R_{t}^{(\eta)}= G_{2t}C_{t-1}^{(\eta)}G'_{2t} + W_{2t} ;$

\item[--] One step ahead prediction:

$ {L}_{t}|\mathbf{D}^{*}_{t-1} \sim N({f}_{t}^{(\eta)},Q_{t}^{(\eta)}), $
\noindent
with ${f}_{t}^{(\eta)} = \bm{F}_{2t}'\bm{a}_{t}^{(\eta)}$ and $Q_{t}^{(\eta)}= \bm{F}_{2t}'R_{t}^{(\eta)}\bm{F}_{2t}+V_t^{(2)}$, 

\item[--] Posterior distribution at time $t$: $\boldsymbol{\eta}_{t}|\textbf{D}^{*}_{t} \sim N(\bm{m}_{t}^{(\eta)},C_{t}^{(\eta)}),$
\noindent
with $\bm{m}_{t}^{(\eta)} =\bm{a}_{t}^{(\eta)}+{A}_{t}^{(\eta)}{e}_{t}^{(\eta)}$ and $C_{t}^{(\eta)}=R_{t}^{(\eta)}-A_{t}^{(\eta)}Q_{t}^{{-1}^{(\eta)}}A^{'(\eta)}_{t}$ and $A_{t}^{(\eta)}=R_{t}^{(\eta)}\bm{F}_{2t}Q_{t}^{{-1}^{(\eta)}}$, ${e}^{(\eta)}_{t}={L}^{*}_{t}-{f}_{t}^{(\eta)}$.
\end{itemize}
Analogously to Appendix \ref{ApA1}, having completed the forward filtering over time, backward sampling is computed from the posterior $p(\boldsymbol{\eta}_{1:T} \mid \mathbf{D}^{*}_{T})$ with $t=1, \ldots, T$.\\

\textit{Backward Sampling equations:}
\begin{itemize}
\item[--] This step is computed retrospectively, using the following decomposition:
\end{itemize}
$$p\left(\boldsymbol{\eta}_0,...,\boldsymbol{\eta}_T|\textbf{D}_T^{*}\right)=p(\boldsymbol{\eta}_T|\textbf{D}_T^{*})\prod_{t=0}^{T}p(\boldsymbol{\eta}_t|\boldsymbol{\eta}_{t+1},\textbf{D}_t^{*}).$$

From Bayes Theorem, for $t=T-1,...,0$:
$$ p\left(\boldsymbol{\eta}_t|\boldsymbol{\eta}_{t+1},\textbf{D}_{T}^{*}\right) \propto p(\boldsymbol{\eta}_{t+1}|\boldsymbol{\eta}_{t},\textbf{D}_T^{*})p(\boldsymbol{\eta}_t|\textbf{D}_{T}^{*}),  $$
with $\boldsymbol{\eta}_t|\boldsymbol{\eta}_{t+1},\textbf{D}_{T}^{*} \sim N(\bm{h}_t,H_t)$,  $\bm{h}_t = \bm{m}_{t}^{(\eta)}+C_{t}^{(\eta)} G{'}_{2,t+1}R_{t+1}^{-1}({\boldsymbol{\eta}_{t+1}}-{\bm{a}^{(\eta)}_{t+1}})$, $H_{t} = C_{t}^{(\eta)} -C_{t}^{(\eta)}G{'}_{2t+1} R^{(\eta)}_{t+1}{}^{-1}G_{2t+1}C_{t}^{(\eta)}$  and $\bm{h}^{(\eta)}_T = \bm{m}_{T}^{(\eta)}$ e $H_T = C_{T}^{(\eta)}$, the initial values. For $\boldsymbol{\omega}_{2t}$, we use a discount factor $\delta_2 \in (0,1)$ subjectively evaluated, controlling the loss of information.

\section{Simulated data examples}\label{ApSimD}

This appendix presents two simulated examples to investigate the performance of the predictive scoring rules in identifying the data generating model. Both data sets consist of $I = 64$ spatial
locations (5 for testing and 59 for training) and $J = 303$ time points with covariates and latitudes and longitudes based on the ozone dataset presented in \ref{sec:real2}. Priors on parameters are the same used in the application to ozone data. Convergence of chains were tested using the Z statistic of \cite{Gewe92} which computes equality of the means for the first ($10\%$) and last part ($50\%$) of a Markov Chain. The burnin and lag for spacing of the chain were selected so that the effective sample size were around 1,000 samples.

\subsection{Gaussian data}

\begin{figure}[!ht]
\centering
\includegraphics[width=15cm]{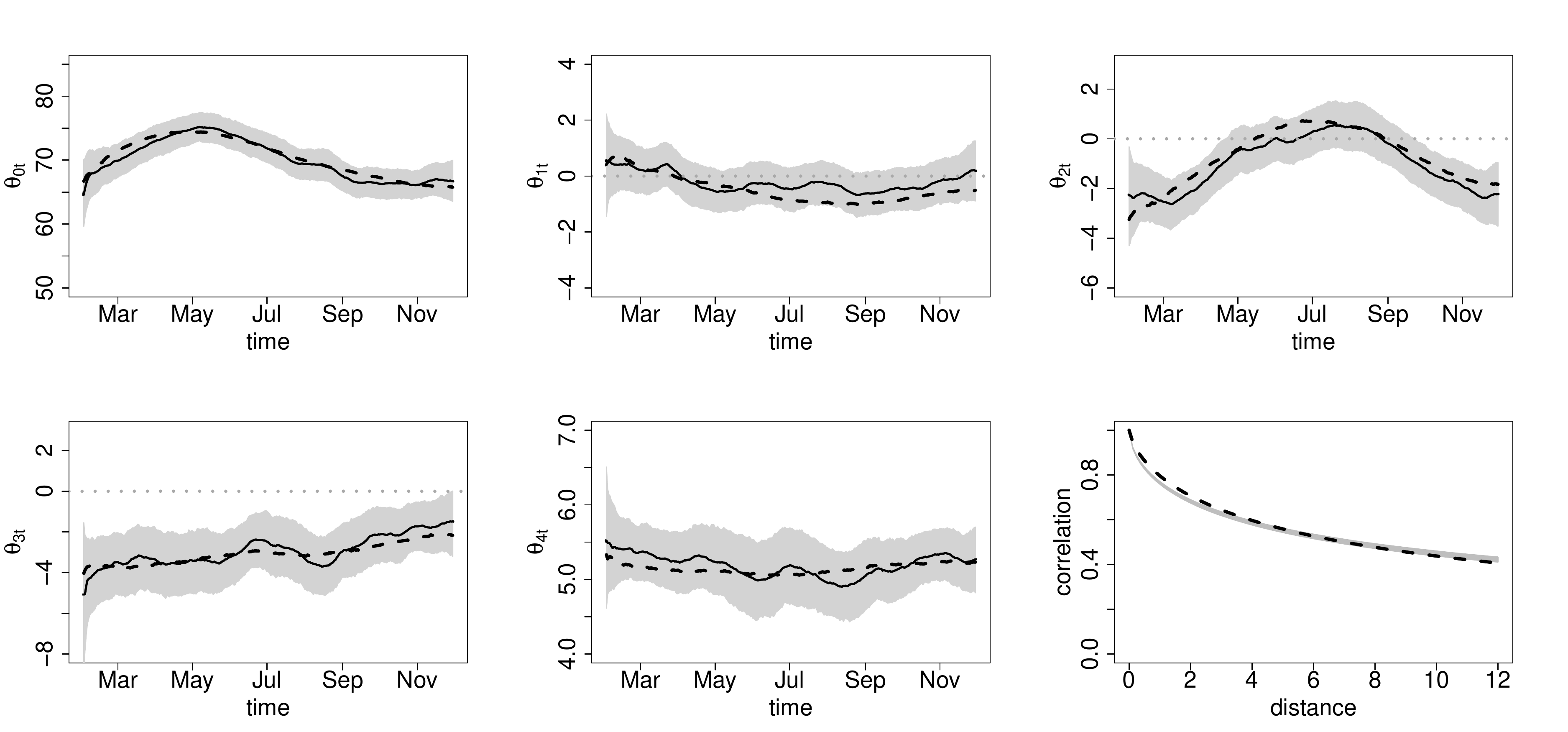} 
\caption{Simulated Gaussian data: Posterior median (solid line), $95\%$ credible intervals and true values (dashed line) for the dynamic coefficients and correlation function estimated assuming the Gaussian (G) model.\label{secGa}}
\end{figure}

\begin{figure}
\centering
\begin{tabular}{ccc}
\includegraphics[width=4.6cm]{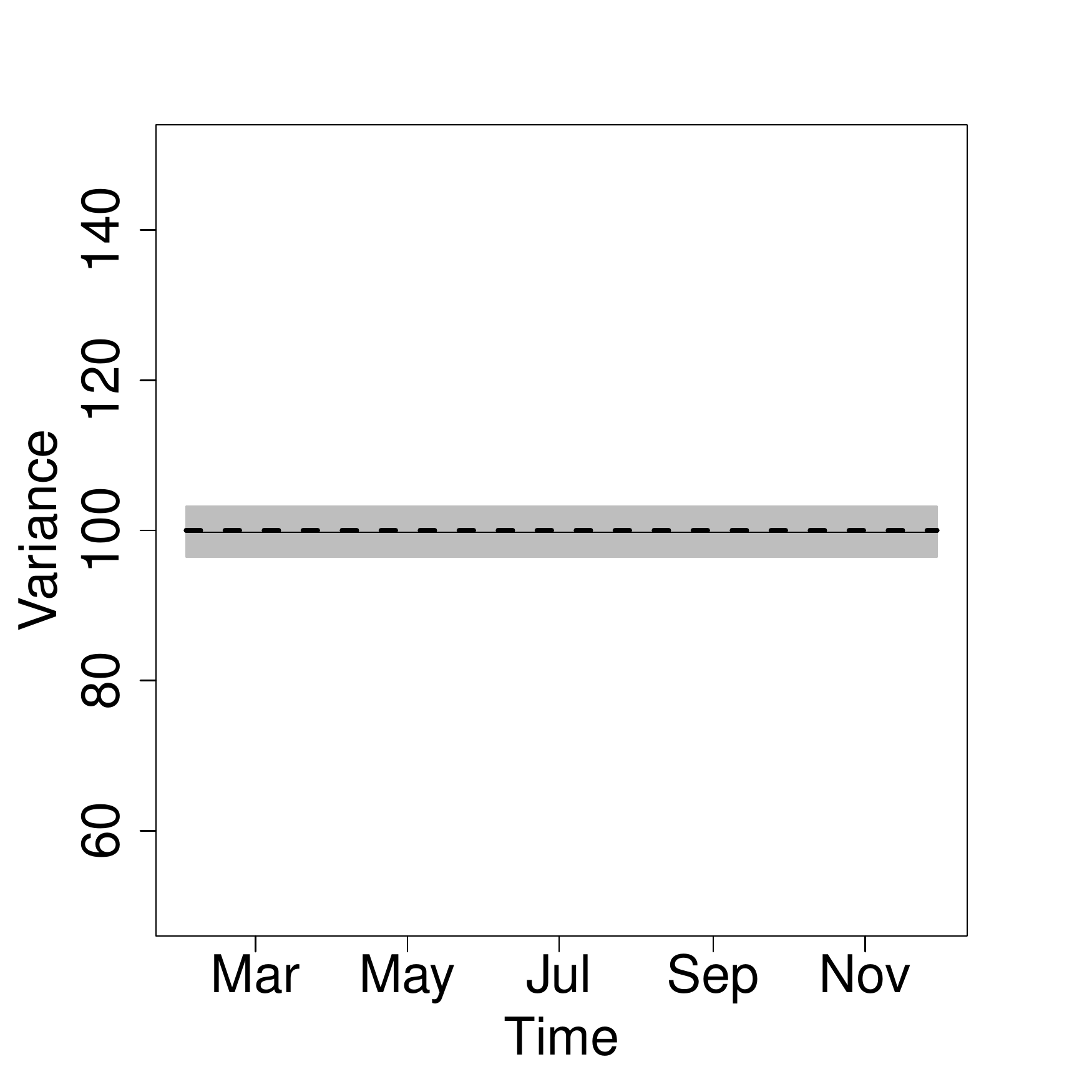} &
\includegraphics[width=4.6cm]{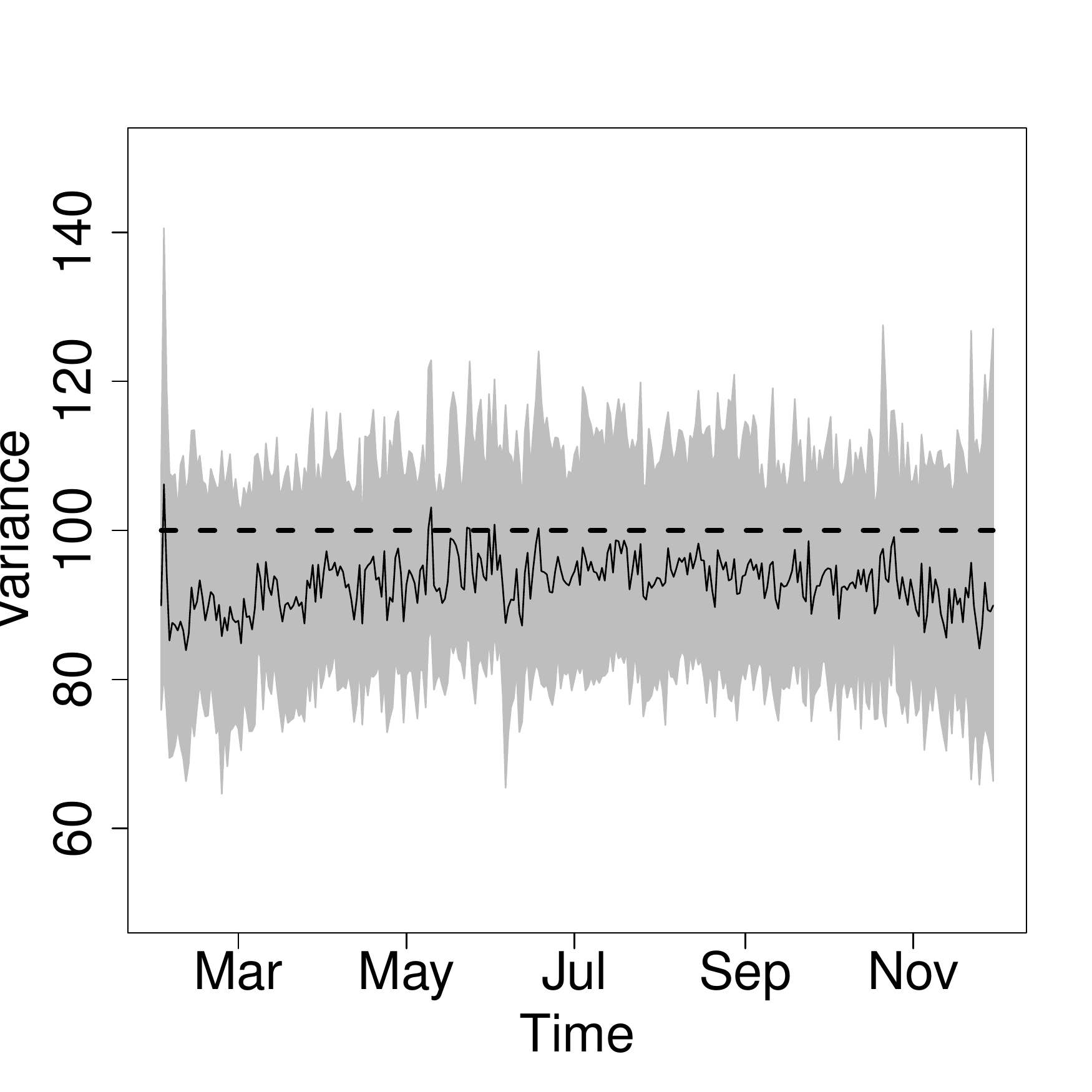}
&
\includegraphics[width=4.6cm]{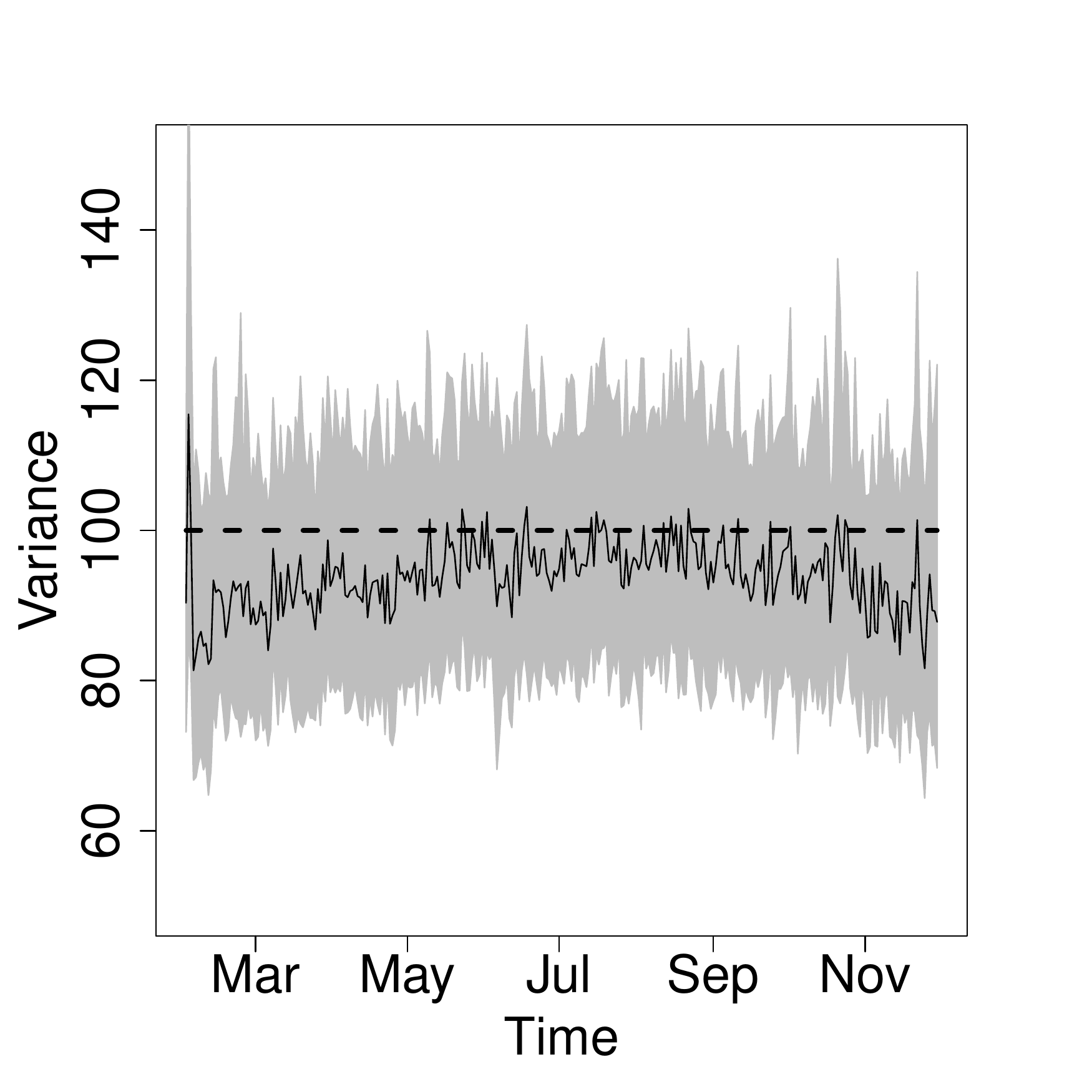}
\\
(a) Gaussian & (b)  CovDyn & (c) CovDynGLG \\
\end{tabular}
\caption{Simulated Gaussian data: Posterior median (full line), 95\% credible intervals and true values (broken line) for the temporal variance estimated assuming the Data Generating model (G), the Dynamical Gaussian-Log-Gaussian model (CovDyn) and the Dynamical Gaussian-Log-Gaussian model (CovDynGLG) .\label{secGb}}

\end{figure}

The first data set was simulated from a Gaussian process $\{Z_t({\bm s}) : {\bm s} \in D, t \in T \}$ assuming the mean function $m_t(\mathbf{s}) = \theta_{0t} + \theta_{1t}\thinspace lat(\mathbf{s}) + \theta_{2t}\thinspace long (\mathbf{s}) + \theta_{3t} \thinspace temp_t(\mathbf{s})+ \theta_{4t} \thinspace wind_t(\mathbf{s})$, and the spatial correlation function $c(s,s')=(1+||s-s'||/\phi)^{-\alpha}$, which were defined based on the posterior point estimates obtained for the parameters in the ozone data application. Priors on parameters are the same used in the fitting of the ozone data. All covariates were the real variables from the ozone data application. 
Figure \ref{secGa} shows that the dynamic coefficients and the correlation function are well recovered by the Gaussian model. This is an expected result, since the data are generated
from a Gaussian process. Figure \ref{secGb} illustrates the temporal variance estimated for the Gaussian, the CovDyn and the CovDynGLG models. As we can see, the CovDyn and the CovDynGLG models can recover the true variance that is constant over time with smooth peaks of variability (due to the covariates structure in the mean of the variance process). For prediction purposes, Table \ref{tabsimulated1} shows model comparison based on scoring rules criteria indicating that all fitted models have similar performances, mainly when we compare the interval score which highlights the similar predictive uncertainty recovered by all competing models. This was expected as the non-Gaussian alternatives have the Gaussian process as a limiting case. 
 In particular, the parameters $\nu_1$, $\nu_2$ and $\boldsymbol{\eta}_t$ tends to small values, retrieving Normal tails. 

 \begin{table}
     \caption{Model comparison based on the Interval Score (IS), the Log Predictive Score (LPS) and the Variogram Score of order 0.25 (VS-0.25) criteria for the predicted observations at the out-of-sample locations under all fitted models for the simulated Gaussian dataset.\label{tabsimulated1}}
  \centering   
 \fbox{   \begin{tabular}{cccccccc}
    \hline
  &  G  & ST & GLG & CovDyn & DynGLG & CovDynGLG & Full\\
    \hline
     IS & 20  &20 & 20& 20&  20& 20 &20\\
 LPS &  4290  &4283  & 4286 & 4286  & 4287 & 4289&4285\\
VS-0.25 & 4902  &4885 & 4886 & 4883  & 4884 & 4887&4898 \\
         \hline
    \end{tabular}}
\end{table}



\newpage
\subsection{Non-Gaussian data}

The second data set was simulated by modifying the first dataset by assuming  $Z_t({\bm s})=m_t({\bm s})+\epsilon_t({\bm s})/\sqrt{\lambda_1({\bm s})\lambda_{2t}}$. The mixing variables were simulated as $ln(\lambda_{2t})\sim N(-\nu_2/2+\mu_{t},\nu_2)$ and $\lambda_1({\bm s})$ from a Gaussian process with mean function $-\nu_1/2$ and spatial correlation function $c_1(s,s')=\exp\{-||s-s'||/\gamma\}$, which were defined based on the posterior point estimates obtained for the parameters in the ozone data application. The parameter $\mu_t$ was defined as $\mu_t=0.5\; sin(t\pi /J)+0.5\; cos(2t\pi/J)$, so that the data generating model is closest to the DynGLG or CovDynGLG which can accommodate patterns in the spatiotemporal variance. Indeed, both the DynGLG and CovDynGLG models can recover the true variance (Figure \ref{simulNGvar}). The Gaussian, Student-t and GLG models estimate constant variances over time. This has a direct impact on the predictive performance as presented in Table \ref{tabSimulapred}. The best models according to the predictive measures are the DynGLG and CovDynGLG. Note that both the temporal and spatial components are important in the variance estimation, with the GLG and CovDyn, which are purelly spatial and temporal respectively, having worse performances when compared to the models with both components. Moreover, note that the full model that includes a regression in $\lambda_1(\bm{s})$ is not selected as the best model. Indeed, this was expected as we did not include covariates to simulate the spatial mixing process. The $95\%$ credibility intervals for these regression coefficients indicate that they are non-significant with $IC(95\%,\beta_1)=(-0.0056,  0.0005)$ and $IC(95\%,\beta_2)=(-0.0855, 0.0265)$.

 

\begin{figure}
\centering
\includegraphics[width=15cm]{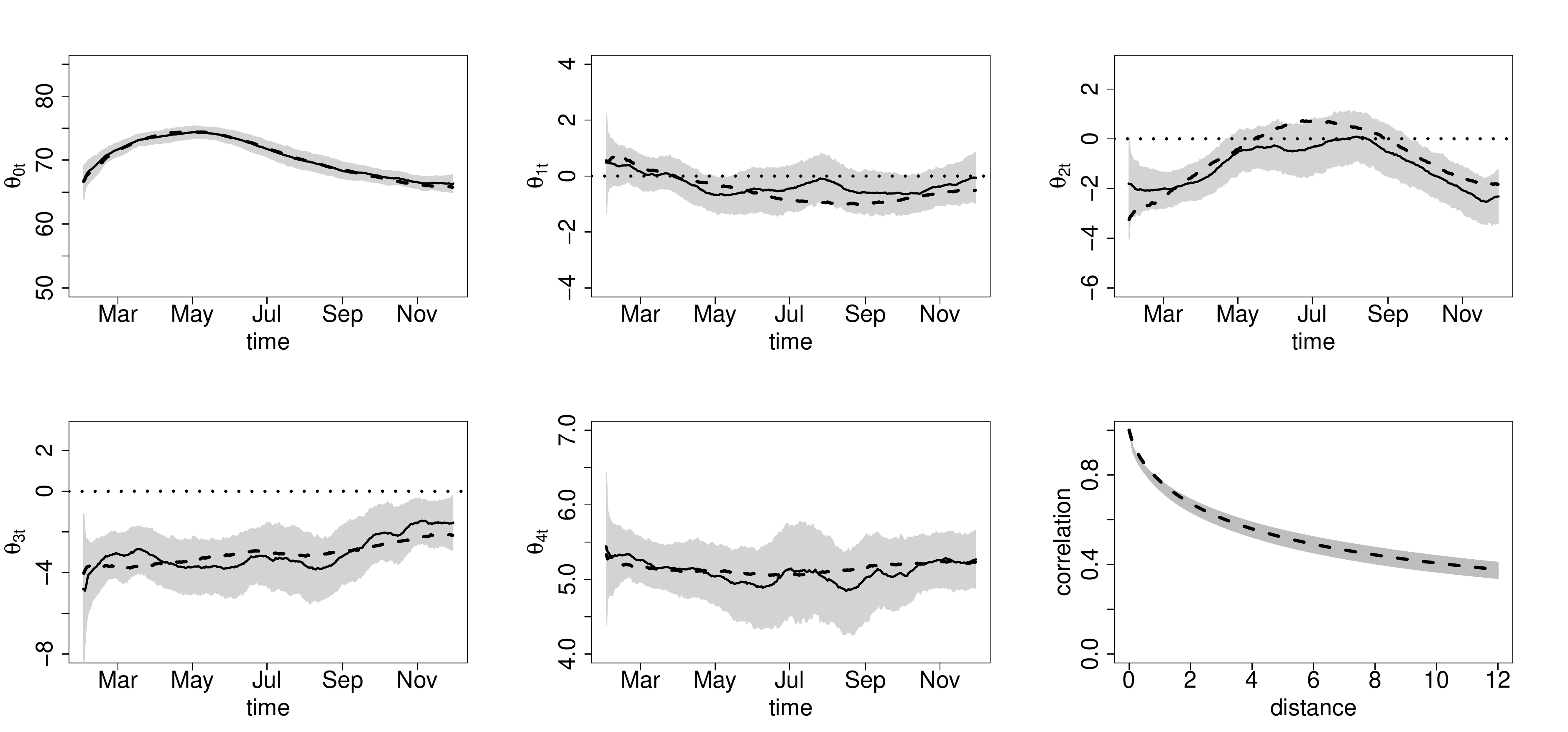}
\caption{Simulated Non-Gaussian data: Posterior median (solid line), $95\%$ credible intervals and true values (dashed line) for the dynamic coefficients and correlation function, estimated assuming the DynGLG model.}
\label{secNG1}
\end{figure}

\begin{figure}
\centering
\begin{tabular}{ccc}
\includegraphics[width=4.6cm]{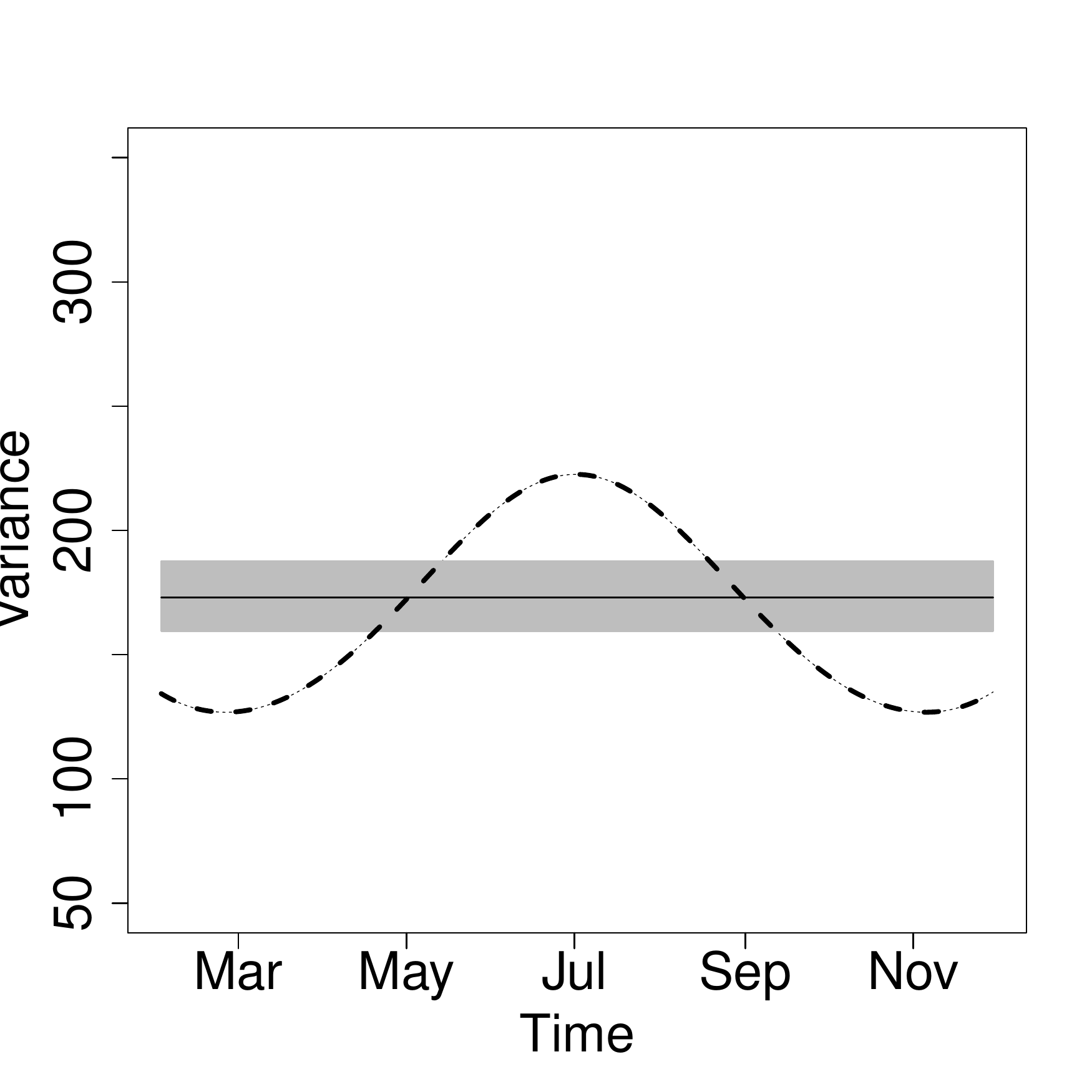}& 
\includegraphics[width=4.6cm]{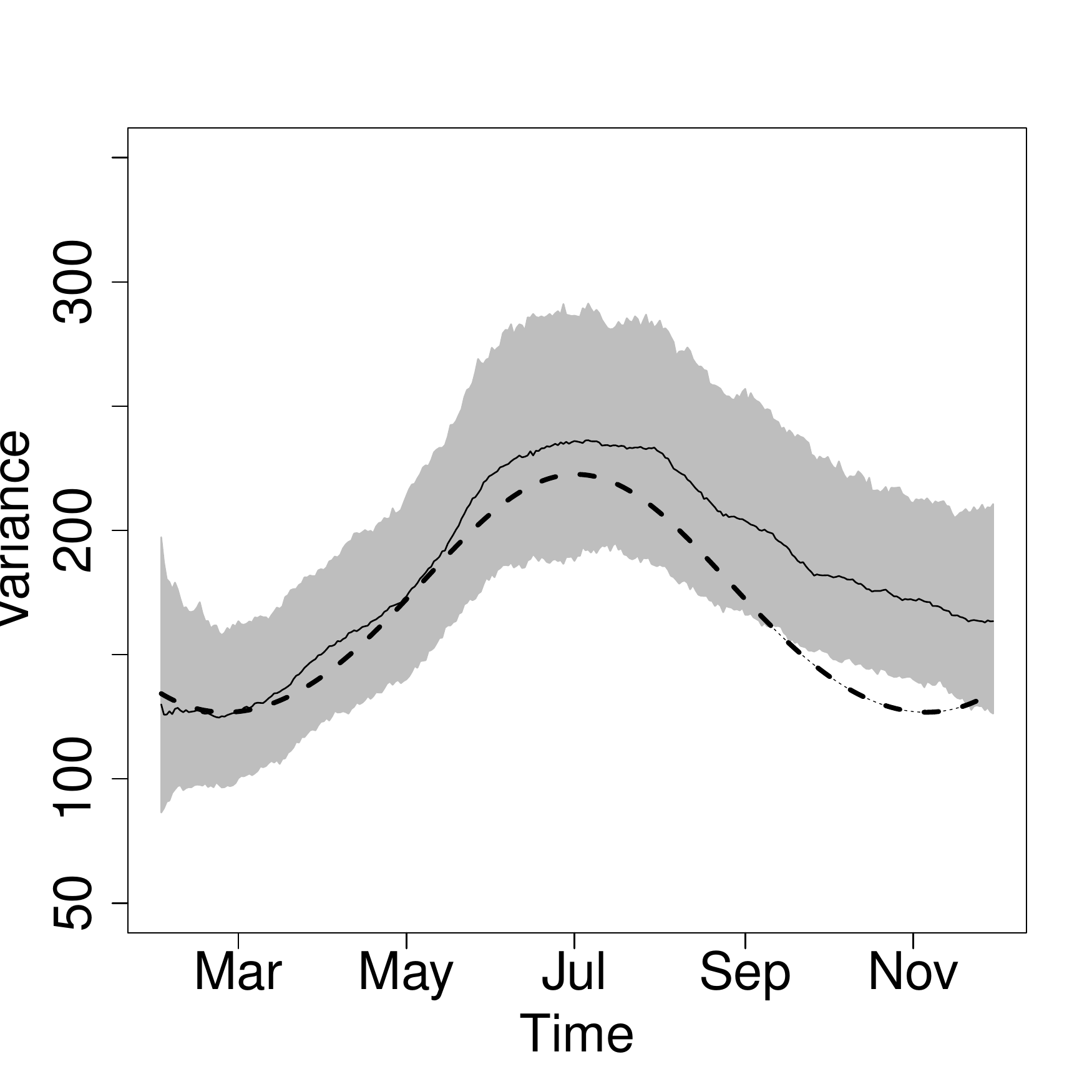} & \includegraphics[width=4.6cm]{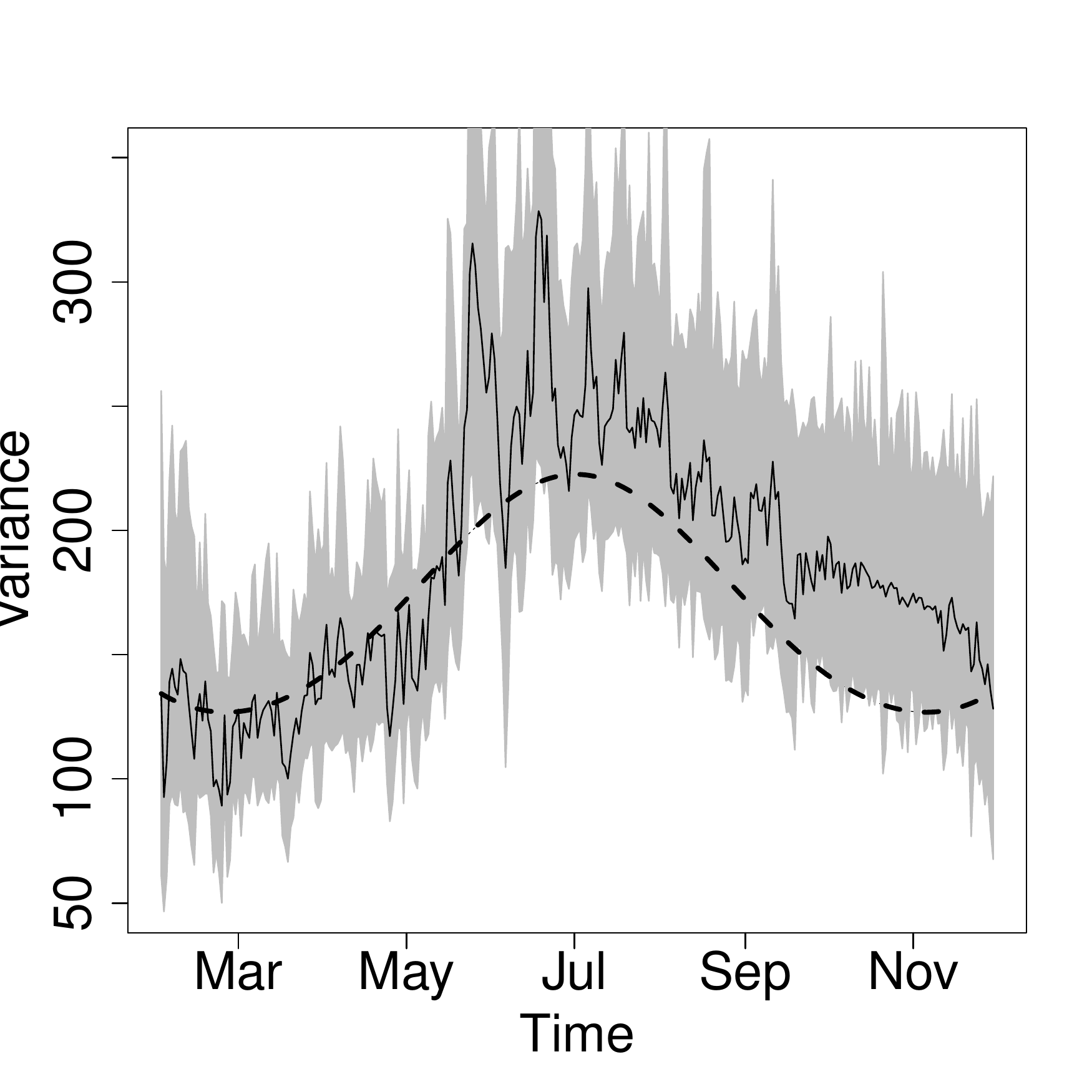}\\
(a) Gaussian & (b) DynGLG & (c) CovDynGLG
\end{tabular}
\caption{Simulated Non-Gaussian data: Posterior median (solid line), $95\%$ credible intervals and true values (dashed line) for the temporal variance, estimated assuming the Gaussian model, the DynGLG Model and the CovDynGLG Model.}
\label{simulNGvar}
\end{figure}

\begin{table}
     \caption{Model comparison based on the Interval Score (IS), the Log Predictive Score (LPS) and the Variogram Score of order 0.25 (VS-0.25) criteria for the predicted observations at the out-of-sample locations under all fitted models for the simulated Non-Gaussian dataset (DynGLG). The smallest measures are highlighted in boldface. \label{tabSimulapred}}
       \centering
 \fbox{   \begin{tabular}{cccccccc}
    \hline
  &  G  & ST & GLG & CovDyn & DynGLG & CovDynGLG & Full\\
    \hline
 IS &  54 & 52 & 46 & 49  &   { \bf 44} & {\bf 44} & 45\\
 LPS & 5407  & 5352 & 5047 & 5249  & { 4865} & {\bf 4860}& 4880\\
VS-0.25 & 7890  & 7959 & { 7539} & 7813  & { 7585}  &  {\bf 7537}& 7610\\
         \hline
    \end{tabular}}
\end{table}

\clearpage
\newpage
\bibliography{references}
\bibliographystyle{apalike}
\end{document}